\documentclass[iop,apj]{emulateapj}
\pdfoutput=1 
\usepackage{ifthen}
\usepackage{thumbpdf}
\usepackage{amsmath,amssymb}
\usepackage{graphicx,mathptmx}
\usepackage{natbib}
\usepackage[usenames]{color}
\usepackage{ifpdf}
\usepackage{xcolor}[2007/01/21]
\usepackage{refcount}
\usepackage{textcase}
\usepackage{epstopdf}

\newcommand{\beq}{
\begin{equation}
}
\newcommand{\eeq}{
\end{equation}
}
\newcommand{\beqa}{
\begin{eqnarray}
}
\newcommand{\eeqa}{
\end{eqnarray}
}

\newcounter{emulateapj}
\makeatletter
\@ifclassloaded{emulateapj}
{
\setcounter{emulateapj}{1}
}
{
\setcounter{emulateapj}{0}
}
\makeatother

\newcommand{\units}[1]  {\ensuremath{\mathrm{\ {#1}}}}
\newcommand{\msun}     {\ensuremath{{{M}}_{\scriptscriptstyle \odot}}}

\newcommand{\kms}      {\ensuremath{\mathrm{km\ s^{-1}}}}

\newcommand{\ergs}     {\ensuremath{\mathrm{erg\ s^{-1}}}}

\newcommand{\msigma}   {\ensuremath{M}{--}\ensuremath{\sigma}}

\newcommand{\mbh}      {\ensuremath{M_{\mathrm{BH}}}}

\newcommand{\lhardedd}{\ensuremath{L_{2\textrm{--}10} / L_\mathrm{Edd}}}
\newcommand{\lhard}{\ensuremath{L_{2\textrm{--}10}}}
\newcommand{\ledd}{\ensuremath{L_\mathrm{Edd}}}
\newcommand{\lbol}{\ensuremath{L_\mathrm{Bol}}}
\newcommand{\lallband}{\ensuremath{L_{0.3\textrm{--}10}}}
\newcommand{\lulx}{\ensuremath{L_\mathrm{ULX}}}
\providecommand{\ion}[2]{#1$\;$\textsmaller{\@Roman{#2}}}

\def\spose#1{\hbox to 0pt{#1\hss}}
\newcommand{\lta}{\mathrel{\spose{\lower 3pt\hbox{$\mathchar"218$}}
      \raise 2.0pt\hbox{$\mathchar"13C$}}}
\newcommand{\gta}{\mathrel{\spose{\lower 3pt\hbox{$\mathchar"218$}}
      \raise 2.0pt\hbox{$\mathchar"13E$}}}
\def\simlt{\mathrel{\rlap{\lower 3pt\hbox{$\sim$}}\raise 2.0pt\hbox{$<$}}}
\def\simgt{\mathrel{\rlap{\lower 3pt\hbox{$\sim$}} \raise 2.0pt\hbox{$>$}}}

\newcommand{\combh}{2009ApJ...706..404G}

\definecolor{KayhanCiteColor}{rgb}{0,0.08,0.35}
\definecolor{KayhanURLColor}{rgb}{0,0.08,0.35}
\definecolor{KayhanLinkColor}{rgb}{0,0.08,0.35}
\definecolor{KayhanPageColor}{rgb}{0,0.08,0.35}
\definecolor{medred}{rgb}{0.75,0.0,0.0}

\shorttitle{$L_\mathrm{X}$ Survey of SMBHs with Measured \mbh}
\shortauthors{G\"{u}ltekin et al.}

\submitted{Accepted by The Astrophysical Journal}

\usepackage[pdftitle={A Chandra Survey of Supermassive Black Holes
  with Dynamical Mass Measurements}, pdfauthor={Kayhan Gultekin},
  pdfsubject={Astrophysics}, pdfkeywords={accretion --- accretion
  disks --- black hole physics --- galaxies: general ---
  galaxies:nuclei --- X-rays: galaxies ---X-rays: general},
  linktocpage, colorlinks=true, linkcolor={KayhanLinkColor},
  urlcolor={KayhanURLColor}, citecolor={KayhanCiteColor},
  hyperfootnotes=false]{hyperref}
\usepackage[figure,figure*]{hypcap}
\usepackage{atveryend}
\usepackage[
  open,
  openlevel=3,
  atend
]{bookmark}[2010/04/08]
\begin{document}

\label{firstpage}
 
\title{A \emph{Chandra} Survey of Supermassive Black Holes with Dynamical Mass Measurements}

\author{Kayhan G\"{u}ltekin\altaffilmark{1}}
\author{Edward M. Cackett\altaffilmark{2}}
\author{Jon M. Miller\altaffilmark{1}}
\author{Tiziana Di Matteo\altaffilmark{3}}
\author{Sera Markoff\altaffilmark{4}}
\author{Douglas O. Richstone\altaffilmark{1}}
\affil{\altaffilmark{1}Department of Astronomy, University of Michigan, Ann Arbor, MI, 48109.  Send correspondence to \href{mailto:kayhan@umich.edu}{kayhan@umich.edu}.}
\affil{\altaffilmark{2}Institute of Astronomy, University of
 Cambridge, Madingley Rd, Cambridge, CB3 0HA, UK; Wayne State
 University, Department of Physics \& Astronomy, 666 W Hancock St,
 Detroit, MI 48201, USA}
\affil{\altaffilmark{3}McWilliams Center for Cosmology, Physics Department, Carnegie Mellon University, Pittsburgh, PA, 15213.}
\affil{\altaffilmark{4}Astronomical Institute `Anton Pannekoek,' Science Park 904, 1098XH Amsterdam, the Neterlands.}

\begin{abstract}
\hypertarget{abstract}{}%
We present \emph{Chandra} observations of 12 galaxies that contain
supermassive black holes with dynamical mass measurements.  Each
galaxy was observed for 30 ksec and resulted in a total of 68 point
source detections in the target galaxies including supermassive black
hole sources, ultraluminous X-ray sources, and extragalactic X-ray
binaries.  Based on our fits of the X-ray spectra, we report fluxes,
luminosities, Eddington ratios, and slope of the power-law spectrum.
Normalized to the Eddington luminosity, the 2--10 keV band X-ray
luminosities of the SMBH sources range from $10^{-8}$ to $10^{-6}$,
and the power-law slopes are centered at $\sim2$ with a slight trend
towards steeper (softer) slopes at smaller Eddington fractions,
implying a change in the physical processes responsible for their
emission at low accretion rates.  We find 20 ULX candidates, of which
six are likely ($>90\%$ chance) to be true ULXs.  The most promising
ULX candidate has an isotropic luminosity in the 0.3--10 keV band of
$1.0_{-0.3}^{+0.6} \times 10^{40}\ \ergs$.

\bookmark[ rellevel=1,
keeplevel, dest=abstract ]{Abstract}
\end{abstract}

\section{Introduction}
\label{intro}

The prevalence of large black holes at the centers of galaxies
\citep[e.g.,][]{richstoneetal98} and their role as the central engines
of active galactic nuclei \citep[AGNs; e.g.,][]{rees84} has been well
appreciated.  The possibility of co\"evolution of black holes and
their host galaxies, particularly through self-regulated growth and
feedback from accretion-powered outflows \citep{sr98,
1999MNRAS.308L..39F}, has focused recent observational, theoretical,
and computational effort \citep[e.g.,][]{2007MNRAS.382.1415S, dsh05}.

The detailed microphysics at play in accretion-powered feedback,
however, is not yet understood.  Accretion onto a black hole is
thought to proceed via an accretion disk, which launches relativistic
jets and outflows from its inner regions \citep[e.g.,][]
{1978PhyS...17..185L, 1982MNRAS.199..883B, 1977MNRAS.179..433B}.  In
the same way that accretion disk properties appear mostly to scale
with black hole mass, so do the length- and time-scales of jets from
stellar-mass black holes to supermassive black holes (SMBHs).

The radio emission, X-ray emission, and mass of an accreting black
hole are empirically related through what is sometimes called ``the
fundamental plane of black hole accretion'' \citep{merlonietal03,
fkm04}.  Radio emission coming from synchrotron emission in the jets
clearly must depend on the amount of matter accreted towards the black
hole, at which point it is turned into an outflow.  X-ray emission has
several possible origins, including from the accretion disk corona, or
from jets.  The X-ray emission depends on the accretion rate and the
source compactness, itself a function of the size of and thus mass of
the black hole.  Thus, it is natural to expect some form of mutual
covariance among these three quantities.  The relatively small scatter
in the relation, which spans over 8 orders of magnitude in black hole
mass, however, suggests that accretion and outflows are self-regulated
in a similar way across in all black holes.

Since the discovery of the fundamental plane \citep{merlonietal03,
fkm04}, there has been a concerted effort to understand it, both from
an observational perspective, which focuses on the universality and
extent of the relation, and also from a theoretical perspective, which
has focused on understanding the mechanisms of jet production,
constraining which radiative mechanism (or mechanisms) drive the
correlation and at what efficiency for jets and accretion inflow
processes \citep{merlonietal06, 2006A&A...456..439K,
2006ApJ...645..890W, 2008ApJ...688..826L, 2009ApJ...703.1034Y,
2011arXiv1105.3211P, 2011MNRAS.tmp..835D}.

In \citet{\combh}, we performed an archival \emph{Chandra} analysis of
all SMBHs with primary, direct black-hole-mass measurements and
available radio data.  We make a distinction between mass measurements
that are ``primary and direct'' such as stellar dynamical
\citep[e.g.,][]{2009ApJ...695.1577G}, gas dynamical
\citep[e.g.,][]{2001ApJ...555..685B}, and megamaser measurements
\citep[e.g.,][]{2011ApJ...727...20K} and reverberation mapping
measurements \citep[e.g.,][]{bentzetal06b}.  The reverberation mapping
technique is direct in that it directly probes a black hole's
surrounding environment but is secondary because it must be normalized
to the quiescent population through host-galaxy scaling relations
\citep{onkenetal04}.  There are several advantages to using only
sources with primary black hole masses.  First, statistical techniques
that make use of measurement errors can be faithfully used with the
actual errors on black hole mass rather than on an inferred scatter to
a host-galaxy scaling relation \citep[e.g.,][]{2009ApJ...698..198G}.
Second, by focusing on black holes with known, dynamical masses, we
can measure true Eddington fractions and understand how accretion
processes depend on this.  Third, and perhaps most importantly, the
results can be used to directly calibrate an estimator for black hole
mass based on nuclear X-ray and radio measurements.

In this paper we make the first step in completing the X-ray and radio
survey of black holes with primary, direct mass measurements with 12
new \emph{Chandra} observations.  This paper will be complimented with
one using Extended Very Large Array (EVLA) observations of the same
sources.  While the sample was designed to study the fundamental
plane, a \emph{Chandra} survey of SMBHs with dynamical mass
measurements provides an interesting look at the X-ray properties of
low luminosity AGNs (LLAGNs).  LLAGNs do not appear to be scaled-down
versions of more luminous Seyferts and quasars, but instead display
very different accretion physics \citep{2008ARA&A..46..475H}.  X-ray
emission is generally considered a probe of accretion power, wherever
it is ultimately produced.  Because of the expected differences in
physics at low accretion rates, X-ray studies of LLAGNs should provide
as direct a look as possible.

In this and other X-ray observational work on LLAGNs, a key measure of
the difference in their energetics is the how spectral shape changes
with accretion rate.  One simple, direct, and purely observational
diagnostic is the change in slope of hard X-ray emission (2--10 keV)
with hard X-ray Eddington fraction.  This has been examined in both
X-ray binaries \citep{2006ApJ...636..971C} and AGNs
\citep{2008ApJ...682...81S, 2009MNRAS.399..349G, 2009ApJ...690.1322W,
2009ApJ...705.1336C, 2011A&A...530A.149Y}.  One potential limitation
of the AGN studies is the use of secondary mass estimates in
determining Eddington fraction.  With our sample, we are not limited
by this, and we are able to test down to very low accretion rates.

In Section \ref{obs} we describe the sample selection, observations,
data reduction, and spectral fitting.  In Section \ref{concl} we
detail the results of the SMBH sources as well as serendipitous
detection of ultra luminous X-ray sources (ULXs) and other point
sources in the target galaxies.

\section{Observations and Spectral Fitting} 
\label{obs}
In this section we describe our experimental method from sample
selection, though data reduction, point source detection, and spectral
extraction and fitting.  All data reduction was done with CIAO version
4.3 and calibration databases (CALDB) version 4.4.3, and spectral
fitting was done with XSPEC version 12 \citep{1996ASPC..101...17A}.
Reduction was done with the distributed Level 2 event files, which
were processed at different times but all between 2010 Apr 14
(NGC 4486A) and 2010 Dec 12 (NGC 4291).

\subsection{Sample Selection}
\label{sampleselection}
In \citet{\combh}, we analyzed archival \emph{Chandra} data of all
SMBHs with dynamical masses.  The parent sample was the list of
``secure'' black hole mass detections in \citet{2009ApJ...698..198G}.
Of the approximately 50 black holes in that list, 21 had only limits
on their X-ray luminosity or else were not observed with
\emph{Chandra} at all.  Of these 21, we identified 13 whose nuclear
flux could potentially be determined with a 30 ks \emph{Chandra}
observation.  The others had large amounts of contaminating hot gas or
existing archival data indicated that they were too faint to be
observed with 30 ks exposures.  We were granted 12 of these
observations with joint Extended Very Large Array time.  The rest
should be observed with \emph{X-ray Multi-mirror Mission--Newton},
which has a larger effective area, or are extremely faint objects that
have not been detected with deep \emph{Chandra} exposures.  Table
\ref{t:xray} summarizes target galaxies and masses of the central
black holes.  All but one galaxy are within 30 Mpc, and the black hole
masses span a wide range: $4.1 \times 10^6$--$8.0 \times 10^8\ \msun$.
The only galaxy with a nuclear activity classification is NGC 5576,
which is classified as a Low Ionization Narrow Emission Region (LINER)
with broad Balmer lines \citep{1997ApJS..112..391H,
2006A&A...455..773V}.

One potential selection effect that arises from requiring a dynamical
mass measurement for inclusion in our sample is that it is biased to
low Eddington rates.  In general, the methods used to measure black
holes in this sample are hampered by contamination from a strong AGN
contribution in the same bands used for the mass measurement.  For
this reason, the results and conclusions we draw from this study only
apply to the very low Eddington rates ($\sim
10^{-5}\textrm{--}10^{-9}$) that our data adequately cover.

    \begin{deluxetable}{lrlrlrlr}
    \tablecaption{\emph{Chandra} Observations}
    \tablewidth{0pt}
    \tablehead{
    \colhead{Obs ID} & \colhead{Exp.} & \colhead{Galaxy} & \colhead{Dist.} & \colhead{Ref.} & \colhead{\mbh} & \colhead{Ref.}& \colhead{$N_H$} \\
     \colhead{(1)} & \colhead{(2)} & \colhead{(3)} & \colhead{(4)}  & \colhead{(5)} & \colhead{(6)} & \colhead{(7)} & \colhead{(8)}}
    
    \startdata
     11775 & 30.04 &  NGC 1300  & 20.1 & D1 & $7.1_{-3.4}^{+6.7} \times 10^{7}$ & M1 & 2.53 \\
     11776 & 30.05 &  NGC 2748  & 24.9 & D1 & $4.7_{-3.9}^{+3.8} \times 10^{7}$ & M1 & 1.55 \\
     11777 & 29.55 &  NGC 2778  & 24.2 & D2 & $1.6_{-1.0}^{+0.9} \times 10^{7}$ & M2 & 1.61 \\
     11782 & 29.04 &  NGC 3384  & 11.7 & D2 & $1.8_{-0.2}^{+0.1} \times 10^{7}$ & M2 & 2.94 \\
     11778 & 30.16 &  NGC 4291  & 25.0 & D2 & $3.2_{-2.4}^{+0.8} \times 10^{8}$ & M2 & 2.87 \\
     11784 & 30.18 &  NGC 4459  & 17.0 & D3 & $7.4_{-1.4}^{+1.4} \times 10^{7}$ & M3 & 2.67 \\
     11783 & 29.05 &  NGC 4486A & 17.0 & D3 & $1.3_{-0.4}^{+0.4} \times 10^{7}$ & M4 & 2.00 \\
     11785 & 31.38 &  NGC 4596  & 18.0 & D4 & $8.4_{-2.5}^{+4.0} \times 10^{7}$ & M3 & 1.43 \\
     11779 & 33.08 &  NGC 4742  & 16.4 & D2 & $1.4_{-0.5}^{+0.4} \times 10^{7}$ & M5 & 3.43 \\
     11780 & 29.05 &  NGC 5077  & 44.9 & D5 & $8.0_{-3.3}^{+5.1} \times 10^{8}$ & M2 & 3.05 \\
     11781 & 30.05 &  NGC 5576  & 27.1 & D2 & $1.8_{-0.4}^{+0.3} \times 10^{8}$ & M7 & 2.44 \\
     11786 & 29.04 &  NGC 7457  & 14.0 & D2 & $4.1_{-1.6}^{+1.3} \times 10^{6}$ & M2 & 4.60 

    \enddata
    \label{t:xray}
    \tablecomments{We summarize the observations and the targets.  The
    columns are (1) the \emph{Chandra} observation identification
    number; (2) the exposure time in units of ks; (3) the targeted
    galaxy; (4) the distance of the galaxy in units of Mpc; (5) a
    reference code for the distance; (6) the mass of the central black
    hole in units of \msun\ with 1$\sigma$ uncertainties; (7) a
    reference code for the mass measurement; (8) the Galactic column
    towards each source in units of $10^{20}\ \mathrm{cm^{-2}}$
    \citep{2005A&A...440..767B}.  The distances have been scaled from
    the indicated literature values to a common Hubble parameter of $h
    = 0.7$, and NGC 4459 and NGC4486A have been set at a Virgo
    distance of 17.0 Mpc.}
    \tablerefs{ 
      (D1) \citealt{2005MNRAS.359..504A},
      (D2) \citealt{2001ApJ...546..681T},
      (D3) \citealt{2007ApJ...655..144M} adopted to 17.0 Mpc,
      (D4) \citealt{1988ngc..book.....T},
      (D5) \citealt{1989ApJS...69..763F} group distance,
      (M1) \citealt{2005MNRAS.359..504A}, 
      (M2) \citealt{2003ApJ...583...92G}, 
      (M3) \citealt{2001ApJ...550...65S}, 
      (M4) \citealt{nowaketal07}, 
      (M5) G\"ultekin et al.\ in preparation, 
      (M6) \citealt{2008A&A...479..355D}, 
      (M7) \citealt{2009ApJ...695.1577G}. 
    }
    \end{deluxetable}
\bookmarksetup{color=[rgb]{0,0,0.54}} 
\bookmark[
rellevel=1,
keeplevel,
dest=table.\getrefnumber{t:xray}
]{Table \ref*{t:xray}: Summary of observations and targets}
\bookmarksetup{color=[rgb]{0,0,0}}

\subsection{Point source detection}

To detect points sources in the field of each galaxy, we used the
{wavdetect} tool on the whole image in the whole \emph{Chandra} band
at full resolution, run with the {large\_detect.pl} wrapper script
provided by the ACIS team\footnote{See \href{http://goo.gl/hk7e6}
{http://goo.gl/hk7e6}}.  The wrapper script splits the image into
multiple, overlapping sub-images and runs wavdetect\footnote{It also
runs celldetect, but we only use the wavdetect result.} on each
region, identifying multiply detected sources in the overlapping
portions to minimize edge effects in the wavlet algorithm. Wavdetect
was used default settings for detection and background rejection
thresholds and with searches in wavelet radii (i.e., the ``scales''
parameter) of 1, 2, 4, 8, and 16 pixels. The whole image was used
rather than just the region near the galaxy because

\begin{figure*}[tbh]
\hypertarget{images}{}%
\centering
\includegraphics[width=0.90\columnwidth]{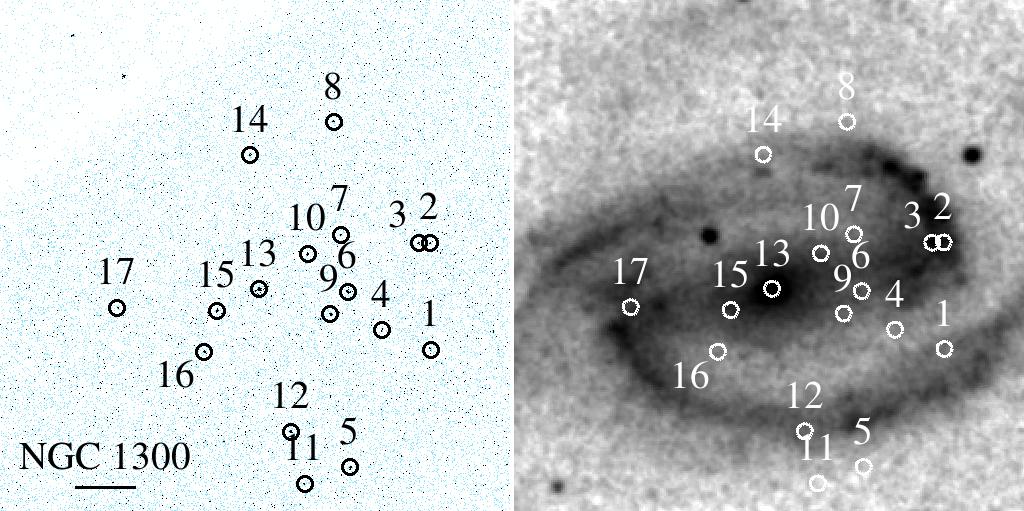}
\includegraphics[width=0.90\columnwidth]{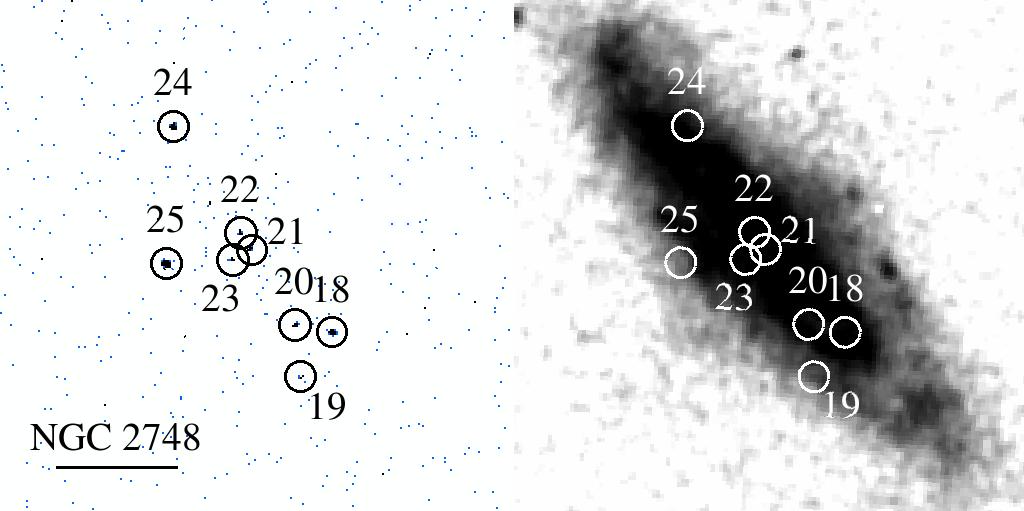}
\includegraphics[width=0.90\columnwidth]{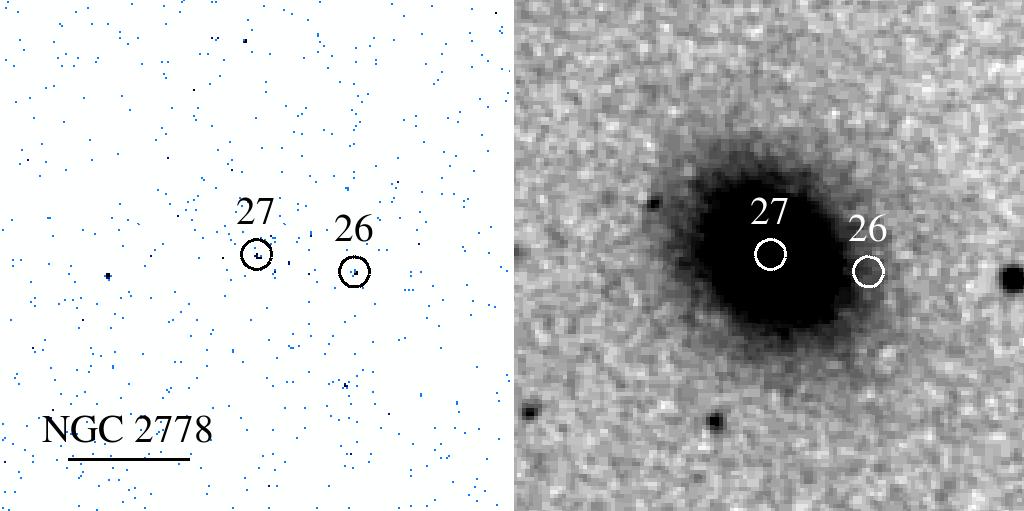}
\includegraphics[width=0.90\columnwidth]{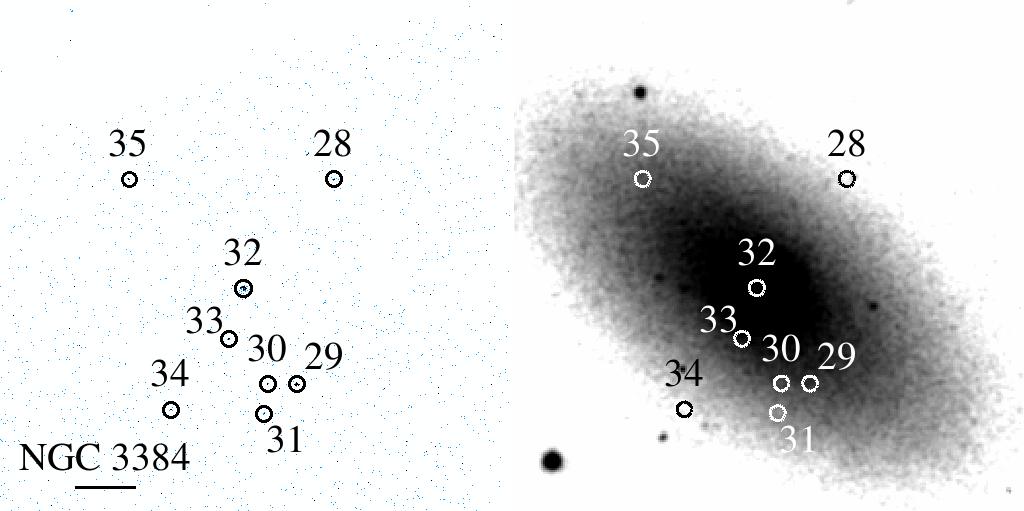}
\includegraphics[width=0.90\columnwidth]{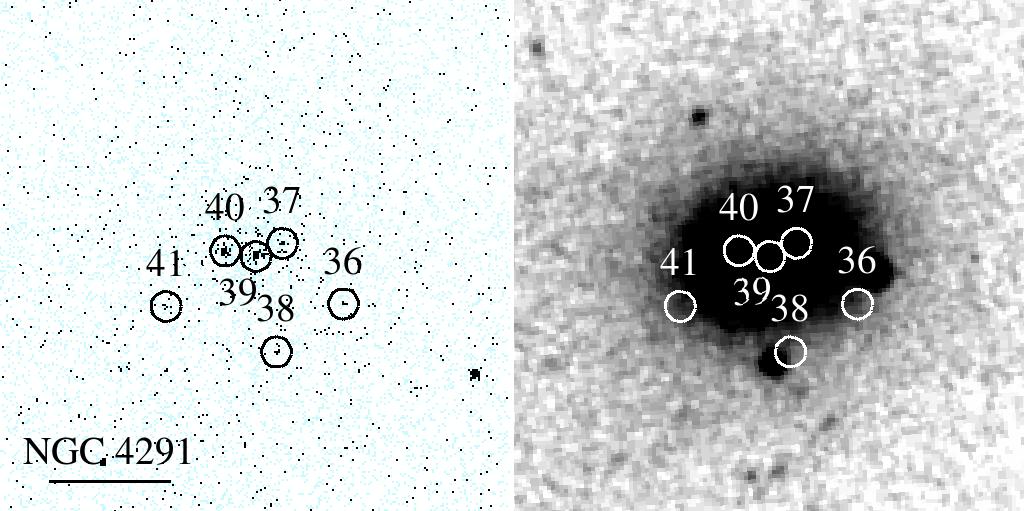}
\includegraphics[width=0.90\columnwidth]{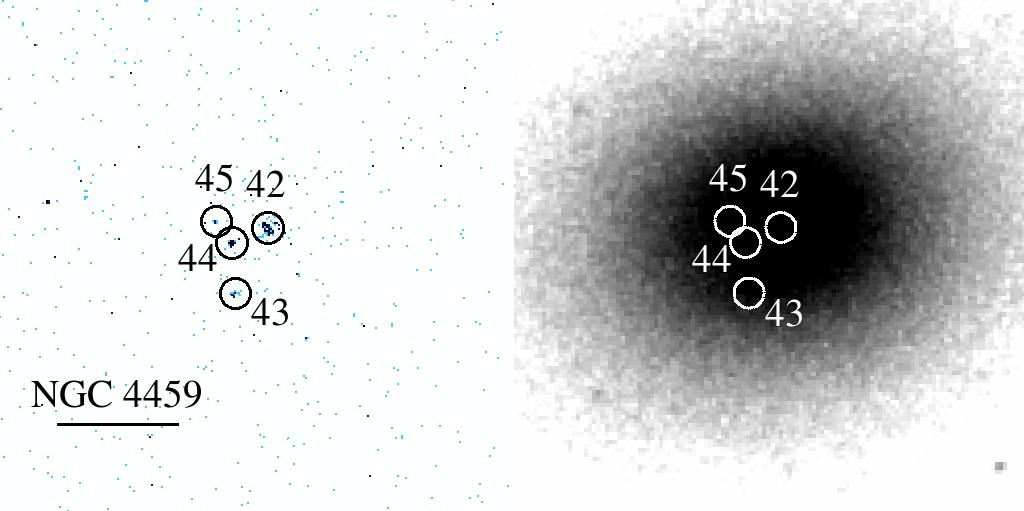}
\includegraphics[width=0.90\columnwidth]{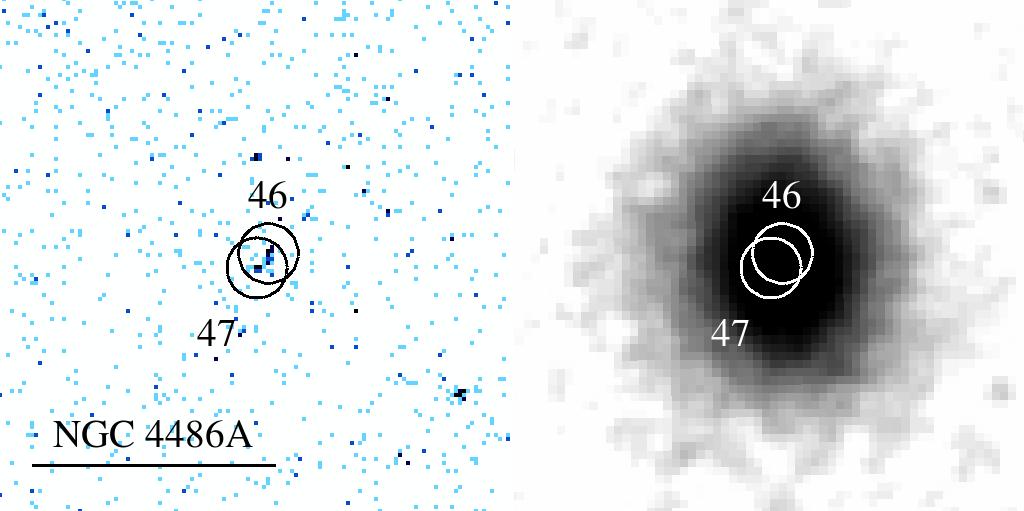}
\includegraphics[width=0.90\columnwidth]{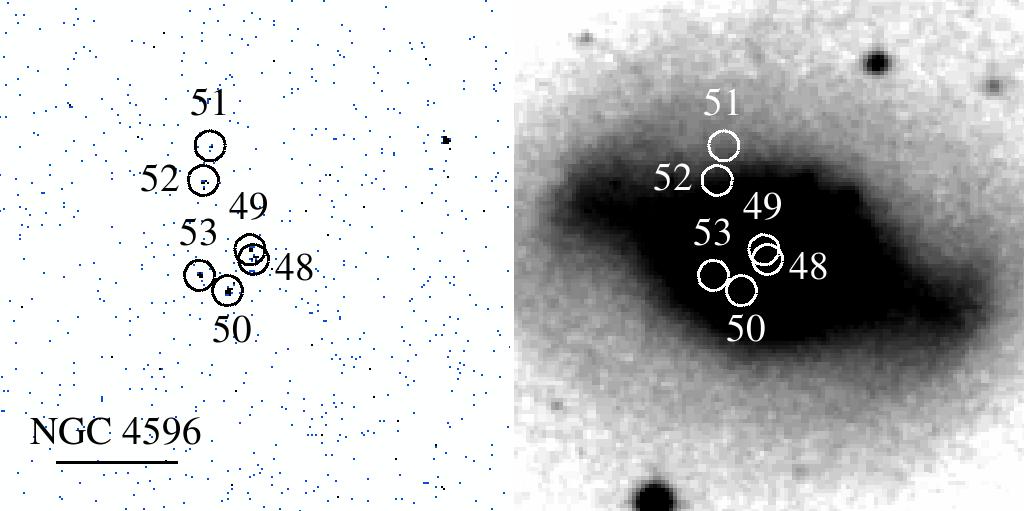}
\includegraphics[width=0.90\columnwidth]{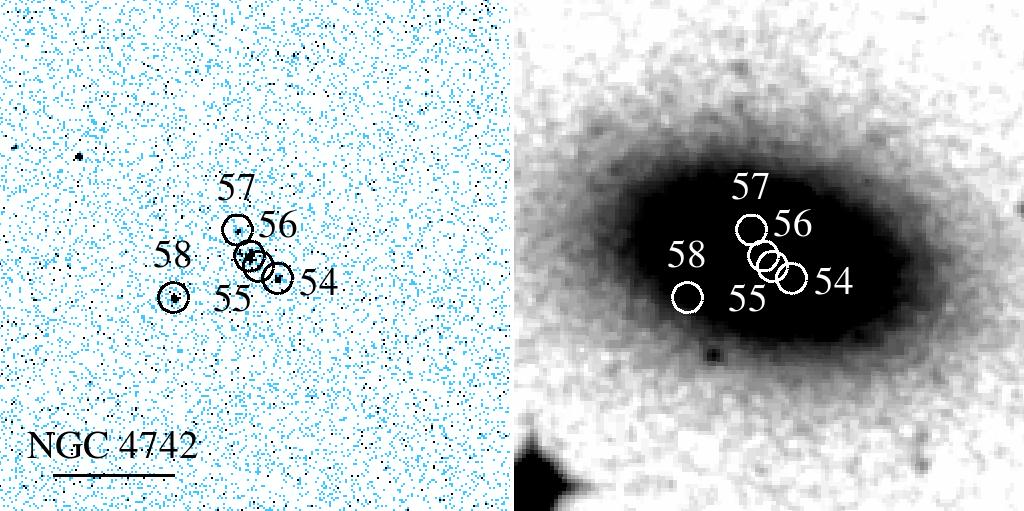}
\includegraphics[width=0.90\columnwidth]{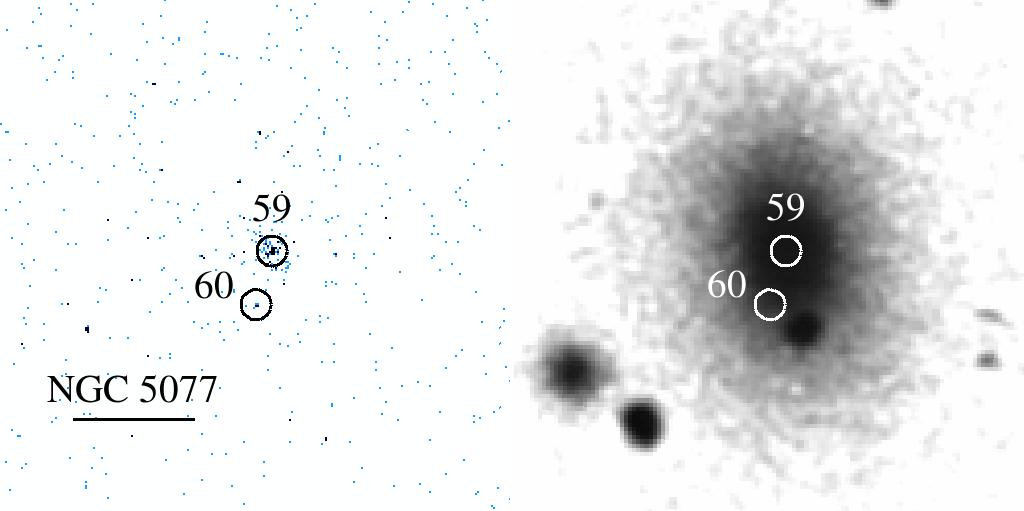}
\includegraphics[width=0.90\columnwidth]{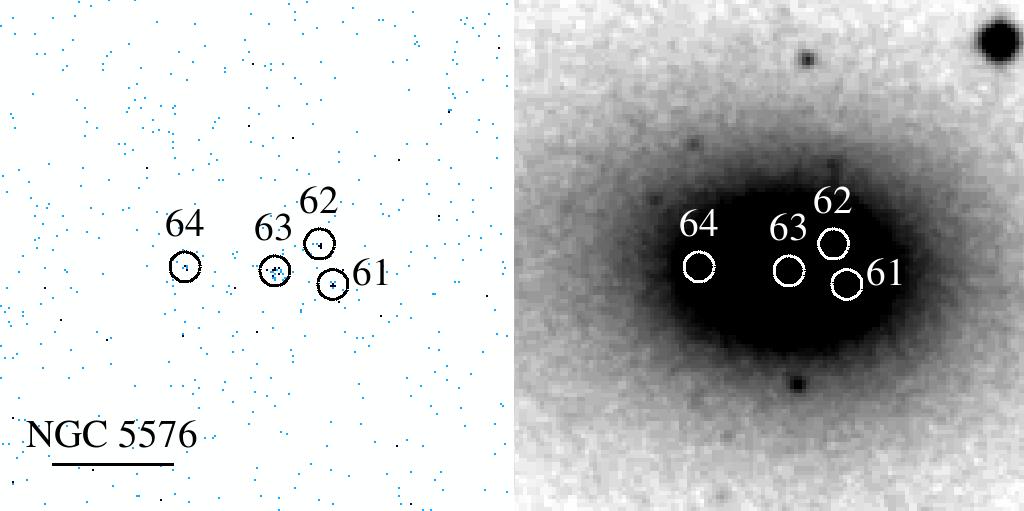}
\includegraphics[width=0.90\columnwidth]{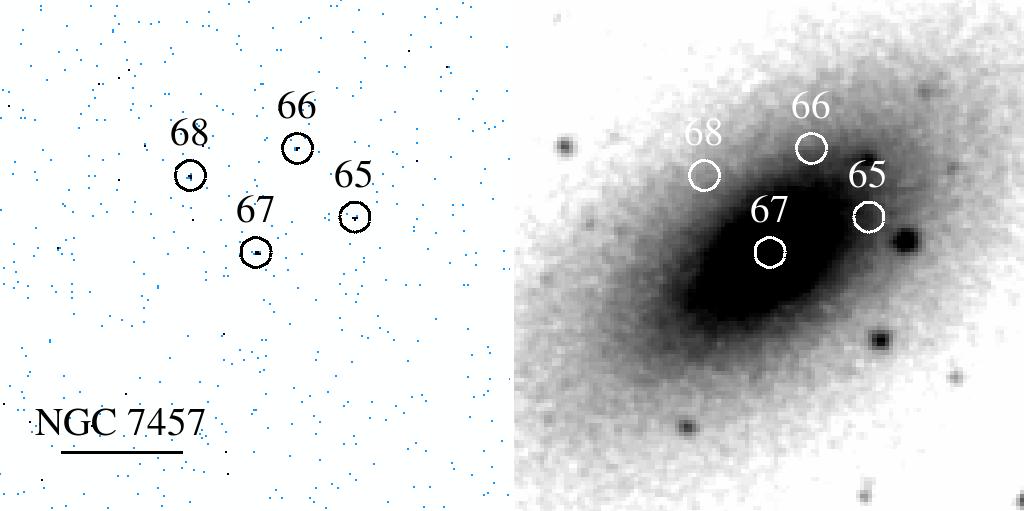}
\caption{Each pair of plots shows the \emph{Chandra} image
(\emph{left}) with detected point sources labeled with their running
identification number and the DSS image (\emph{right}) with locations
of \emph{Chandra} point sources labeled.  The circles are centered at
the location of the extraction region and are drawn with an
arbitrarily sized radius for clarity rather than indicating the size
of the extraction region.  The bar under the galaxy name is 30\arcsec\ 
long.}
\label{figimages}
\end{figure*}

 For each \emph{Chandra} image we considered only the point sources
that were located on the galaxy as determined by comparison to 2MASS
and DSS images of the galaxy.  The size of the galaxy was determined
manually for each galaxy by looking at the images and is roughly
comparable to using isophotes of $25\ \mathrm{mag\ arcsec^{-2}}$ in
the $B$-band or $20\ \mathrm{mag\ arcsec^{-2}}$ in the $K$ band.  The
range of values for diameters at a particular surface brightness in a
particular band for each galaxy could vary by as much as 10\%,
motivating our choice for a simple, manual estimate.  In total, we
found 68 point sources over the 12 images.  We list all point sources
detected in table \ref{t:ptsrcdetec} with a running identification
number, IAU approved source name, J2000 coordinates, and net count
rates in the 0.3--1, 1--2, and 2--10 keV bands.  We show positions of
all sources in the \emph{Chandra} images as well as locations on the
DSS images of the galaxy in Figure \ref{figimages}.

\bookmarksetup{color=[rgb]{0.54,0,0}}
\bookmark[rellevel=1,keeplevel,dest=images]{Fig \ref*{figimages}: X-ray and optical images.}
\bookmarksetup{color=[rgb]{0,0,0}}

\subsection{Astrometry}
Because we are principally interested in the nuclear X-ray source of
each galaxy, we have taken extra care in identifying which point
source is the nuclear source.  For most galaxies it was unambiguous:
there was only one point source consistent with the optical or
infrared center of the galaxy.  In three galaxies (NGC 4486A, NGC
4596, and NGC 4742) there were two X-ray point sources that were
consistent with the position of the galaxy's center.  In a fourth
galaxy (NGC 3384), there was faint emission that was not detected by
wavdetect as a point source, but which we considered for potential
confusion.  Typical separation of the confusing sources was 1 to
3\arcsec, which requires only modest absolute astrometric corrections.
For the four galaxies, we registered the \emph{Chandra} images to
Sloan Digital Sky Survey (SDSS) or Deep Near Infrared Survey (DENIS)
coordinates using background AGNs that appeared in both the X-ray and
optical/infrared images.  Each image had between 3 and 6 sources for
registration.  The optical/infrared center was obvious in each image
so that we could unambiguously determine which source was closer to
the galaxy center.  In the end, we were able to determine that the
\emph{Chandra} coordinates were sufficient for determining the nuclear
sources.  The nuclear source in each galaxy is identified in Table
\ref{t:ptsrcdetec} in the classification column as ``Nuc.''

One possible source of confusion is if an X-ray binary was located at
the position of the nuclear center.  
To estimate this amount of contamination, we adopt the
\citet{2003MNRAS.339..793G} universal cumulative luminosity function
of high-mass X-ray binaries (HMXBs) in a given galaxy as a function of
star formation rate, $\mathrm{SFR}$:
\beq
N(>L) = 5.4 (\mathrm{SFR} / \msun\ \units{yr^{-1}}) \left[ \left(L / 10^{38}\ \ergs\right)^{-0.61} - 0.038 \right],
\eeq
where the second term in the brackets comes from assuming an upper
limit of $L_X = 2.1 \times 10^{40}\ \ergs$ for a HMXB and can be
neglected for our calculations.  We take the luminsoity of each
nuclear source; e.g., for the dimmest source, $L_X = 3 \times 10^{38}\
\ergs$ so that $N(L > 3 \times 10^{38}\ \ergs) = 2.8 (\mathrm{SFR} /
\msun\ \units{yr^{-1}})$.
 We calculated the star formation rate, $\mathrm{SFR}$, assuming the
 \citep{2002MNRAS.332..283R} estimator,
\beq 
\mathrm{SFR} = 4.5 \times 10^{-44}\ \msun\ \units{yr^{-1}}
(L_\mathrm{FIR} / \ergs),
\eeq
where the far-infrared luminosity is calculated as $L_\mathrm{FIR} =
1.7 \lambda F_\lambda$ at $\lambda = 60\ \mu\mathrm{m}$.  This
estimator is based on the \citet{1998ARA&A..36..189K} prescritpion and
agrees with other, similar prescriptions
\citep[e.g.,][]{2003ApJ...586..794B}.  Using far-infrared may
overestimate the star-formation rate in early-type galaxies but is
conservative for our argument.  We based our calculations on the the
IRAS $60\ \mu\mathrm{m}$ flux density \citep{1989ApJS...70..329K} for
all galaxies except NGC 5077 for which we substituted the MIPS $70\
\mu\mathrm{m}$ flux density \citep{2009ApJ...695....1T}.  The galaxy
with the highest star-formation rate in our sample is NGC 2748 with
$\mathrm{SFR} = 2.0\ \msun\ \units{yr^{-1}}$, and all but 4 have
$\mathrm{SFR} < 0.1\ \msun\ \units{yr^{-1}}$.  If we assume that the
X-ray binaries follow the light in the galaxy, then we can estimate
the number of potential contaminating sources.  This is an
overestimate of central contamination probability for galaxies like
NGC 1300 for which X-ray binaries follow the high rate of star
formation in the spiral arms.  The starlight in the central 1\arcsec,
roughly our astrometric uncertainty, ranges from approximately 10\%
for NGC 1300 to less than 1\%.  Assuming 10\% for all galaxies, we can
calculate the expected number of contaminating sources at the center
each galaxy.  The values range from 0.04 for NGC 2748 to less than
0.01 for the majority.  These numbers are conservative because the
actual amount of light at the center of each galaxy is less 10\%, and
the star formation rate may be over-estimated in the early-type
galaxies.  

For early-type galaxies, a greater concern is the chance positioning
of a low-mass X-ray binary (LMXB), which generally traces an older
population \citep{2004ApJ...611..846K}.  In this case, we use the
results of \citet{2004ApJ...611..846K}, who found that the cumulative
number of LMXBs greater than a luminosity, $L_X$ to scale as $N(>L_X)
\sim L_X^{-1}$, with a total luminosity of the LMXB population scaling
with the galaxy luminosity $L_K$ as $L_\mathrm{pop} = (0.2 \pm 0.08)
\times 10^{30}\ \ergs L_K / L_{K{\scriptscriptstyle \odot}}$.  This is
the most concern for NGC 5077, which has the largest total $K$-band
luminosity of $L_K = 8.5 \times 10^{10}\ L_{K{\scriptscriptstyle
\odot}}$ \citep{2006AJ....131.1163S}, and for NGC 4486A, which has the
lowest nuclear luminosity ($L_X = 2.5 \times 10^{38}\ \ergs$) and a
$K$-band luminosity of $L_K = 1.5 \times 10^{10}\
L_{K{\scriptscriptstyle \odot}}$ \citep{2006AJ....131.1163S}.  Again
assuming that LMXBs follow the light, the expected number of LMXBs
that would be as bright or brighter than the identified nuclear
sources is less than 0.005 and 0.04 for NGC 5077 and NGC 4486A,
respectively.  It is lower for all other galaxies.  Because each
galaxy type is predominantly affected by contamination from only one
of HMXBs or LMXBs and not both, the combined expected contamination by
both are then all less than 0.05.  Thus we expect not to have
incorrectly identified an X-ray binary as an SMBH source.

\subsection{Spectral reduction}
We followed the standard pipeline in reduction of all data sets, using
the most recent \emph{Chandra} data reduction software package (CIAO
version 4.3) and calibration databases (CALDB version 4.4.3).  There
was no significant background flaring, so there was no need for
filtering.  We used the CIAO tool psextract to extract the
point-source spectra.  Since our observations all used the Advanced
CCD Imaging Spectrometer (ACIS), we ran psextract with the mkacisrmf
tool to create the response matrix file (RMF) and with mkarf set for
ACIS ancillary response file (ARF) creation.  Source regions were
circles with centers at the coordinates given by wavdetect.  The radii
of the regions were the semimajor axes of the wavdetect error
ellipses.  This accounts for two effects: (1) the uncertainty in the
location of the source and (2) the degradation in the point spread
function off axis.  For background regions, we used annuli with inner
radii just larger than the source region radius and outer radii that
was typically 14 pixels larger.

\subsection{Spectral fitting}
We modeled the reduced spectra using XSPEC12
\citep{1996ASPC..101...17A}.  For some sources we used $\chi^2$
statistics and for others $C$-stat statistics
\citep{1979ApJ...228..939C}.  The decision on which statistics to use
was based on the number of photons available for binning in energy
bands.  If binning the spectra in energy so that each bin contained a
minimum of 20 counts resulted in five or more bins, we used $\chi^2$
statistics; otherwise we used $C$-stat statistics with unbinned
spectra.  There is necessarily a loss of information when binning the
spectra, but we found when using both types of statistics on sources
that had enough counts to support it, the resulting parameter
estimates were always consistent within 1$\sigma$.  We report the
results from $\chi^2$ statistics on account of the intuitive nature of
the goodness-of-fit with $\chi^2$ statistics.

All spectra were modeled with a photoabsorbed power-law model with the
intrinsic flux modeled directly as one of the parameters, i.e., the
XSPEC model used was {\tt phabs(cflux * powerlaw)}, with the {\tt
cflux} component normalized to the 2--10 keV band.  The spectra were
fitted from 0.3 to 10 keV.  While most of the photon counts above
$\sim7$ keV are probably dominated by background, the fitting
methodology takes this into account.  We tested our results for
sensitivity to the adopted upper energy cutoff and found that as long
it was greater than 5 keV, our results were robust in the sense that
they changed by far less than the 1$\sigma$ uncertainties.  When
fitting the spectrum, we set a hard minimum for the absorption to be
the Galactic value for $N_\mathrm{H}$ towards each source
\citep{2005A&A...440..767B}, but in calculating the uncertainty in the
column, we allowed it to drop below this value.  The results of the
spectral fits are presented in Table \ref{t:spectralfits} with
best-fit parameters and 1$\sigma$ (68\%) uncertainties for
$N_\mathrm{H}$, $\log F_{2\textrm{--}10}$, and $\Gamma$ (the power-law
index).  Some sources did not have sufficient counts to produce
reliable fits and are not included in the table.  Fits to three
sources (5, 14, and 62) were unconstrained on at least the spectral
index parameter, $\Gamma$, and we consider these fits approximate.

    \tabletypesize{\scriptsize}
    \def\arraystretch{1.200}
    \begin{deluxetable}{lrlrrrr}
    \tablecaption{Results of spectral fits}
    \tablewidth{0pt}
    \tablehead{
    \colhead{Galaxy} & \colhead{ID} & \colhead{Class.} & \colhead{$N_H$} & \colhead{$\log{F_{2\textrm{--}10}}$} & \colhead{$\Gamma$} & \colhead{$\chi^2/\mathrm{d.o.f.}$} \\
    \colhead{(1)} & \colhead{(2)} & \colhead{(3)} & \colhead{(4)} & \colhead{(5)} & \colhead{(6)} & \colhead{(7)}
    }
    \startdata

NGC1300 & 3 &  &  $0.0_{-0.0}^{+0.3}$ & 
$-15.0_{-0.5}^{+0.3}$ &
$2.6_{-0.6}^{+1.6}$ & 
\dots
\\

\dots & 5 &  &  $4.4_{-0.7}^{+0.4}$ & 
$-14.5_{-0.2}^{+0.2}$ &
$10.0_{\phantom{-0.0}}^{\phantom{+0.0}}$ & 
\dots
\\

\dots & 6 & ULX &  $0.6_{-0.2}^{+0.3}$ & 
$-14.6_{-0.2}^{+0.2}$ &
$4.3_{-0.9}^{+1.1}$ & 
\dots
\\

\dots & 9 & ULX &  $0.9_{-0.8}^{+0.9}$ & 
$-15.5_{-0.9}^{+0.5}$ &
$5.9_{-3.4}^{+4.0}$ & 
\dots
\\

\dots & 10 &  &  $0.1_{-0.1}^{+0.3}$ & 
$-15.0_{-0.6}^{+0.4}$ &
$2.8_{-0.7}^{+1.7}$ & 
\dots
\\

\dots & 12 & ULX &  $0.3_{-0.2}^{+0.3}$ & 
$-15.1_{-0.5}^{+0.4}$ &
$3.8_{-1.2}^{+1.7}$ & 
\dots
\\

\dots & 13 & Nuc. &  $0.0_{-0.0}^{+0.1}$ & 
$-13.0_{-0.1}^{+0.1}$ &
$0.4_{-0.2}^{+0.3}$ & 
4.51/ 3
\\

\dots & 14 &  &  $0.6_{-0.6}^{+0.4}$ & 
$-14.8_{\phantom{-0.0}}^{\phantom{+0.0}}$ &
$4.1_{\phantom{-0.0}}^{\phantom{+0.0}}$ & 
\dots
\\

NGC2748 & 18 & ULX &  $1.6_{-0.5}^{+0.6}$ & 
$-13.2_{-0.1}^{+0.1}$ &
$2.3_{-0.7}^{+0.7}$ & 
2.09/ 3
\\

\dots & 20 &  &  $0.1_{-0.1}^{+0.2}$ & 
$-14.4_{-0.3}^{+0.2}$ &
$1.8_{-0.5}^{+0.9}$ & 
\dots
\\

\dots & 21 & Nuc. &  $1.0_{-0.5}^{+1.2}$ & 
$-15.0_{-0.6}^{+0.5}$ &
$5.0_{-2.0}^{+4.1}$ & 
\dots
\\

\dots & 22 & ULX &  $1.0_{-0.4}^{+0.6}$ & 
$-15.0_{-0.4}^{+0.3}$ &
$5.1_{-1.6}^{+2.3}$ & 
\dots
\\

\dots & 23 & ULX &  $2.3_{-1.0}^{+1.9}$ & 
$-14.4_{-0.4}^{+0.2}$ &
$5.1_{-1.7}^{+6.4}$ & 
\dots
\\

\dots & 24 & ULX &  $1.6_{-0.3}^{+0.4}$ & 
$-13.5_{-0.1}^{+0.1}$ &
$3.0_{-0.5}^{+0.5}$ & 
\dots
\\

\dots & 25 & ULX &  $0.2_{-0.1}^{+0.1}$ & 
$-13.7_{-0.1}^{+0.1}$ &
$2.9_{-0.4}^{+0.4}$ & 
6.23/ 9
\\

NGC2778 & 26 & ULX &  $0.2_{-0.2}^{+0.2}$ & 
$-14.5_{-0.4}^{+0.3}$ &
$2.4_{-0.8}^{+1.1}$ & 
\dots
\\

\dots & 27 & Nuc. &  $0.2_{-0.2}^{+0.3}$ & 
$-15.7_{-0.8}^{+0.6}$ &
$4.6_{-1.4}^{+2.6}$ & 
\dots
\\

NGC3384 & 29 &  &  $0.0_{-0.0}^{+0.2}$ & 
$-13.3_{-0.3}^{+0.2}$ &
$1.0_{-0.3}^{+0.7}$ & 
\dots
\\

\dots & 30 & ULX &  $0.6_{-0.3}^{+0.4}$ & 
$-15.2_{-0.4}^{+0.4}$ &
$4.1_{-1.4}^{+1.9}$ & 
\dots
\\

\dots & 32 & Nuc. &  $0.1_{-0.1}^{+0.2}$ & 
$-13.8_{-0.3}^{+0.2}$ &
$2.0_{-0.4}^{+0.7}$ & 
\dots
\\

\dots & 34 & ULX &  $0.8_{-0.5}^{+0.8}$ & 
$-14.9_{-0.7}^{+0.5}$ &
$3.8_{-1.8}^{+3.3}$ & 
\dots
\\

\dots & 35 & ULX &  $0.3_{-0.2}^{+0.3}$ & 
$-15.2_{-0.4}^{+0.4}$ &
$4.2_{-1.4}^{+1.8}$ & 
\dots
\\

NGC4291 & 36 &  &  $0.1_{-0.1}^{+0.3}$ & 
$-14.5_{-0.5}^{+0.3}$ &
$1.8_{-0.6}^{+1.2}$ & 
\dots
\\

\dots & 37 & ULX &  $1.6_{-0.6}^{+1.1}$ & 
$-14.8_{-0.3}^{+0.2}$ &
$6.1_{-1.7}^{+2.6}$ & 
\dots
\\

\dots & 38 &  &  $0.6_{-0.5}^{+2.2}$ & 
$-14.0_{-0.4}^{+0.3}$ &
$1.4_{-0.9}^{+2.5}$ & 
\dots
\\

\dots & 39 & Nuc. &  $0.1_{-0.1}^{+0.1}$ & 
$-13.9_{-0.1}^{+0.1}$ &
$1.7_{-0.4}^{+0.4}$ & 
\dots
\\

\dots & 40 & ULX &  $0.0_{-0.0}^{+0.1}$ & 
$-13.6_{-0.1}^{+0.1}$ &
$1.4_{-0.3}^{+0.3}$ & 
1.07/ 3
\\

\dots & 41 &  &  $1.1_{-0.7}^{+2.2}$ & 
$-15.6_{-0.7}^{+0.5}$ &
$6.3_{-2.7}^{+3.7}$ & 
\dots
\\

NGC4459 & 42 & Nuc. &  $0.0_{-0.0}^{+0.1}$ & 
$-13.6_{-0.1}^{+0.1}$ &
$1.8_{-0.2}^{+0.3}$ & 
4.28/ 4
\\

\dots & 43 & ULX &  $1.7_{-0.8}^{+0.9}$ & 
$-14.6_{-0.2}^{+0.2}$ &
$4.7_{-1.7}^{+2.1}$ & 
\dots
\\

\dots & 44 &  &  $0.3_{-0.1}^{+0.2}$ & 
$-13.6_{-0.1}^{+0.1}$ &
$1.8_{-0.3}^{+0.4}$ & 
\dots
\\

NGC4486A & 46 &  &  $0.4_{-0.2}^{+0.3}$ & 
$-14.2_{-0.2}^{+0.2}$ &
$2.5_{-0.6}^{+0.8}$ & 
\dots
\\

\dots & 47 & Nuc. &  $0.3_{-0.3}^{+0.6}$ & 
$-14.4_{-0.3}^{+0.4}$ &
$2.0_{-1.0}^{+1.4}$ & 
\dots
\\

NGC4596 & 48 & Nuc. &  $0.0_{-0.0}^{+0.2}$ & 
$-14.4_{-0.3}^{+0.2}$ &
$2.0_{-0.4}^{+0.9}$ & 
\dots
\\

\dots & 49 &  &  $0.0_{-0.0}^{+0.2}$ & 
$-13.9_{-0.2}^{+0.1}$ &
$1.5_{-0.3}^{+0.6}$ & 
\dots
\\

\dots & 50 &  &  $0.1_{-0.1}^{+0.1}$ & 
$-14.0_{-0.2}^{+0.2}$ &
$2.0_{-0.5}^{+0.6}$ & 
\dots
\\

\dots & 53 &  &  $0.0_{-0.0}^{+0.0}$ & 
$-13.9_{-0.2}^{+0.1}$ &
$1.1_{-0.3}^{+0.3}$ & 
\dots
\\

NGC4742 & 54 &  &  $0.3_{-0.1}^{+0.1}$ & 
$-13.8_{-0.1}^{+0.1}$ &
$2.2_{-0.4}^{+0.4}$ & 
\dots
\\

\dots & 55 & ULX &  $1.1_{-0.5}^{+0.7}$ & 
$-14.5_{-0.5}^{+0.3}$ &
$3.5_{-1.2}^{+2.4}$ & 
\dots
\\

\dots & 56 & Nuc. &  $0.1_{-0.1}^{+0.1}$ & 
$-13.6_{-0.1}^{+0.1}$ &
$1.7_{-0.2}^{+0.4}$ & 
7.01/ 4
\\

\dots & 57 & ULX &  $0.4_{-0.3}^{+0.5}$ & 
$-15.4_{-0.6}^{+0.5}$ &
$4.3_{-1.7}^{+2.6}$ & 
\dots
\\

\dots & 58 &  &  $0.0_{-0.0}^{+0.1}$ & 
$-13.5_{-0.2}^{+0.1}$ &
$1.3_{-0.3}^{+0.3}$ & 
\dots
\\

NGC5077 & 59 & Nuc. &  $0.0_{-0.0}^{+0.1}$ & 
$-13.5_{-0.1}^{+0.1}$ &
$1.5_{-0.2}^{+0.2}$ & 
2.18/ 4
\\

\dots & 60 &  &  $0.2_{-0.2}^{+0.4}$ & 
$-15.1_{-0.6}^{+0.5}$ &
$3.1_{-1.2}^{+2.2}$ & 
\dots
\\

NGC5576 & 61 & ULX &  $0.5_{-0.4}^{+0.6}$ & 
$-14.7_{-0.7}^{+0.5}$ &
$3.0_{-1.4}^{+2.9}$ & 
\dots
\\

\dots & 62 &  &  $4.8_{\phantom{-0.0}}^{\phantom{+0.0}}$ & 
$-14.5_{\phantom{-0.0}}^{\phantom{+0.0}}$ &
$9.5_{\phantom{-0.0}}^{\phantom{+0.0}}$ & 
\dots
\\

\dots & 63 & Nuc. &  $0.5_{-0.2}^{+0.2}$ & 
$-14.5_{-0.3}^{+0.2}$ &
$3.5_{-0.8}^{+1.1}$ & 
\dots
\\

\dots & 64 & ULX &  $0.4_{-0.4}^{+0.6}$ & 
$-15.7_{-0.8}^{+0.9}$ &
$5.1_{-2.2}^{+2.8}$ & 
\dots
\\

NGC7457 & 67 & Nuc. &  $0.3_{-0.2}^{+0.2}$ & 
$-14.4_{-0.3}^{+0.3}$ &
$2.5_{-0.8}^{+1.0}$ & 
\dots
\\

\dots & 68 &  &  $0.3_{-0.2}^{+0.3}$ & 
$-14.5_{-0.4}^{+0.3}$ &
$2.8_{-0.9}^{+1.2}$ & 
\dots
\\

    \enddata
    \label{t:spectralfits}
    \tablecomments{This table lists the best-fit parameters with
    $1\sigma$ uncertainties of our spectral fits to
    sources with sufficient counts to warrant fitting.  If we could not
    reliably obtain uncertainties, we omit the listing of errors and consider
    the results approximate.
    Columns list: (1) the name of the galaxy in which the source appears to
    lie; (2) our running identification number; (3) classification of source;
    (4) total absorption column towards the source in units of
    $10^{22}\ \mathrm{cm^{-2}}$; (5) logarithmic
    normalization of the powerlaw in unabsorbed (intrinsic) flux in the
    2--10 keV band in units of $\mathrm{erg\ s^{-1}\ cm^{-2}}$;
    (6) powerlaw slope;  and (7) $\chi^2$ per degrees of freedom if
    there were sufficient counts to use $\chi^2$ statistics.
    }
    \end{deluxetable}
    \bookmarksetup{color=[rgb]{0,0,0.54}} 
    \bookmark[
    rellevel=1,
    keeplevel,
    dest=table.\getrefnumber{t:spectralfits}
    ]{Table \ref*{t:spectralfits}: Parameters of spectral fits}
    \bookmarksetup{color=[rgb]{0,0,0}}

\section{Analysis and Discussion}
\label{concl}

\subsection{SMBHs}
\label{discusssmbh}

    \tabletypesize{\scriptsize}
    \def\arraystretch{1.200}
    \begin{deluxetable*}{llrrrrrrr}
    \tablecaption{Nuclear fluxes}
    \tablewidth{0pt}
    \tablehead{
    \colhead{Galaxy} & \colhead{ID} & \colhead{$F_{0.3\textrm{--}2}$} & \colhead{$L_{0.3\textrm{--}2}$} & \colhead{$F_{2\textrm{--}10}$} & \colhead{$L_{2\textrm{--}10}$}  & \colhead{$F_{0.3\textrm{--}10}$} & \colhead{$L_{0.3\textrm{--}10}$} & \colhead{$\log(L_{2\textrm{--}10} / \ledd)$} \\
    \colhead{(1)} & \colhead{(2)} & \colhead{(3)} & \colhead{(4)} & \colhead{(5)} &     \colhead{(6)} & \colhead{(7)} & \colhead{(8)} & \colhead{(9)}
    }
    \startdata

NGC1300 & 13
 & 
$7.3^{+0.2}_{-2.5}\times10^{-15}$ & $3.8^{+1.1}_{-0.5}\times10^{38}$
 & 
$1.1^{+0.4}_{-0.3}\times10^{-13}$ & $5.2^{+1.2}_{-1.1}\times10^{39}$
 & 
$1.1^{+0.3}_{-0.3}\times10^{-13}$ & $5.6^{+1.2}_{-1.1}\times10^{39}$
&
$-6.3 \pm 0.3$
\\

NGC2748 & 21
 & 
$2.0^{+0.4}_{-1.1}\times10^{-15}$ & $1.9^{+2599.2}_{-1.8}\times10^{40}$
 & 
$7.3^{+27.5}_{-6.0}\times10^{-16}$ & $6.9^{+12.5}_{-4.9}\times10^{37}$
 & 
$2.7^{+0.8}_{-1.3}\times10^{-15}$ & $2.0^{+694.8}_{-1.8}\times10^{40}$
&
$-8.0 \pm 0.7$
\\

NGC2778 & 27
 & 
$4.3^{+1.1}_{-4.2}\times10^{-15}$ & $1.9^{+49.2}_{-1.5}\times10^{39}$
 & 
$1.9^{+68.9}_{-1.8}\times10^{-16}$ & $1.4^{+4.0}_{-1.1}\times10^{37}$
 & 
$4.4^{+1.0}_{-4.3}\times10^{-15}$ & $1.9^{+49.2}_{-1.5}\times10^{39}$
&
$-8.2 \pm 0.8$
\\

NGC3384 & 32
 & 
$1.3^{+0.1}_{-0.4}\times10^{-14}$ & $3.2^{+3.4}_{-0.9}\times10^{38}$
 & 
$1.7^{+1.5}_{-0.8}\times10^{-14}$ & $2.8^{+1.7}_{-1.2}\times10^{38}$
 & 
$3.0^{+0.5}_{-1.1}\times10^{-14}$ & $5.9^{+2.4}_{-0.9}\times10^{38}$
&
$-6.9 \pm 0.2$
\\

NGC4291 & 39
 & 
$5.6^{+0.7}_{-1.6}\times10^{-15}$ & $6.6^{+4.2}_{-2.2}\times10^{38}$
 & 
$1.4^{+0.6}_{-0.4}\times10^{-14}$ & $1.0^{+0.4}_{-0.3}\times10^{39}$
 & 
$1.9^{+0.3}_{-0.5}\times10^{-14}$ & $1.7^{+0.3}_{-0.3}\times10^{39}$
&
$-7.6 \pm 0.4$
\\

NGC4459 & 42
 & 
$1.6^{+0.1}_{-0.6}\times10^{-14}$ & $6.3^{+1.8}_{-0.8}\times10^{38}$
 & 
$2.3^{+0.9}_{-0.7}\times10^{-14}$ & $8.0^{+2.2}_{-2.0}\times10^{38}$
 & 
$3.9^{+0.3}_{-1.4}\times10^{-14}$ & $1.4^{+0.2}_{-0.2}\times10^{39}$
&
$-7.1 \pm 0.1$
\\

NGC4486A & 47
 & 
$1.8^{+0.4}_{-0.6}\times10^{-15}$ & $1.3^{+11.5}_{-0.8}\times10^{38}$
 & 
$4.0^{+4.3}_{-2.4}\times10^{-15}$ & $1.4^{+1.6}_{-0.8}\times10^{38}$
 & 
$5.8^{+2.4}_{-3.5}\times10^{-15}$ & $2.8^{+10.5}_{-0.9}\times10^{38}$
&
$-7.1 \pm 0.4$
\\

NGC4596 & 48
 & 
$4.2^{+0.7}_{-2.9}\times10^{-15}$ & $1.8^{+2.4}_{-0.4}\times10^{38}$
 & 
$3.9^{+4.3}_{-2.2}\times10^{-15}$ & $1.5^{+1.0}_{-0.8}\times10^{38}$
 & 
$8.1^{+0.3}_{-5.1}\times10^{-15}$ & $3.3^{+1.7}_{-0.7}\times10^{38}$
&
$-7.9 \pm 0.3$
\\

NGC4742 & 56
 & 
$1.3^{+0.1}_{-0.3}\times10^{-14}$ & $5.2^{+2.3}_{-1.0}\times10^{38}$
 & 
$2.5^{+0.8}_{-0.6}\times10^{-14}$ & $7.2^{+2.0}_{-1.9}\times10^{38}$
 & 
$3.8^{+0.5}_{-0.8}\times10^{-14}$ & $1.2^{+0.2}_{-0.2}\times10^{39}$
&
$-6.4 \pm 0.2$
\\

NGC5077 & 59
 & 
$1.3^{+0.1}_{-0.5}\times10^{-14}$ & $3.6^{+0.9}_{-0.5}\times10^{39}$
 & 
$2.8^{+0.9}_{-0.7}\times10^{-14}$ & $6.8^{+1.6}_{-1.5}\times10^{39}$
 & 
$4.1^{+0.4}_{-1.2}\times10^{-14}$ & $1.0^{+0.2}_{-0.1}\times10^{40}$
&
$-7.2 \pm 0.2$
\\

NGC5576 & 63
 & 
$4.7^{+0.5}_{-2.1}\times10^{-15}$ & $4.3^{+14.5}_{-2.8}\times10^{39}$
 & 
$2.7^{+5.6}_{-1.8}\times10^{-15}$ & $2.6^{+1.9}_{-1.3}\times10^{38}$
 & 
$7.4^{+1.1}_{-4.7}\times10^{-15}$ & $4.6^{+15.4}_{-2.7}\times10^{39}$
&
$-8.0 \pm 0.3$
\\

NGC7457 & 67
 & 
$3.9^{+0.1}_{-2.4}\times10^{-15}$ & $3.1^{+7.7}_{-1.8}\times10^{38}$
 & 
$4.3^{+16.0}_{-3.4}\times10^{-15}$ & $1.0^{+0.9}_{-0.5}\times10^{38}$
 & 
$8.2^{+1.4}_{-6.1}\times10^{-15}$ & $4.2^{+6.8}_{-1.5}\times10^{38}$
&
$-6.7 \pm 0.3$
\\

    \enddata
    \label{t:nuclearflux}
    \tablecomments{Fluxes and luminosities of nuclear sources, assumed
    to be the SMBH in each galaxy.  For each source, we list the $F$,
    the absorbed (apparent) flux in units of $\mathrm{erg\ s^{-1}\
    cm^{-2}}$, and $L$, the unabsorbed (intrinsic) luminosity in units
    of $\mathrm{erg\ s^{-1}}$, for each of the 0.3--2, 2--10, and
    0.3--10 bands.  Uncertainties are listed as 1$\sigma$ intervals,
    and the final column includes uncertainties in the mass.  Note
    that because of covariances between the model parameters, the full
    band is not simply the sum of the soft and hard bands with
    uncertainties added in quadruture.  The final column lists the
    logarithmic Eddington fraction.  While some of the sources have
    full-band luminosities consistent with zero at about the 3$\sigma$
    level, point sources at the centers of each galaxy at greater than
    4$\sigma$ confidence.  This is because the luminosity depends upon
    an unknown spectral form whereas detection depends on raw count
    rate above the background.  For example, the full-band luminosity
    for source 27 in NGC 2778 is only about 1$\sigma$ above zero, but
    inspection of Table \ref{t:ptsrcdetec} shows that the net count
    rate for this source in the full band (summing uncertainties in
    quadruture) is $(9.6 \pm 2.1) \times 10^{-4}\ \units{cts
    s^{-1}}$.}
    \end{deluxetable*}
    \bookmarksetup{color=[rgb]{0,0,0.54}} 
    \bookmark[
    rellevel=1,
    keeplevel,
    dest=table.\getrefnumber{t:nuclearflux}
    ]{Table \ref*{t:nuclearflux}: Fluxes and luminosities of nuclear sources}
    \bookmarksetup{color=[rgb]{0,0,0}}

The primary scientific goal of our \emph{Chandra} program is to
measure the luminosities of SMBHs with direct, primary mass
measurements.  For each nuclear source we list the flux and luminosity
(assuming isotropic radiation and the distance to each galaxy listed
in Table \ref{t:xray}) for the 0.3--2, 2--10, and 0.3--10 keV bands in
Table \ref{t:nuclearflux}.  The fluxes listed are intrinsic
(unabsorbed).  Note that the fluxes in each band are derived from fits
to the X-ray spectrum using different bands as normalizations, and,
therefore, the uncertainties are correlated.  In Figure
\ref{figspectra} (and \ref{figspectra2}), we show plots of the spectra
of the nuclear sources in all 12 galaxies, along with the best-fit
model spectrum for each source.  The parameters of the best-fit model
spectra are in Table \ref{t:spectralfits}.  The spectra have been
binned for visualization purposes.  The spectra are all well fit by a
power law, and none of the spectra requires a more complicated model.

\begin{figure}[tbh]
\hypertarget{gammahist}{}%
\centering
\includegraphics[width=\columnwidth]{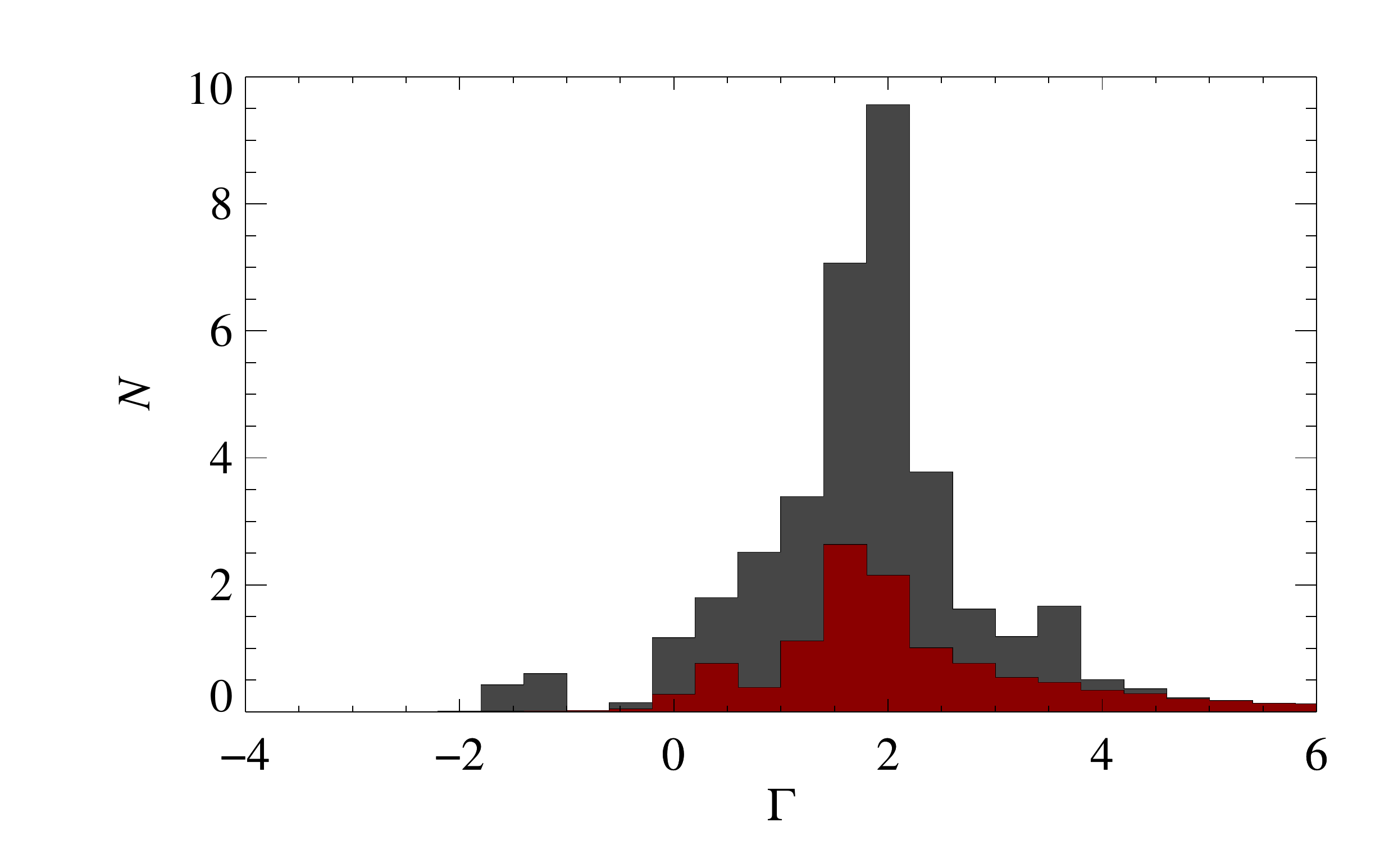}
\caption{Distributed histogram of values of $\Gamma$, the slope of the
power-law.  This is a histogram of the values of $\Gamma$ inferred
from our spectral fits, with each value spread out among the bins
according to the error on the value, assuming a normal distribution.
That is, for each value of $\Gamma$, the histogram in a bin of width
$w$ that is $x\sigma$ away from the central value is $w (\sigma
\sqrt{2\pi})^{-1} \exp(-x^2/2)$, where $\sigma$ is the error of the
measurement.  This method of visualizing the histogram is useful when
the errors vary in magnitude from one measurement to the next and when
the errors are larger than the size of an interesting histogram bin.
The new data from this paper are in red, and the new results plus
those from \citet{2009ApJ...706..404G} are in gray.  The point that
divides the area under the distributed histogram in two is $\Gamma =
1.86$, very close to the canonical $\Gamma = 1.7$ power-law in AGNs
\citep{1984AdSpR...3..157M}.}
\label{figgammahist}
\end{figure}

Figure \ref{figgammahist} is a histogram we call a ``distributed
histogram'' of the values of the power-law exponent ($\Gamma$)
inferred from our spectral fits.  The uncertainties in $\Gamma$ in
some spectral fits are much larger than some of the others and in some
cases much larger than the size of the histogram bins in which we are
interested.  To account for this we assume that the errors in $\Gamma$
are distributed normally and plot the contribution to each bin
according to the value and error.  That is, for each value of
$\Gamma$, the histogram in a bin of width $w$ that is $x\sigma$ away
from the central value is $w (\sigma \sqrt{2\pi})^{-1} \exp(-x^2/2)$,
where $\sigma$ is the error of the measurement.  We also include
analogous data from \citet{\combh}.  There is an obvious peak at
$\Gamma \approx 2$.  Related to the distributed histogram we introduce
a ``distributed median,'' which we define as the point at which the
area under the distributed histogram is half of the total area.  The
distributed median of $\Gamma$ is 1.86, very close to the canonical
1.7 power-law for AGNs \citep[e.g.,][]{1984AdSpR...3..157M}.

\bookmarksetup{color=[rgb]{0.54,0,0}}
\bookmark[rellevel=1,keeplevel,dest=gammahist]{Fig \ref*{figgammahist}: Histogram of Gamma.}
\bookmarksetup{color=[rgb]{0,0,0}}

\begin{figure}[tbh]
\hypertarget{feddhist}{}%
\includegraphics[width=\columnwidth]{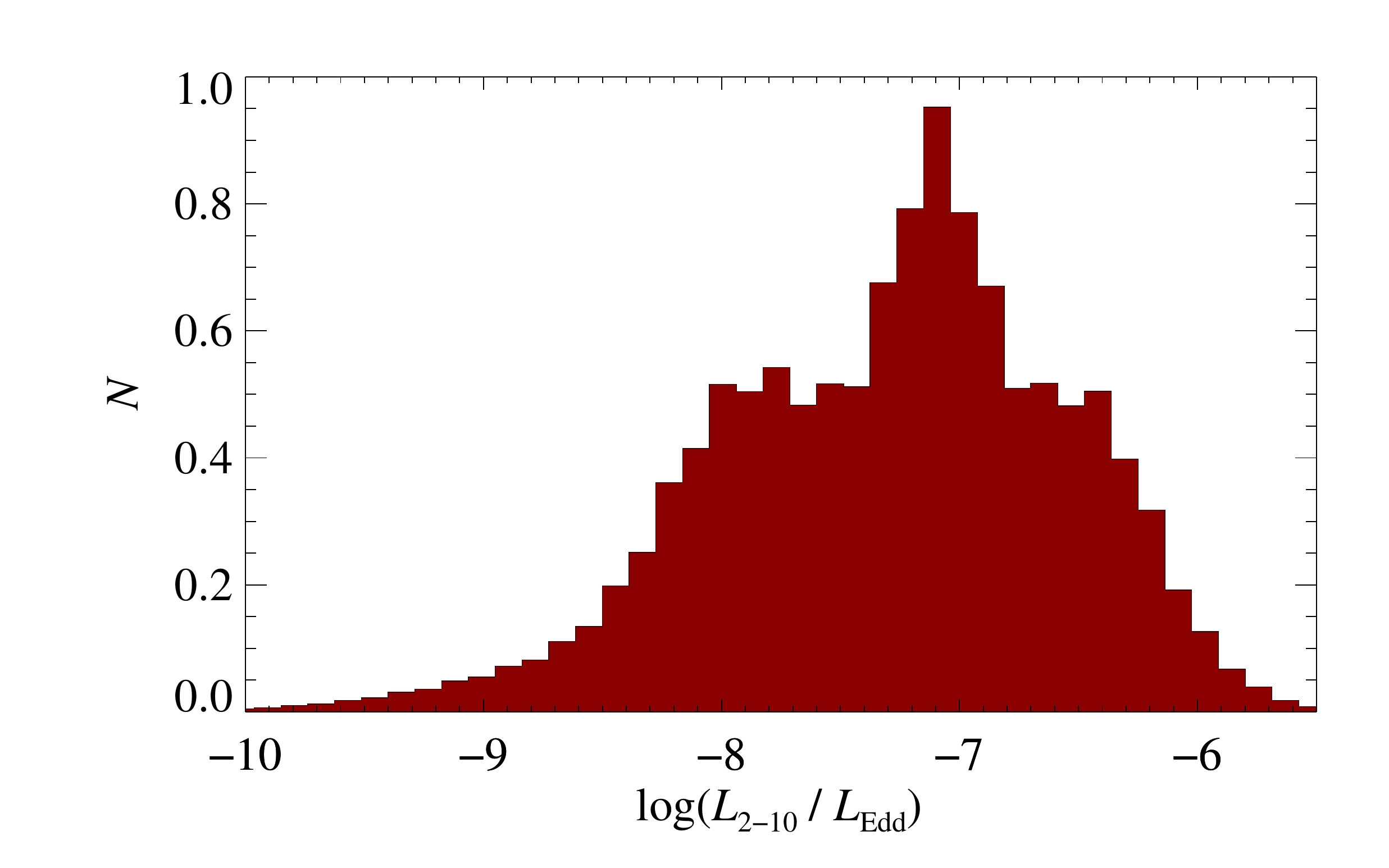}
\caption{Distributed histogram of values of $\log(\lhardedd)$, the
slope of the power-law.  This is a histogram of the values of
\lhardedd\ inferred from our spectral fits, with each value spread out
among the bins according to the error on the value, assuming a normal
distribution.  That is, for each value of $\log(\lhardedd)$ with
measurement error $\sigma$, the histogram in a bin of width $w$ that
is $x\sigma$ away from the central value is $w (\sigma
\sqrt{2\pi})^{-1} \exp(-x^2/2)$.  This method of visualizing the
histogram is useful when the errors vary in magnitude from one
measurement to the next and when the errors are larger than the size
of an interesting histogram bin.  The distributed histogram shows that
the accretion rates that we study in this sample are very low.}
\label{figfeddhist}
\end{figure}

Because we have direct, primary measurements of the mass of the
central black hole in each of these galaxies, we can report
luminosities as true Eddington fractions.  In Figure
\ref{figfeddhist}, we plot a distributed histogram of
$\log(\lhardedd)$.  We include uncertainties in the black hole
mass but instead assume the value given in Table \ref{t:xray}.  As can
be seen in the figure, the hard X-ray Eddington fractions are small,
in the range $10^{-8} < \lhardedd < 10^{-6}$, and the distributed
median of the logarithmic Eddington fraction is $-7.2$.  The reason that
all sources are at such low accretion rates is that the sources were
selected based on having a dynamical mass measurement of the central
black hole.  Most of the black hole masses in
\citet{2009ApJ...698..198G} are based on stellar dynamical models
\citep[e.g.,][]{2009ApJ...695.1577G} and gas dynamical models
\citep[e.g.,][]{2001ApJ...555..685B}, and these methods are best when
the there is no contamination from AGN light.  Typical bolometric
corrections for low-luminosity sources such as these is $\sim10$
\citep{2007MNRAS.381.1235V}, so that the bolometric Eddington
fractions of these sources are in the range $10^{-7}$--$10^{-5}$.

Note that the power of using dynamical mass measurements instead of
secondary mass estimates is evident in our tabulation of \lhardedd\
values.  The median error in $\log(L_{2\textrm{--}10})$ is 0.25 dex,
the median error in $\log\mbh$ is 0.17 dex, and the median error in
secondary mass estimates is 0.51 dex.  Thus the uncertainty in
secondary mass estimates dominates the total uncertainty in $\log
f_\mathrm{Edd}$ and is completely subdominant when using primary
measurements.

\bookmarksetup{color=[rgb]{0.54,0,0}}
\bookmark[rellevel=1,keeplevel,dest=feddhist]{Fig \ref*{figfeddhist}: Histogram of X-ray Eddington fraction.}
\bookmarksetup{color=[rgb]{0,0,0}}

\begin{figure}[tbh]
\hypertarget{gammavsfedd}{}%
\centering
\includegraphics[width=\columnwidth]{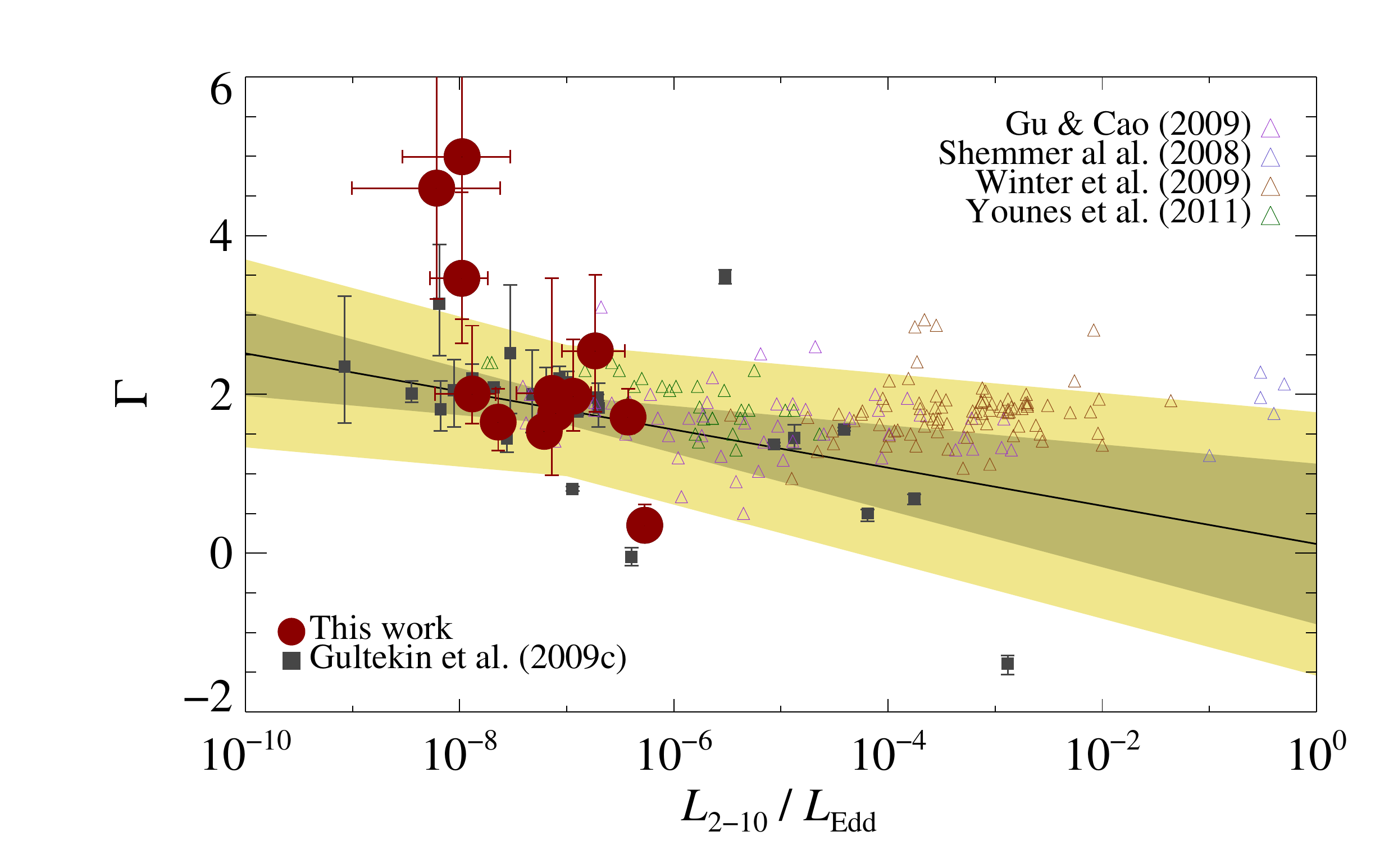}
\caption{Slope of the spectral power-law ($\Gamma$) as a function of
  hard X-ray Eddington fraction (\lhardedd).  Large red circles are
  new results from this paper, and small gray squares are results from
  \citet{2009ApJ...706..404G}.  The slope of the best-fit relation to
  these two data sets, which is drawn as a solid black line, is $0.24
  \pm 0.12$.  The dark shaded region shows the 1$\sigma$ confidence
  band, and the light shaded region shows the 1$\sigma$ confidence
  band plus the rms intrinsic scatter.  Error bars on each point are
  1$\sigma$ uncertainties.  For reference we show similar data based
  primarily on secondary mass estimates from
  \citet{2009MNRAS.399..349G}, \citet{2008ApJ...682...81S},
  \citet{2009ApJ...690.1322W}, and \citet{2011A&A...530A.149Y} as
  indicated in the legend drawn as small open triangles.}
\label{figgammavsfedd}
\end{figure}

In Figure \ref{figgammavsfedd} we plot the spectral power-law slope as
a function of $\log(\lhardedd)$ for the current data and for the data
in \citet{\combh}.  We also fit for a linear relation between these
two quantities.  There is one source, Cen A, with $\lhardedd >
10^{-3}$ and $\Gamma < 0$, that is likely Compton thick so that a
power-law index at energies below 10 keV possibly probes different
energetics and/or accretion physics than the rest of the sources.
Because of this and the potential lever arm it could have on the fit,
we fit both with and without this source, but it had no effect on our
results.  Since there is a strong covariance between $\Gamma$ and
$\lhard$, we do not include measurement uncertainties on $\lhard$ but
instead bootstrap (resample with replacement) the sample to estimate
uncertainties, expressing our result as the median with 68\% interval.
A histogram of the bootstrap results showed roughly Gaussian
distributions, indicating reliability of results.  Finally we include
a scatter term, assumed normally distributed in $\Gamma$ because the
data clearly deviate from any single line by more than their
measurement uncertainties.  The fitting method and code are the same
as that used and described in \citet{2009ApJ...698..198G}.  The method
uses a generalized maximum likelihood method that is capable of
including upper limits, arbitrary error distributions, and arbitrary
form of intrinsic scatter.  Here we assume Gaussian errors in $\Gamma$
and Gaussian scatter in the $\Gamma$ direction.  We plot the best-fit
linear relation, which is $\Gamma = 1.8 \pm 0.2 - (0.24 \pm 0.12)
\log(\lhard / 10^{-7} \ledd)$ with an rms intrinsic scatter of $0.65
\pm 0.20$ (uncertainties are 1$\sigma$ here and throughout this
paper).  Our results are inconsistent with a slope of zero or larger
at the 2$\sigma$ level.  

To illustrate what the additional data in this program add as well as
how sensitive these results are to the use of dynamical masses, we
perform two exercises.  First, we repeat the fit without the new data,
yielding a slope of $-0.33 \pm 0.25$.  So while this is consistent
with our full results, it is far less conclusive as to whether the
X-ray spectral properties at very low Eddington rates is different
from those at high Eddington rates.  Second, we fit with the full
sample but with masses and uncertainties generated from the \msigma\
relation:
\beq
\log(\mbh / \msun) = \alpha + \beta \log(\sigma / 200\ \kms),
\eeq
where $\sigma$ is the velocity dispersion of the host galaxy, and
$\alpha = 8.12 \pm 0.08$ and $\beta = 4.24 \pm 0.41$ are the best-fit
\msigma\ parameters \citep{2009ApJ...698..198G}.  Propagating
uncertainties, $\epsilon_x$, for each quantity, $x$, the variance in
logarithmic mass due to random errors is
\beq
\epsilon_{\log\mbh}^2 = \epsilon_\alpha^2 + \left[\log\left(\sigma / 200\ \kms\right)\right]^2 \epsilon_\beta^2 + \frac{\beta^2}{\sigma^2} \epsilon_\sigma^2 + \epsilon_0^2,
\eeq
where $\epsilon_0 = 0.44$ is the observed intrinsic scatter in the
\msigma\ relation.  When fitting with these new masses and
uncertainties, the slope is $-0.27 \pm 0.13$.  So it appears that
without direct masses, we would have come to the same conclusion,
though by using masses derived from the \msigma\ relation for the very
galaxies from which \msigma\ was derived, we necessarily underestimate
the amount of systematic errors introduced from using a secondary
quantity.

As mentioned in section \ref{sampleselection}, one consequence of our
sample selection is that we only cover SMBHs emitting at very low
Eddington rates.  A direct consequence of this is that our X-ray
observations are more susceptible to contamination from a fixed amount
of hot gas than would be an X-ray source that was intrinsically
brighter.  Since hot gas emission typically peaks at an energy below 1
keV, any contamination hot gas in our spectral data would tend to make
the spectrum softer and thus we would infer a larger value for
$\Gamma$.  This is particularly concerning since our results indicate
larger values of $\Gamma$ at lower Eddington rates, as this
selection effect and contamination would manifest.
We have mitigated this as much as possible with our selection of
background regions, which are annuli that surround the source regions.
This effectively removes the contribution of hot gas that is seen in
the background region.  If, however, the diffuse emission is more
centrally concentrated than the background annulus, then there will
still be some contamination from diffuse gas.  
To take this into account we refit all nuclear sources with an
additional astrophysical plasma emission code
\citep[APEC][]{2001ApJ...556L..91S} component (with abundance fixed to
solar).  For all but three the results for $\lhard$ and $\Gamma$ are
consistent at the 1$\sigma$ or better, and the remaining 3 are
consistent at about the 2$\sigma$ level.  This lack of significant
change strongly suggests that our results are not affected by the
contamination of diffuse gas.

\bookmarksetup{color=[rgb]{0.54,0,0}}
\bookmark[rellevel=1,keeplevel,dest=gammavsfedd]{Fig \ref*{figgammavsfedd}: Gamma vs. hard X-ray Eddington fraction.}
\bookmarksetup{color=[rgb]{0,0,0}}

Our results are consistent with previous works looking at the
anti-correlation between $\Gamma$ and Eddington fraction.  Using a
sample of 55 LLAGNs with a mixture of primary and secondary
black-hole-mass measurements, \citet{2009MNRAS.399..349G} fit a linear
relation assuming a constant $\lbol / \lhard = 30$.  For $\lbol/\ledd
< 10^{-1}$, they inferred $\Gamma = 1.55 \pm 0.07 - (0.09 \pm 0.03)
\log(\lbol / \ledd)$.  Note that the differences in intercepts between
the fits are all a result of defining the independent variable
differently.  While their results are consistent with ours at about
the 1.2$\sigma$ level, the absence of a scatter term in the fitting
function may skew the results since the residuals are larger than
would be expected just from measurement errors.  As part of the
\emph{Chandra} Multiwavelength Project, \citet{2009ApJ...705.1336C}
studied 107 LLAGNs with \emph{Chandra} observations.  They found a
strong anticorrelation between $\Gamma$ and $\lbol / \ledd$, where
they assumed $\lbol = 16 \lhard$ and used the \msigma\ relation to
estimate black hole masses.  Their Spearman rank correlation
coefficient was $-0.75$, and their fit to the relation was $\Gamma =
0.98 \pm 0.13 - (0.27 \pm 0.04) \log(\lbol / \ledd)$.  The
goodness-of-fit, however, was $\chi^2 = 274$ for 105 degrees of
freedom, indicating that it was very unlikely that the data could come
from a scatter-free model.  In a study of 153 AGNs detected by the
\emph{Swift} Burst Alert Telescope hard X-ray instrument,
\citet{2009ApJ...690.1322W} found a no correlation between $\Gamma$
and \lhardedd.  It is possible that their sample, which is primarily
at $\lhardedd > 10^{-4}$ and reaches up to $\lhardedd \approx 0.04$ is
actually measuring an energetically different mode of accretion than
our sample.  In any case, our study agrees with
\citet{2009ApJ...690.1322W} that the spectra do not harden with
increasing accretion rates.  In a study of 13 LINERs with
\emph{XMM-Newton} and/or \emph{Chandra} observations and assuming very
different \msigma\ relation \citep[due to ][]{2011MNRAS.412.2211G},
\citet{2011A&A...530A.149Y} found $\Gamma = 0.11 \pm 0.40 - (0.31 \pm
0.06) \log(\lhardedd)$.  A similar anti-correlation is suggested in
X-ray binary data \citep{2006ApJ...636..971C}.  Given the difference
in assumptions and fitting techniques, we consider all of these
results consistent with each other in the following conclusions: (1)
there is an anti-correlation between the hard X-ray photon index,
$\Gamma$, and Eddington fraction for LLAGNs, and (2) the correlation
between $\Gamma$ and $\log(\lhardedd)$ has a slope of roughly $-0.2$.

The anticorrelation between $\Gamma$ and $\lhardedd$ is in contrast to
AGNs emitting at higher Eddington fractions.  For example,
\citet{2008ApJ...682...81S} found from fits to a sample of 35 radio
quiet, luminous, and high-Eddington-fraction ($\lbol / \ledd \sim 0.01$--1)
AGNs that there was a strong positive correlation.  From their fits,
they found $\log(\lbol / \ledd) = -2.4 \pm 0.6 + (0.9 \pm 0.3)
\Gamma$, corresponding to $\Gamma = 2.6 + 1.1 (\lbol / \ledd)$.
Thus our evidence leads us to conclude that we are seeing a change in
the physical processes responsible for emission at low Eddington
rates.

The softening of the spectrum with decreasing Eddington fractions at
low mass accretion rates as we find in our sample is predicted by
advection dominated accretion flow (ADAF) models in stellar-mass X-ray
binaries \citep{1997ApJ...489..865E}, which should reasonably
translate to the low accretion rates seen in our SMBH sources.
\citet{1997ApJ...489..865E} models predict a steepening of the 1--10
keV band index of $\Gamma \approx 1.7$ to about 2.2 over mass
accretion rates of $\log(\dot{m}) = -2.5$ to $-4$, or a slope of
approximately 0.33.  The comparison is not direct, but our results are
fully consistent with this prediction, which is attributed to the fact
that at lower accretion rates bremsstrahlung emission becomes
relatively more important compared to Comptonizaion.

\subsection{ULXs}
Ultraluminous X-ray sources (ULXs) are a category of X-ray emitting
point sources that are too bright to be explained by isotropic
emission resulting from sub-Eddington accretion onto stellar-mass ($M
\la 10\ \msun$) black holes and are also non-nuclear so that they are
unlikely to be the galaxy's central SMBH.  ULXs are an intriguing
class of source because if (1) they are emitting roughly isotropically
(i.e., not strongly beamed towards our line of sight), so that we may
correctly infer their luminosity, and (2) they are not accreting well
above the Eddington limit, so that we may robustly infer a lower limit
to the mass, then the most natural explanation is a black hole of mass
$\sim 10^2$--$10^5\ \msun$.  These intermediate-mass black holes
(IMBHs) would fill the gap in mass between stellar-mass black holes
and SMBHs.  IMBHs are interesting because their formation in the local
universe requires a non-standard path \citep{mh02a, gmh04, gmh06,
pzetal04}.  There are alternative interpretations to ULXs other than
IMBHs, including beamed, non-isotropic emission \citep{kingetal01}.
This would change the luminosity inferred from the flux so that it is
consistent with sub-Eddington accretion onto stellar-mass black holes.
There are at least a few sources \citep{kwz04, pm01} where emission
from the surrounding medium argues in favor of roughly isotropic
emission.  Another alternative is to have super-Eddington accretion,
which has been invoked in a number of different ways
\citep{begelman02, begelman06}.  The very bright source ESO 243-49
HLX-1 is similarly difficult to interpret without invoking an IMBH
\citep{2009Natur.460...73F}.

    \tabletypesize{\scriptsize}
    \def\arraystretch{1.200}
    \begin{deluxetable}{lrrrrrr}
    \tablecaption{Ultraluminous X-ray Source Candidates}
    \tablewidth{0pt}
    \tablehead{
    \colhead{Galaxy} & \colhead{ID} &  \colhead{$L_{0.3\textrm{--}10}$} & \colhead{$P(L>L_\mathrm{ULX})$} & \colhead{$L_{F}$} & \colhead{$p$} & \colhead{$\langle N_\mathrm{BG}\rangle$}  \\
    \colhead{(1)} & \colhead{(2)} & \colhead{(3)} & \colhead{(4)} & \colhead{(5)} & \colhead{(5)} & \colhead{(6)}}
    
    \startdata
NGC1300 & 6 & $8.9^{+40.0}_{-6.0}\times10^{39}$ & $0.92$ & $7.8^{+1.3}_{-1.1}\times10^{38}$ & $0.07$ & $0.13$
\\
\dots & 9 & $2.8^{+1621.1}_{-2.8}\times10^{40}$ & $0.50$ & $2.3^{+0.9}_{-0.7}\times10^{38}$ & $0.83$ & $0.13$
\\
\dots & 12 & $1.5^{+14.2}_{-1.1}\times10^{39}$ & $0.43$ & $3.5^{+0.8}_{-0.7}\times10^{38}$ & $0.74$ & $0.13$
\\
NGC2748 & 18 & $1.4^{+2.0}_{-0.5}\times10^{40}$ & $>0.99$ & $7.0^{+1.1}_{-0.9}\times10^{39}$ & $0.002$ & $0.02$
\\
\dots & 22 & $2.8^{+101.3}_{-2.5}\times10^{40}$ & $0.88$ & $5.2^{+2.0}_{-1.4}\times10^{38}$ & $0.05$ & $0.02$
\\
\dots & 23 & $9.2^{+1490.0}_{-8.8}\times10^{40}$ & $0.91$ & $7.3^{+3.2}_{-2.4}\times10^{38}$ & $0.03$ & $0.02$
\\
\dots & 24 & $1.9^{+2.3}_{-0.9}\times10^{40}$ & $>0.99$ & $4.6^{+0.8}_{-0.7}\times10^{39}$ & $0.002$ & $0.02$
\\
\dots & 25 & $1.0^{+0.6}_{-0.3}\times10^{40}$ & $>0.99$ & $5.8^{+0.4}_{-0.4}\times10^{39}$ & $0.002$ & $0.02$
\\
NGC2778 & 26 & $8.0^{+12.5}_{-2.8}\times10^{38}$ & $0.16$ & $6.3^{+2.1}_{-1.5}\times10^{38}$ & $0.79$ & $0.01$
\\
NGC3384 & 30 & $5.7^{+94.9}_{-4.7}\times10^{38}$ & $0.31$ & $6.0^{+2.1}_{-1.6}\times10^{37}$ & $0.16$ & $0.03$
\\
\dots & 34 & $5.6^{+498.0}_{-4.5}\times10^{38}$ & $0.37$ & $9.4^{+6.1}_{-3.9}\times10^{37}$ & $0.39$ & $0.03$
\\
\dots & 35 & $7.0^{+88.5}_{-5.4}\times10^{38}$ & $0.33$ & $1.3^{+0.3}_{-0.3}\times10^{38}$ & $0.49$ & $0.03$
\\
NGC4291 & 37 & $2.9^{+15.9}_{-2.7}\times10^{41}$ & $0.99$ & $7.8^{+1.9}_{-1.6}\times10^{38}$ & $0.008$ & $0.01$
\\
\dots & 40 & $2.8^{+0.6}_{-0.5}\times10^{39}$ & $0.96$ & $2.5^{+0.4}_{-0.4}\times10^{39}$ & $0.60$ & $0.01$
\\
NGC4459 & 43 & $1.7^{+73.0}_{-1.6}\times10^{40}$ & $0.77$ & $3.3^{+1.3}_{-1.0}\times10^{38}$ & $0.03$ & $0.01$
\\
NGC4596 & 52 & $1.5^{+34.1}_{-1.5}\times10^{41}$ & $0.85$ & $2.4^{+0.9}_{-0.6}\times10^{38}$ & $0.27$ & $0.01$
\\
NGC4742 & 55 & $1.6^{+53.0}_{-1.2}\times10^{39}$ & $0.46$ & $2.5^{+1.2}_{-0.9}\times10^{38}$ & $0.16$ & $0.00$
\\
\dots & 57 & $9.6^{+419.3}_{-8.2}\times10^{38}$ & $0.41$ & $1.3^{+0.5}_{-0.3}\times10^{38}$ & $0.72$ & $0.00$
\\
NGC5576 & 61 & $1.3^{+51.6}_{-0.8}\times10^{39}$ & $0.42$ & $5.7^{+4.6}_{-2.5}\times10^{38}$ & $0.57$ & $0.03$
\\
\dots & 64 & $5.9^{+759.1}_{-5.3}\times10^{39}$ & $0.63$ & $4.5^{+2.0}_{-1.2}\times10^{38}$ & $0.70$ & $0.03$

    \enddata
    \label{t:ulxs}
    \tablecomments{A listing of the ULX candidates identified in this
    survey.  Columns list: (1) the name of the galaxy in which the
    source appears to lie; (2) our running identification number; (3)
    the unabsorbed (intrinsic) luminosity in the 0.3--10 keV band; (4)
    the probability, based on the flux parameter uncertainty, that the
    given source's luminosity is above the definition of a ULX
    ($L_{0.3\textrm{--}10} > 2 \times 10^{39} \mathrm{\ erg\
    s^{-1}}$); (5) the 0.3--10 keV luminosity inferred when fixing the
    absorption column to the Galactic value towards that source; (6)
    the $p$-value result of our simulations to calculate the
    significance of the improvement of including $N_H$ as a free
    parameter; and (7) the expected number of background sources in
    the galaxy with fluxes $F_{0.5\textrm{--}2} > 2 \times 10^{39}
    \mathrm{\ erg\ s^{-1}} (4 \pi D^{2})^{-1}$.  The probability of
    having at least one background source in the galaxy of such flux
    is $1 - \exp(-\langle N_\mathrm{BG}\rangle) \sim \langle
    N_\mathrm{BG}\rangle$.  All uncertainties are listed as 1$\sigma$
    intervals.  }
    \end{deluxetable}
    \bookmarksetup{color=[rgb]{0,0,0.54}} 
    \bookmark[
    rellevel=1,
    keeplevel,
    dest=table.\getrefnumber{t:ulxs}
    ]{Table \ref*{t:ulxs}: ULX candidates}
    \bookmarksetup{color=[rgb]{0,0,0}}

Although our survey was not targeted at ULXs, we are sensitive to them
and present a list of ULX candidates in table \ref{t:ulxs}.  We adopt
the definition of \citep{iab03} that ULXs are sources with $\lallband
> L_\mathrm{ULX} \equiv 2 \times 10^{39}\ \ergs$, which avoids
contamination from bright, massive stellar-mass X-ray binaries.  We
assume that all sources are isotropically emitting at the distance of
the host galaxy.  Because of measurement uncertainties, there is
always a finite probability that a source that appears to have
$\lallband > \lulx$, is actually intrinsically too dim to be a ULX.
Additionally, sources that are just below \lulx\ are still viable ULX
candidates.  For these reasons, we list all sources that are at least
1$\sigma$ consistent with being a ULX and list $P(\lallband > \lulx)$,
the probability of the source having $\lallband > \lulx$.  This is
calculated by finding $\Delta C$, the change in fit statistic, when
setting $F_{0.3\textrm{--}10} = 2 \times 10^{39}\ \ergs\ (4\pi
D^2)^{-1}$ and refitting.  Then we calculate $P(\lallband > \lulx) =
0.5 \pm 0.5\;\mathrm{erf}\;\left[\left(\Delta C /
2\right)^{0.5}\right]$, taking the top and bottom sign for when the
best-fit luminosity is above or below \lulx, respectively.

One source of contamination is background AGNs that appear to be in
the galaxy.  We have checked the catalog of
\citet{2010A&A...518A..10V} and found no known AGNs at the locations
of our ULX candidates, but it is clearly possible for a previously
unknown AGN to be located at this position. To quantify this effect, we
calculate $\langle N_\mathrm{BG}\rangle$, the expected number of
background AGNs in the region of the galaxy target with fluxes greater
than $2 \times 10^{39}\ \ergs\ (4\pi D^2)^{-1}$.  To do this we use
the fits to the background AGN density as a function of flux from
\citep{2001ApJ...551..624G}.  Unfortunately, they only provide fits
for fluxes in the 0.5--2 and 2--10 keV bands.  Since most of the flux
in a power-law source comes from the soft band, we use the softer band
and assume that all of the flux comes from this band.  Since the
relations are approximately linear, if only half of the emission comes
from this portion of the band, then the average number of background
sources would roughly double, still a small number.  It is important
to use the whole area of the galaxy and the smallest possible flux for
the calculation of $\langle N_\mathrm{BG}\rangle$ because we would
have listed any source above the threshold flux apparently within the
galaxy as a ULX candidate.  The expected background for each galaxy is
listed in Table \ref{t:ulxs}, and it is small.  The probability of a
galaxy having at least one confusing background source is $1 -
\exp(-\langle N_\mathrm{BG}\rangle)$, and a lower limit for the
probability of an individual ULX candidate's truly being a ULX is
$P(\mathrm{ULX}) = P(\lallband > \lulx)\exp(-\langle
N_\mathrm{BG}\rangle)$.

Note that the probability of ULX luminosity that we calculate is only
accurate if our spectral model is a reasonably good model.  In Figure
\ref{figulxspectra} we plot spectra of the six sources (6, 18, 23, 24,
25, and 40) with $P(\lallband > \lulx) > 0.9$.  Sources 18, 23, and 24
all have $N_\mathrm{H} > 10^{22} \units{cm^{-2}}$ and thus while their
apparent flux is rather modest, the inferred, intrinsic
absorption-corrected luminosity is quite high.  This inference depends
strongly on the correct estimation of $N_\mathrm{H}$, which derives
from the assumption of a power-law spectral form.  We have tried
fitting with other spectral forms for these three sources but were not
able to come up with acceptable fits.  In particular, we tried more
complicated models with combined disk blackbody and power-law spectral
forms, but the disk blackbody temperature normalization was either
unphysically high or its normalization was so low as to make the
component irrelevant.

To address the issue of a ULX classification's sensitivity to the
unknown intrinsic absorption column, we refit all ULX candidate
spectra with $N_\mathrm{H}$ fixed to the Galactic value in Table
\ref{t:xray} for each galaxy.  The resulting luminosities ($L_F$) are
listed in Table \ref{t:ulxs}.  We also calculated $p$-values of a
likelihood ratio statistic using Monte Carlo simulations using the
method of \citet{2002ApJ...571..545P}.  Each simulation synthesized
1000 realizations of the ULX spectra based on the best-fit
fixed-$N_\mathrm{H}$ model.  We then fitted each synthetic data set
with a fixed-$N_\mathrm{H}$ model and one in which $N_\mathrm{H}$ was
a free parameter, and calculated a likelihood ratio for each spectrum.
This procedure generated a distribution of our likelihood ratio
statistic according to the null hypothesis that $N_\mathrm{H}$ fixed
at the Galactic value is the correct model.  We adopt $p < 0.01$ as
our level of significance for requiring an additional free parameter.
Using the fixed $N_\mathrm{H}$ spectral model, only 4 sources (18, 24,
25, and 40) are bright enough to be considered ULXs.  We note that
source 25 has a luminosity of $\lallband = 1.03 \times 10^{40}\
\ergs$, with $P(\lallband > \lulx) > 0.99$ and is very well fit by the
absorbed power-law model ($p < 0.002$).

Finally, we note that NGC 4291 is also represented in the
\emph{XMM-Newton} catalog of ULXs due to \citet{2011MNRAS.tmp.1147W}.
In NGC 4291 we find 2 ULX candidates (sources 37 and 39).  Given the
best-fit absorption column, both of these are likely to have a flux
bright enough to be true ULXs ($\P(\lallband > \lulx) > 0.96$), and
source 40 has $L_F = 2.5^{+1.2}_{-0.9} \times 10^{39}$.  Neither
source, however, is listed as a ULX in \citet{2011MNRAS.tmp.1147W}.
These two sources are within 9\arcsec of the nuclear source and would
not be reliably resolved with \emph{XMM-Newton}, and
\citet{2011MNRAS.tmp.1147W} did not consider bright sources within
15\arcsec\ of the center of elliptical galaxies or any sources within
7\farcs5 of the center of the galaxy.  On the other hand,
\citet{2011MNRAS.tmp.1147W} list 2XMM J122012.5$+$752204 as a ULX
candidate, using a definition of $\lulx = 1 \times 10^{39}\ \ergs$.
This \emph{XMM} source is consistent with the position of our source
36 (CXOU J122012.1$+$752203) for which we measure a luminosity
$\lallband = 4.0^{+3.0}_{-1.6} \times 10^{38}\ \ergs$.  The
\citet{2011MNRAS.tmp.1147W} classification of this source as a ULX is
based on the 2XMMS serendipitous source catalog
\citep{2009A&A...493..339W} 0.2--12 keV flux of $(5.5 \pm 1.0) \times
10^{-14} \units{erg\ s^{-1}\ cm^{-2}}$, which, at our adopted
distance, corresponds to an isotropic luminosity of $(4.1 \pm 0.8)
\times 10^{39}$.  The 2XMMS flux measurement assumes a spectral form
of an absorbed power-law with $\Gamma = 1.7$ and $N_\mathrm{H} = 3
\times 10^{20} \units{cm^{-2}}$, very close to the best-fit values we
find from our spectral fit.  If we fix $\Gamma = 1.7$ and
$N_\mathrm{H} = 3 \times 10^{20} \units{cm^{-2}}$ and fit to our
\emph{Chandra} data, we get a 0.2--12 keV flux of $(4.9 \pm 1.5)
\times 10^{-15} \units{erg\ s^{-1}\ cm^{-2}}$.  Given that ten years
elapsed between the \emph{XMM} observation (MJD 51671) and the
\emph{Chandra} observation (MJD 55541), we conclude that the source is
variable on these scales.

\subsection{Off-nuclear sources}
\begin{figure}[tbh]
\hypertarget{colorintensity}{}%
\centering
\includegraphics[width=\columnwidth]{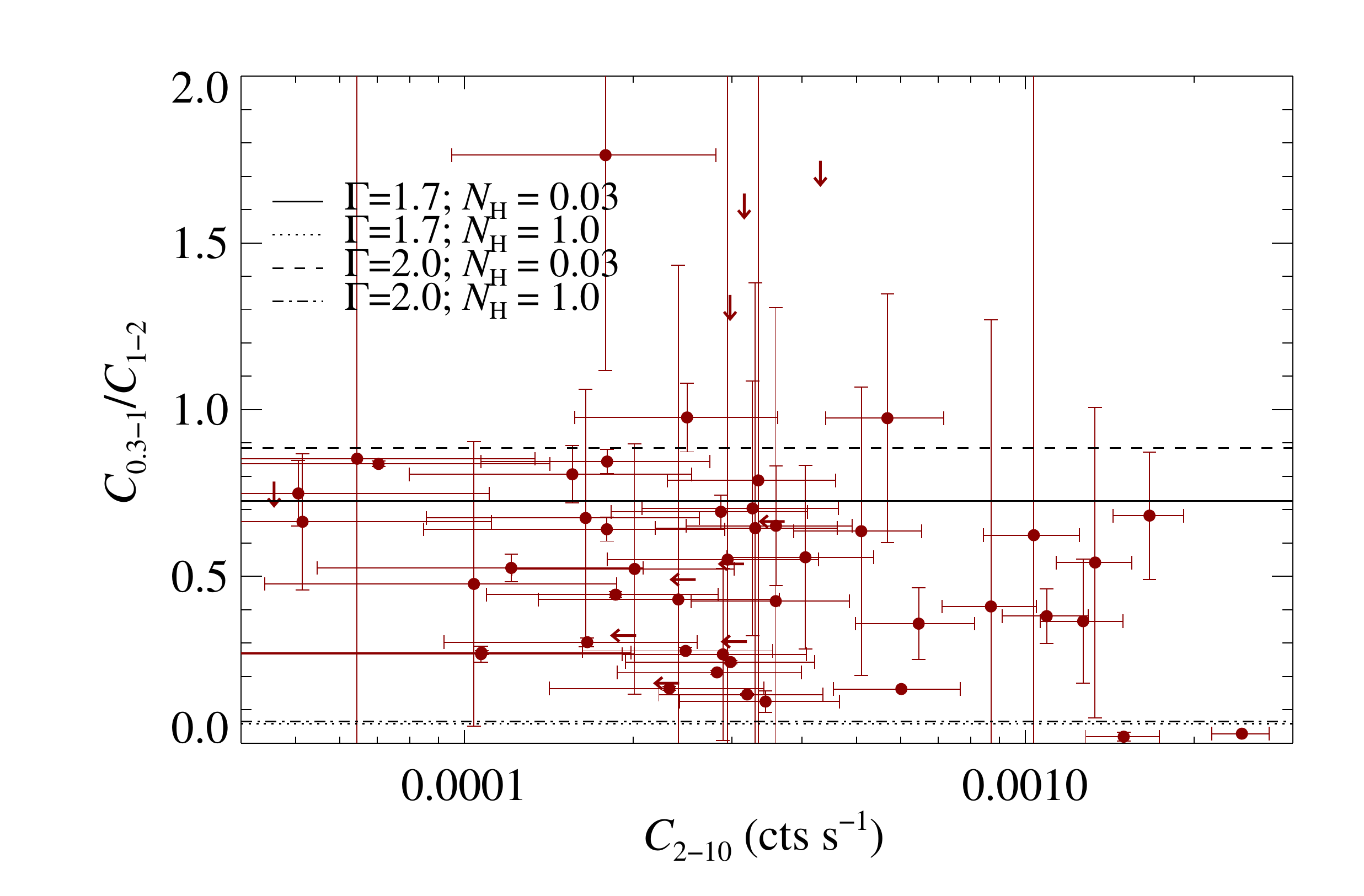}
\caption{Color-intensity plot of off-nuclear sources.  As a function
of $C_{2\textrm{--}10}$, the 2--10 keV count rate, we plot the ratio
of the count rates in the 0.3--1 to 1--2 keV bands.  Limits are given
with arrows at the 3$\sigma$ limit.  We do not plot source 4, which
only had upper limits in all three bands.  The horizontal lines are
color-intensity curves for an absorbed power-law model with the
parameters and indicated in the legend ($N_\mathrm{H}$ in units of
$10^{22}\units{cm^{-2}}$).  With only 2 data points (one color and one
intensity), it is not possible to break the degeneracy between
$N_\mathrm{H}$ and $\Gamma$, but the range of parameters needed to
reproduce most of the data points is reasonable.}
\label{figcolorintensity}
\end{figure}

In addition to the nuclear and ULX candidate sources, we detected 36
off-nuclear sources whose position on the sky is consistent with being
in the galaxy.  As the flux goes down, the probability for a source to
be a background AGN increases.  Thus we cannot be certain that the
sources are intrinsic to the galaxy, but, generally, the clustering of
point sources within the galaxies' optical extent is much higher than
over the entire \emph{Chandra} image.  To summarize the off-nuclear
sources, in Figure \ref{figcolorintensity} we plot
$C_{0.3\textrm{--}1} / C_{1\textrm{--}2}$, the ratio of the count
rates in the 0.3--1 to 1--2 keV bands, as a function of
$C_{2\textrm{--}10}$, the 2--10 keV count rate.  With only 2 data
points (one color and one intensity), it is not possible to break the
degeneracy in an absorbed power-law model between $N_\mathrm{H}$ and
$\Gamma$, but the range of parameters needed to reproduce most of the
data points is reasonable.

\bookmarksetup{color=[rgb]{0.54,0,0}}
\bookmark[rellevel=1,keeplevel,dest=colorintensity]{Fig \ref*{figcolorintensity}: Color-intensity plot of off-nuclear sources.}
\bookmarksetup{color=[rgb]{0,0,0}}

\section{Summary}
\label{summary}

In this paper we have presented an X-ray survey of 12 galaxies with
central black hole mass measurements.  The observations were designed
to characterize the nuclear source of each galaxy but were
sensitive to much more.
\begin{enumerate}
\item Each galaxy was observed for 30 ksec with \emph{Chandra}, and in
  total we detected 68 point sources in the region of the sky occupied
  by these galaxies.
\item We detected all 12 nuclear sources with sufficient count rates
  to model their spectra and determine their X-ray luminosities and
  Eddington ratios.  The sources all were found to be emitting in the
  2--10 keV band at $10^{-8}$--$10^{-6}$ of Eddington.  
\item When fitting with an absorbed power-law spectral model, we found
  $\Gamma$, the photon spectral index.  Fitting for $\Gamma$ as a
  function of Eddington fraction, we found a negative correlation,
  consistent with several earlier reports on low luminosity AGNs.  Our
  best fit was $\Gamma = 1.8 \pm 0.2 - (0.24 \pm 0.12) \log(\lhard /
  10^{-7} \ledd)$ with an rms intrinsic scatter of $0.65 \pm 0.20$,
  which is consistent with predictions from ADAF models, which expect
  bremsstrahlung emission to become more important at lower accretion
  rates.
\item Our observations were also sensitive to ULXs in the target
  galaxies.  We found 20 ULX candidates.  Based on considerations of
  the probability distribution of their intrinsic fluxes and the
  probability of having a background AGN of sufficient brightness to
  appear as a ULX, we concluded that 6 of these candidates are likely
  ($>90\%$ chance) to be true ULXs.  The most promising ULX candidate
  is in NGC 2748 and has an isotropic luminosity of $\lallband =
  1.03_{-0.27}^{+0.57} \times 10^{40}\ \ergs$.
\item We also present a color-intensity plot of the remaining point
  sources, most of which are likely to be X-ray binaries local to the
  galaxy.
\end{enumerate}

This work will be followed up with EVLA observations of the nuclear
sources.  When combining radio and X-ray data, we will be able to
provide a complete mass-calibrated fundamental plane that will allow
for the estimation of black hole masses using X-ray and radio
observations.  We will also, with the possible addition of archival
sub-millimeter data, broad-band spectral energy distribution modeling.

\hypertarget{ackbkmk}{}
\acknowledgements We thank Elena Gallo, Dominic Walton, Jimmy Irwin,
Aneta Siemiginowska, and the anonymous referee for helpful discussions
and comments.  We thank the \emph{Chandra} ACIS Team for advice on
data analysis.

K.G.\ acknowledges support provided by the National Aeronautics and
Space Administration through Chandra Award Number GO0-11151X issued by
the Chandra X-ray Observatory Center, which is operated by the
Smithsonian Astrophysical Observatory for and on behalf of the
National Aeronautics Space Administration under contract NAS8-03060.
K.G.\ also thanks the Aspen Center for Physics for their hospitality.
S.M. acknowledges support from a Netherlands Organization for
Scientific Research (NWO) Vidi Fellowship, as well as The European
Community's 7th Framework Program (FP7) under grand agreement number
ITN 215212 ``Black hole Universe.''  
D.O.R.\ thanks the Institute for Advanced Study and acknowledges
support of a Corning Glass Works Foundation Fellowship.

This work made use of the NASA's Astrophysics Data System (ADS), and
the NASA/IPAC Extragalactic Database (NED), which is operated by the
Jet Propulsion Laboratory, California Institute of Technology, under
contract with the National Aeronautics and Space Administration.  This
research has made use of the VizieR catalogue access tool, CDS,
Strasbourg, France.

\bookmark[level=0,dest=ackbkmk]{Acknowledgements}

\bibliographystyle{apjads}
\hypertarget{refbkmk}{}%
\bookmark[level=0,dest=refbkmk]{References}
\bibliography{gultekin}

\clearpage
\begin{figure*}[tbh]
\hypertarget{spectra}{}%
\centering
\includegraphics[width=0.45\columnwidth]{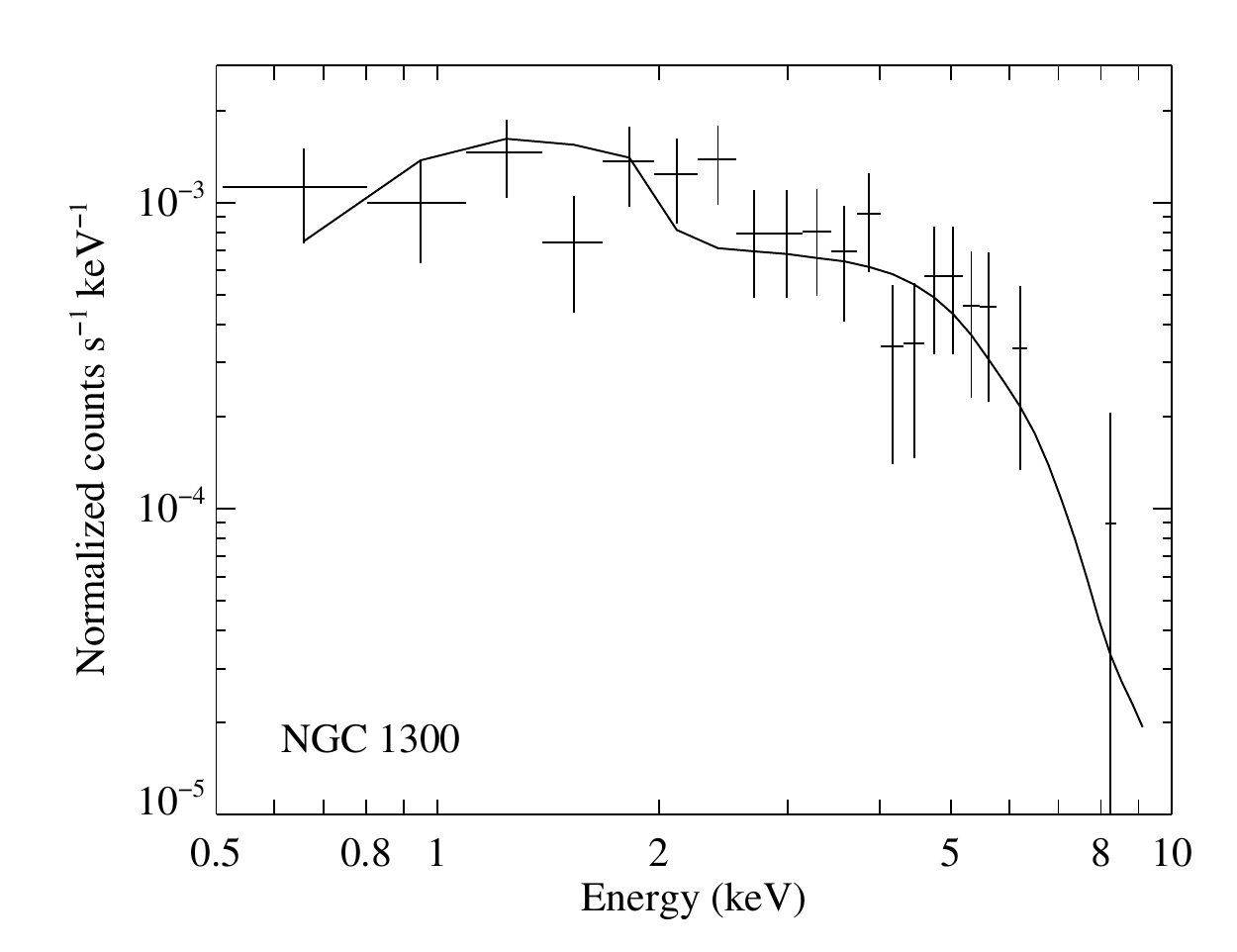}
\includegraphics[width=0.45\columnwidth]{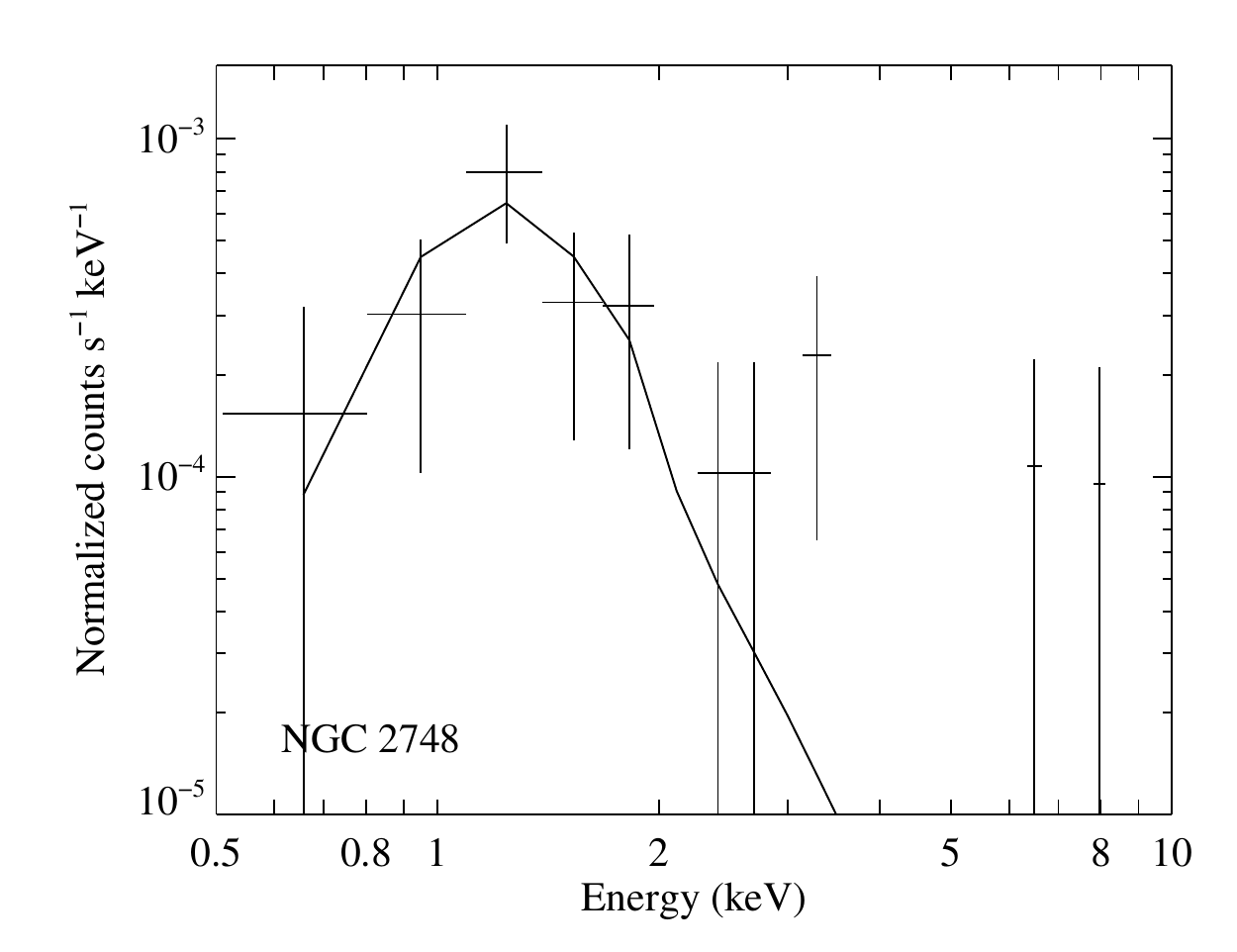}
\includegraphics[width=0.45\columnwidth]{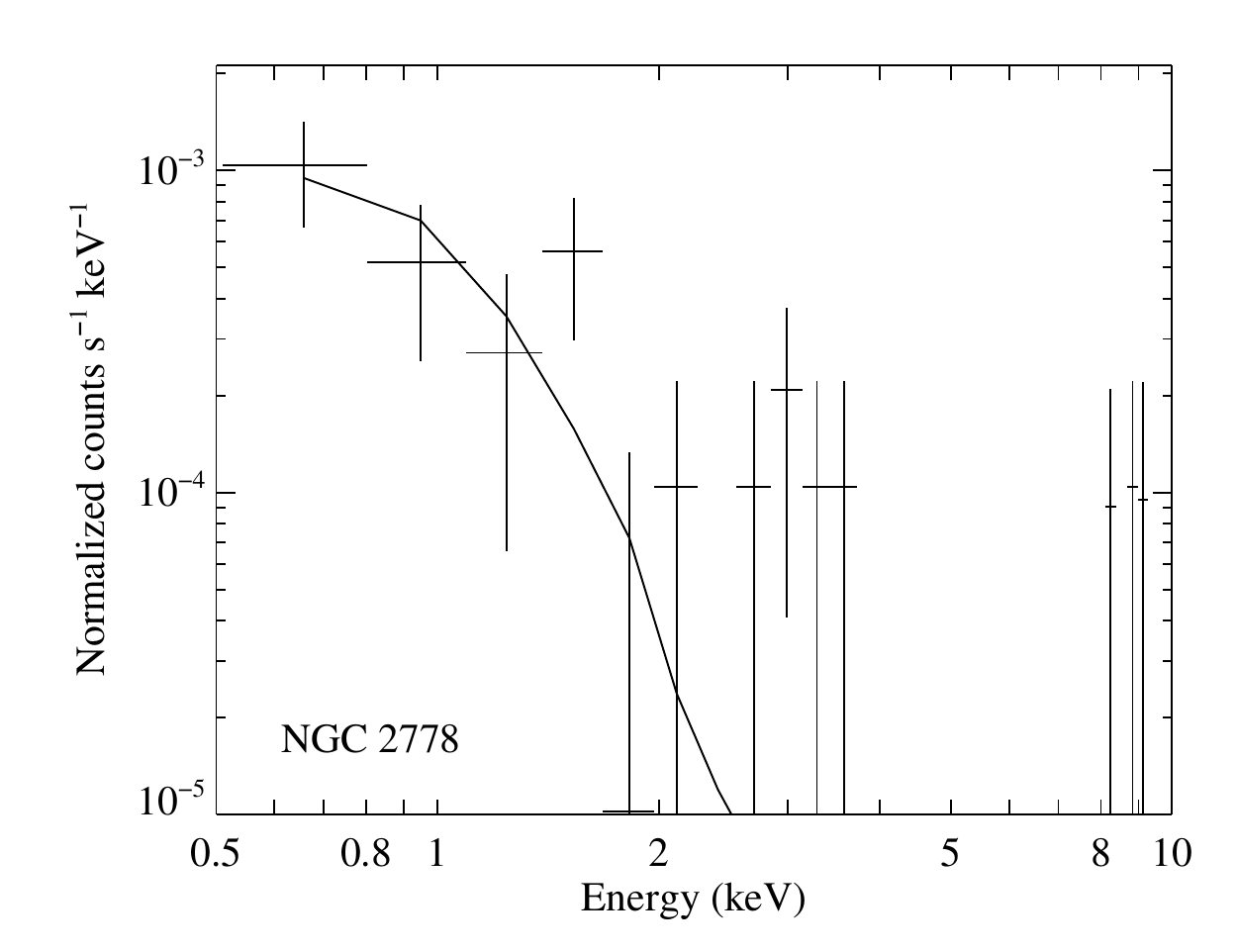}
\includegraphics[width=0.45\columnwidth]{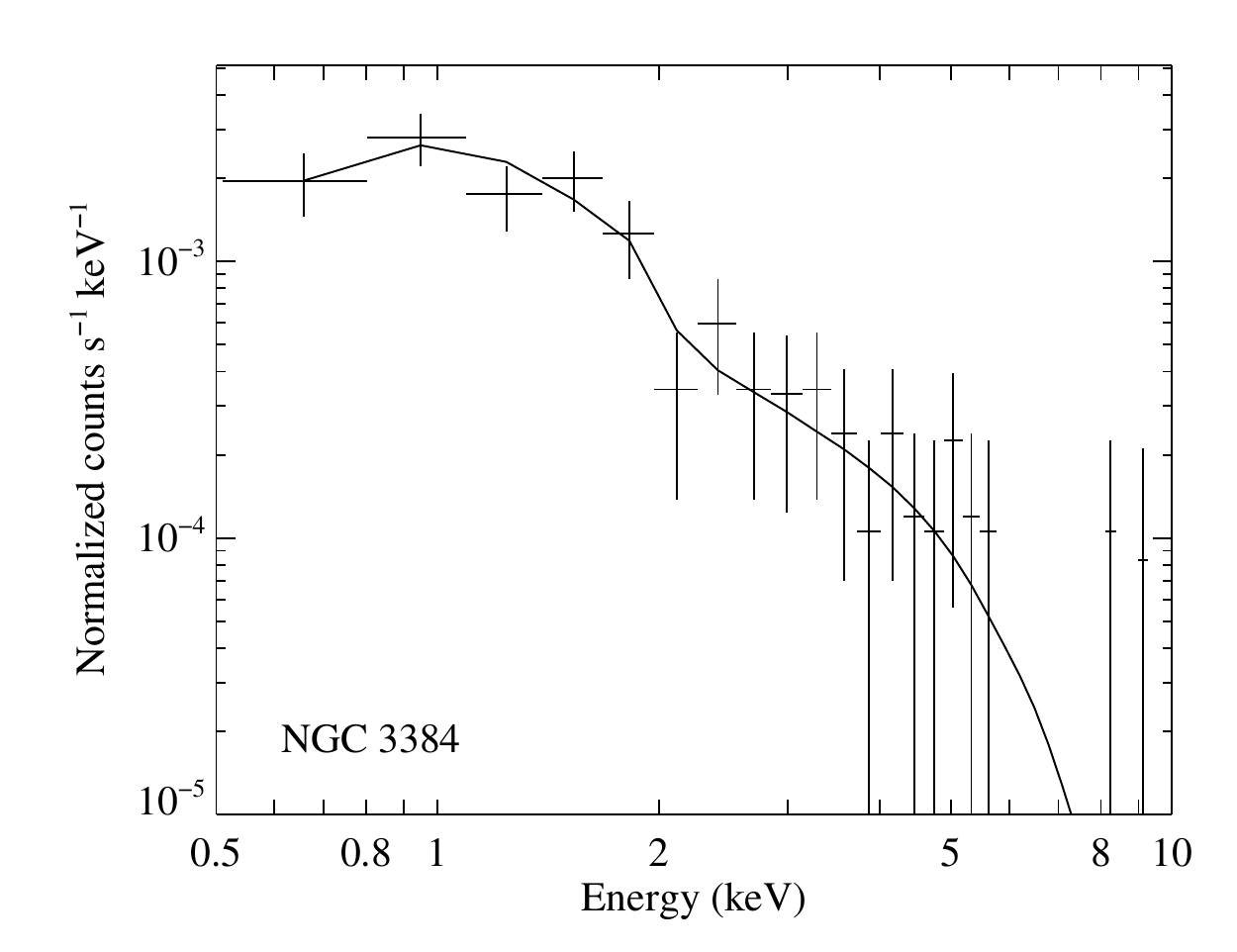}
\includegraphics[width=0.45\columnwidth]{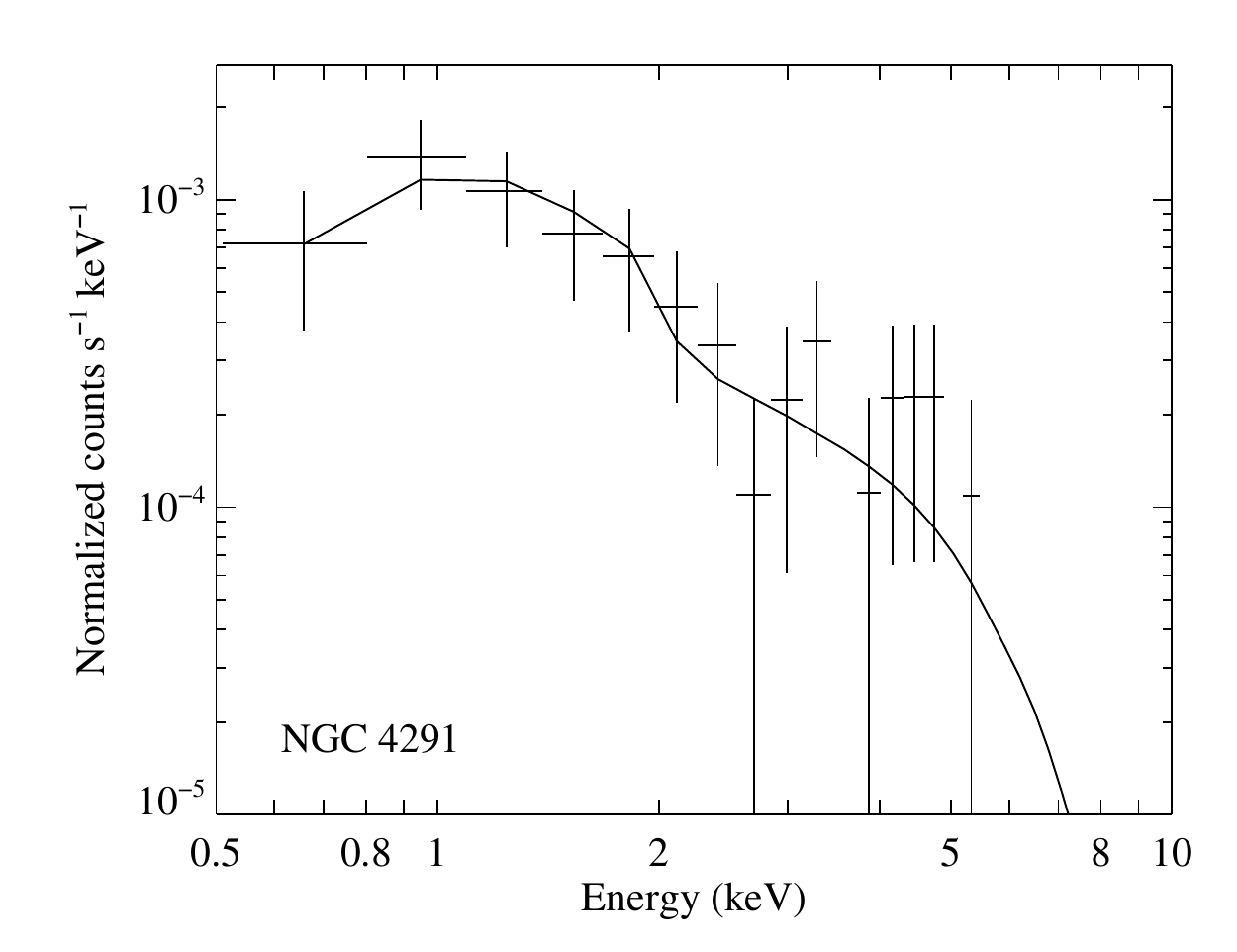}
\includegraphics[width=0.45\columnwidth]{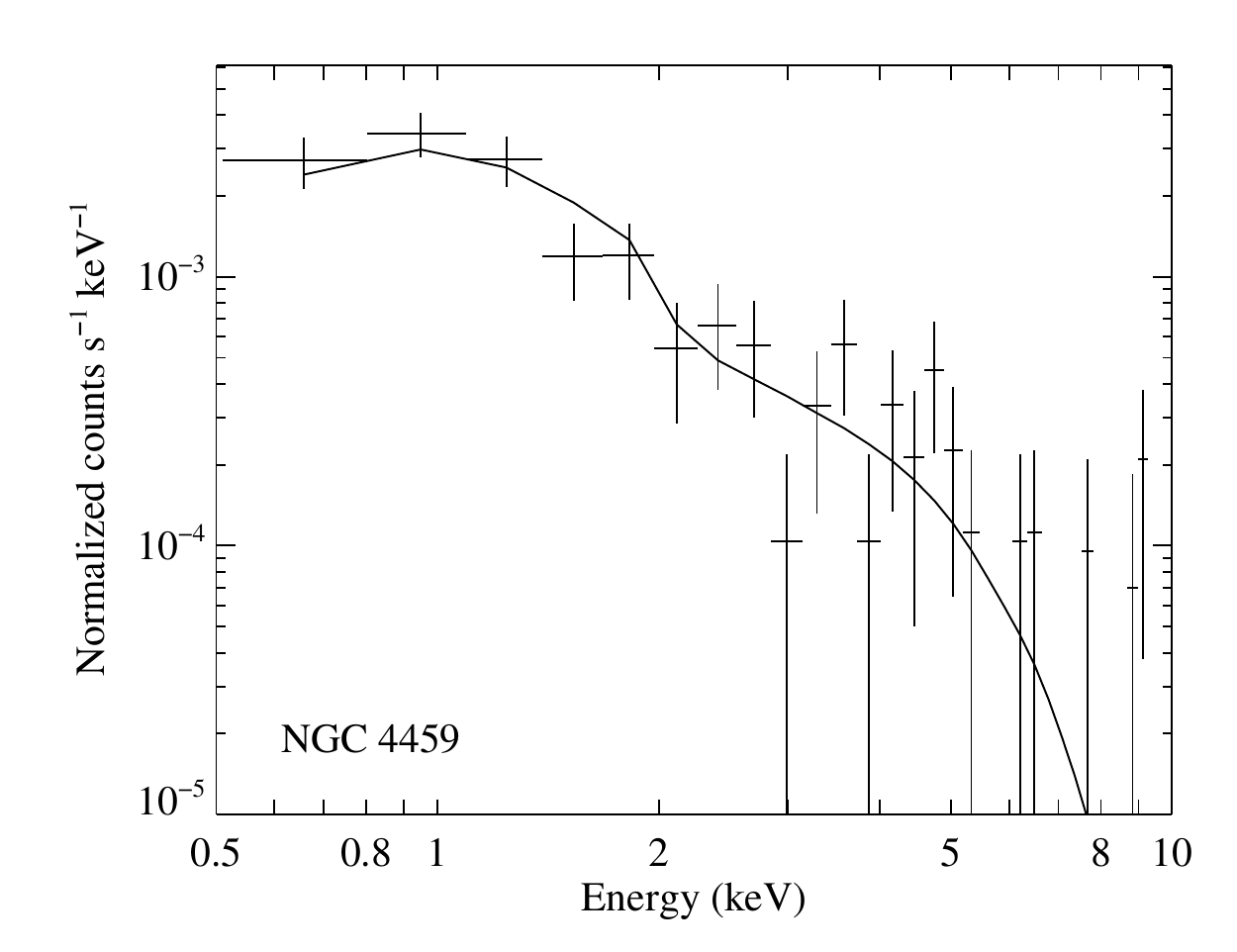}
\caption{X-ray spectra, folded through the instrument response, of the
first half of the sample.  The spectra are binned for visualization
purposes her.  The solid line shows the best-fit model.  The second
half of the sample is in Figure \ref{figspectra2}.  Error bars show 1$\sigma$ uncertainties.}
\label{figspectra}
\end{figure*}
\bookmarksetup{color=[rgb]{0.54,0,0}}
\bookmark[rellevel=1,keeplevel,dest=spectra]{Fig \ref*{figspectra}: Spectra of nuclear sources.}
\bookmarksetup{color=[rgb]{0,0,0}}

\begin{figure*}[tbh]
\hypertarget{spectra2}{}%
\centering
\includegraphics[width=0.45\columnwidth]{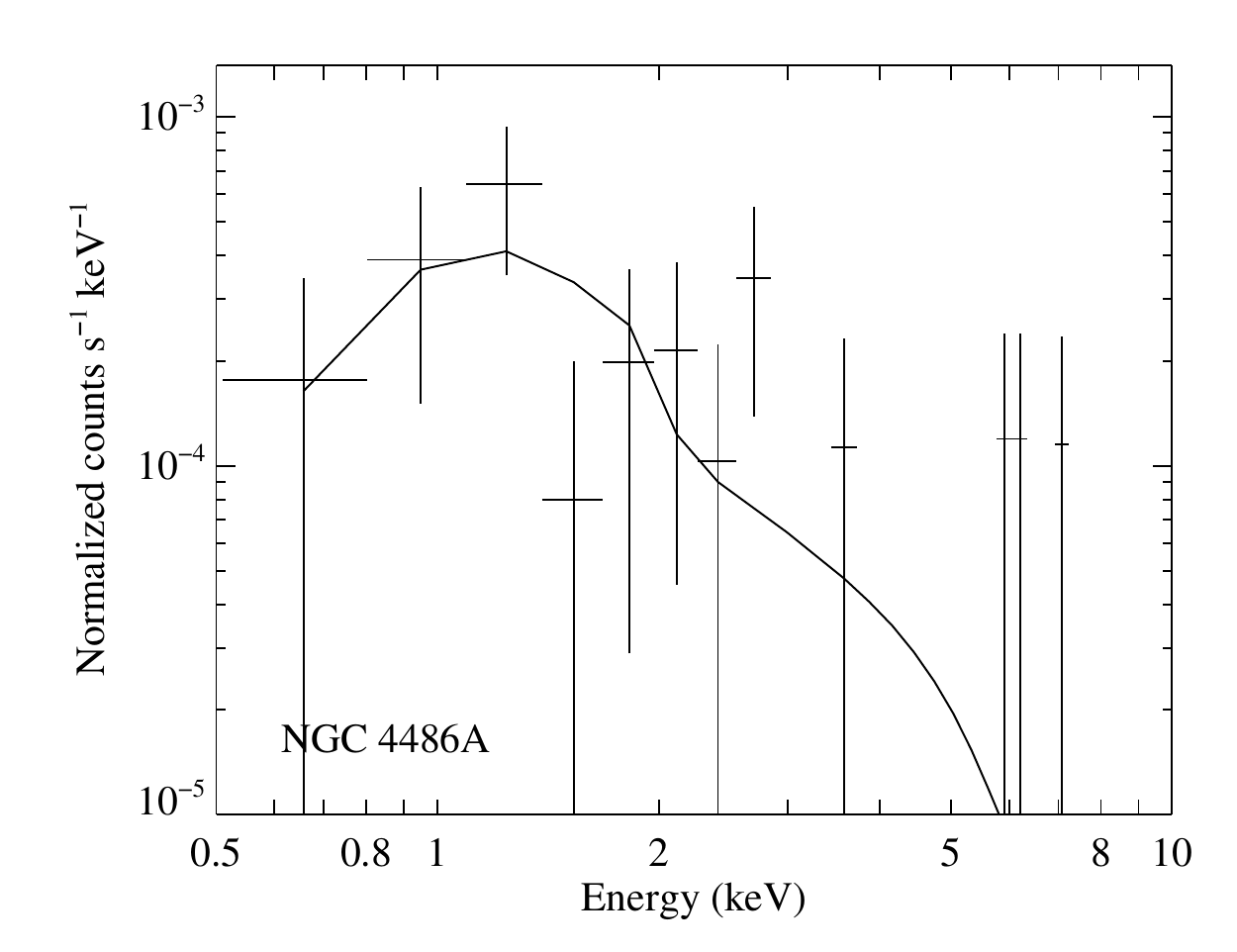}
\includegraphics[width=0.45\columnwidth]{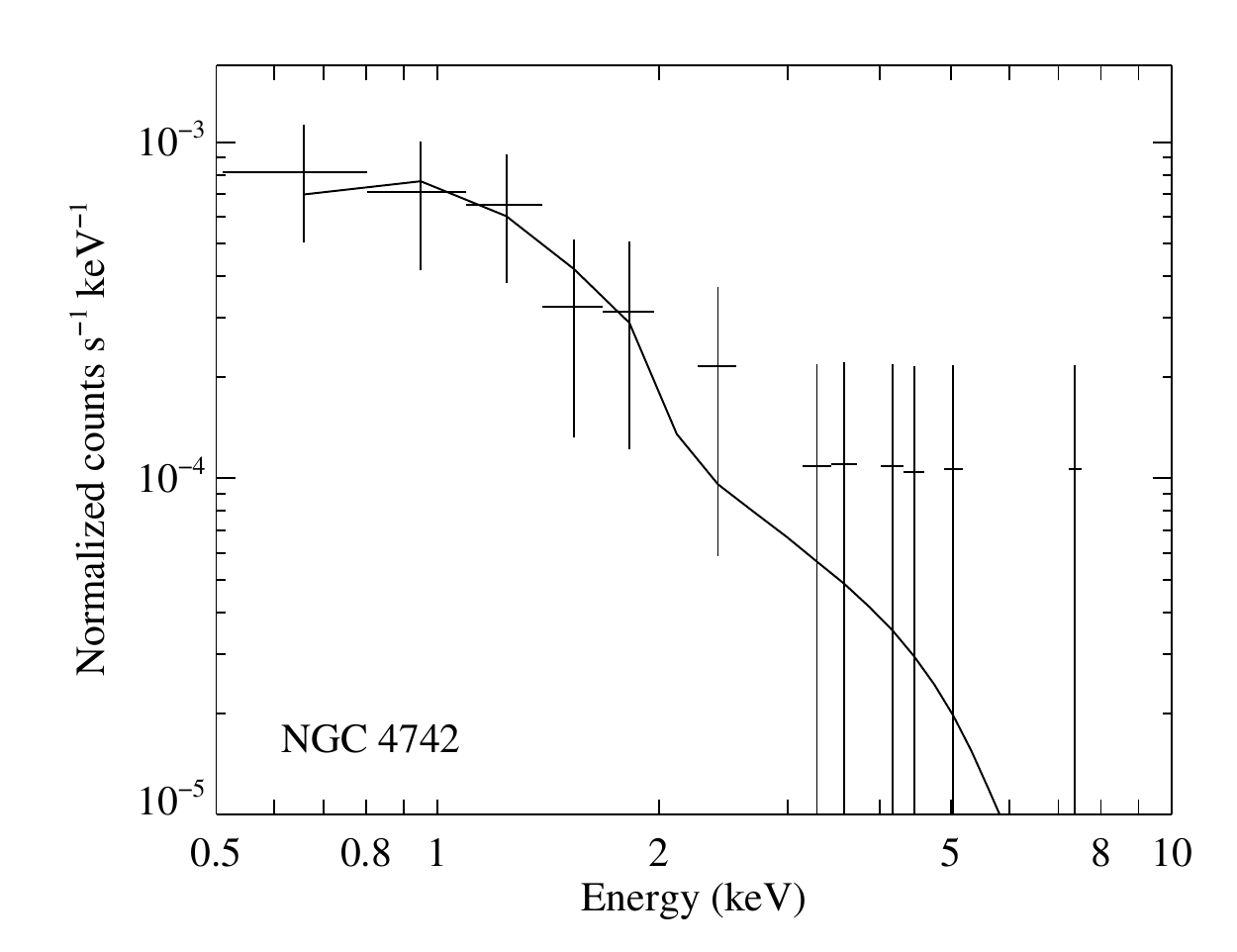}
\includegraphics[width=0.45\columnwidth]{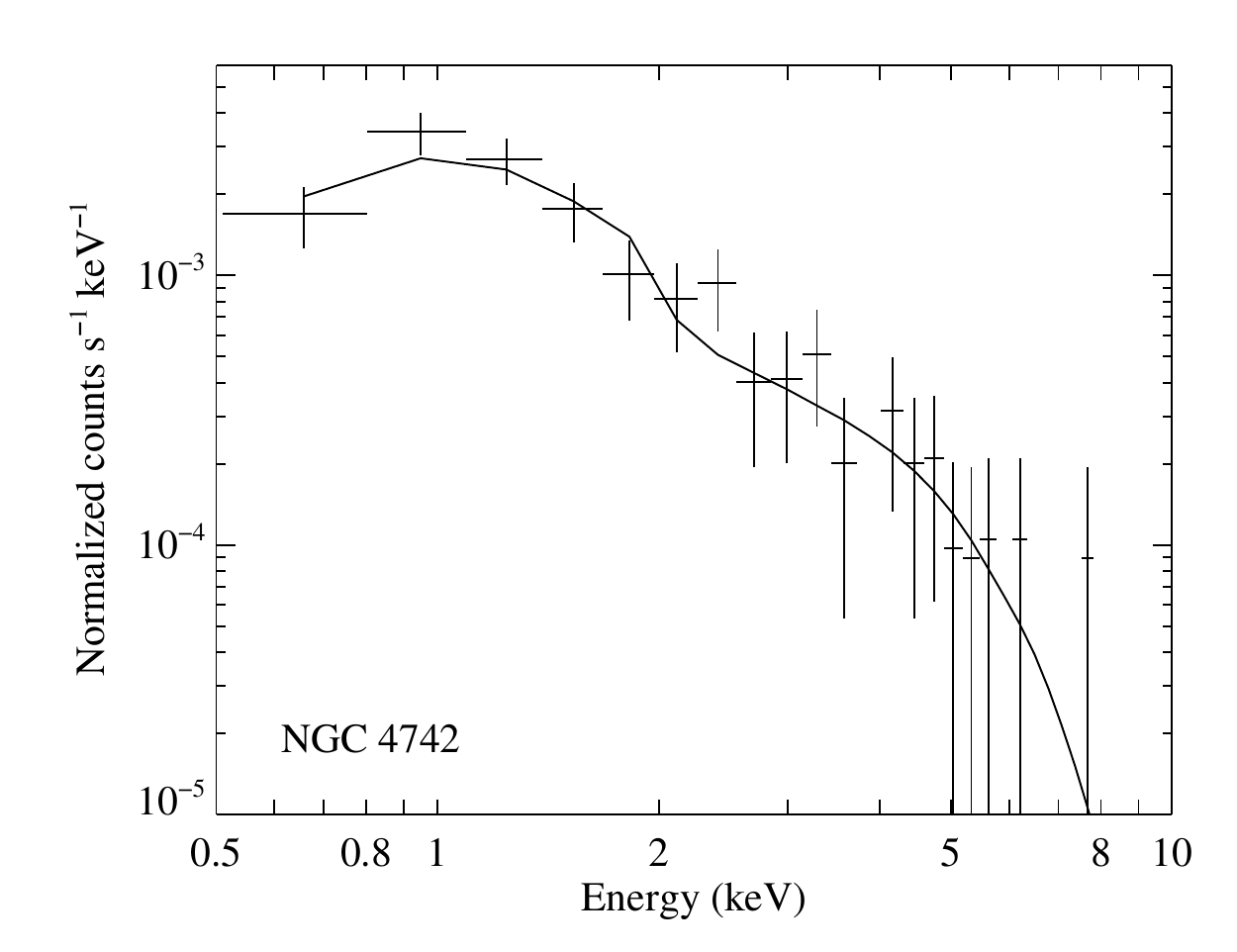}
\includegraphics[width=0.45\columnwidth]{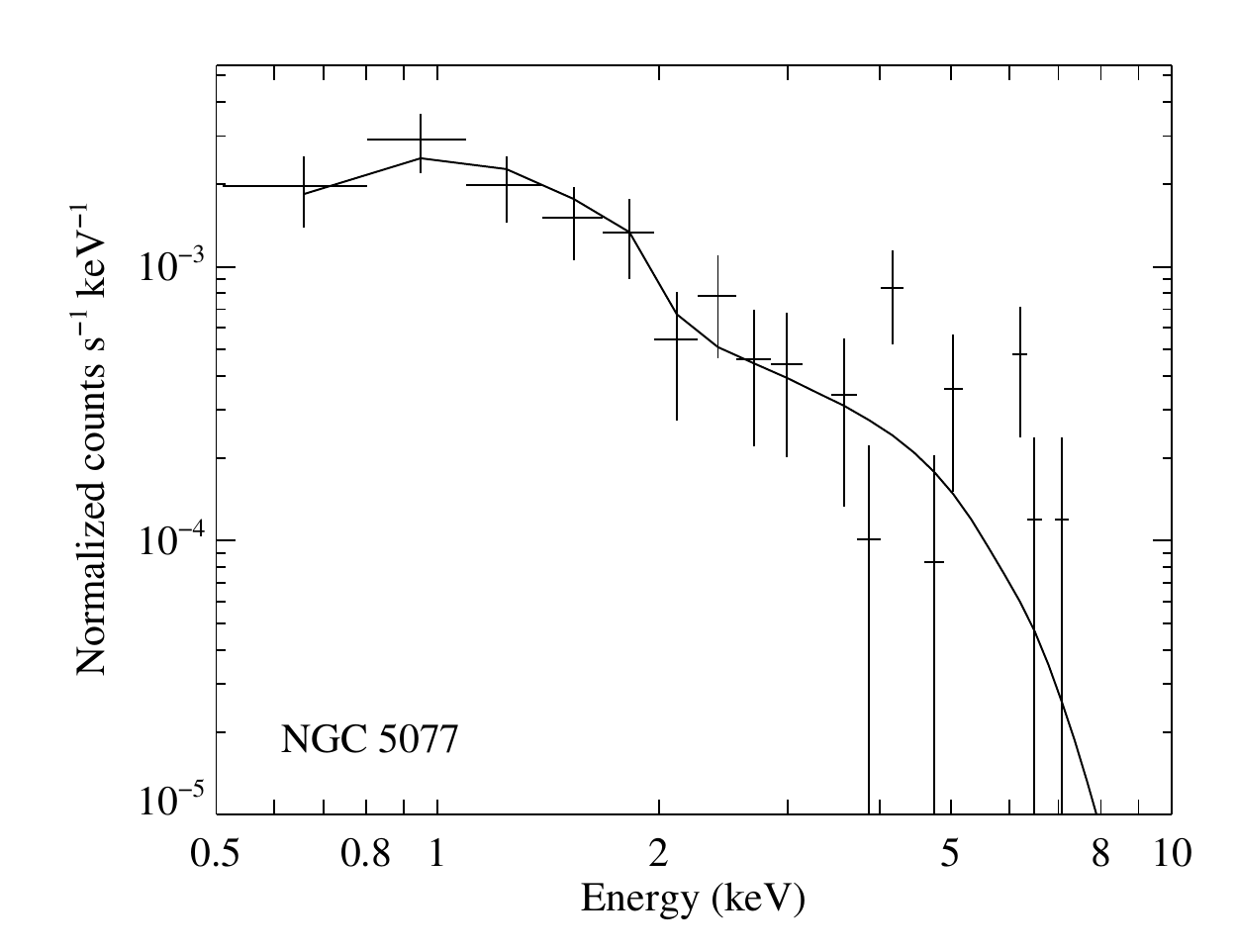}
\includegraphics[width=0.45\columnwidth]{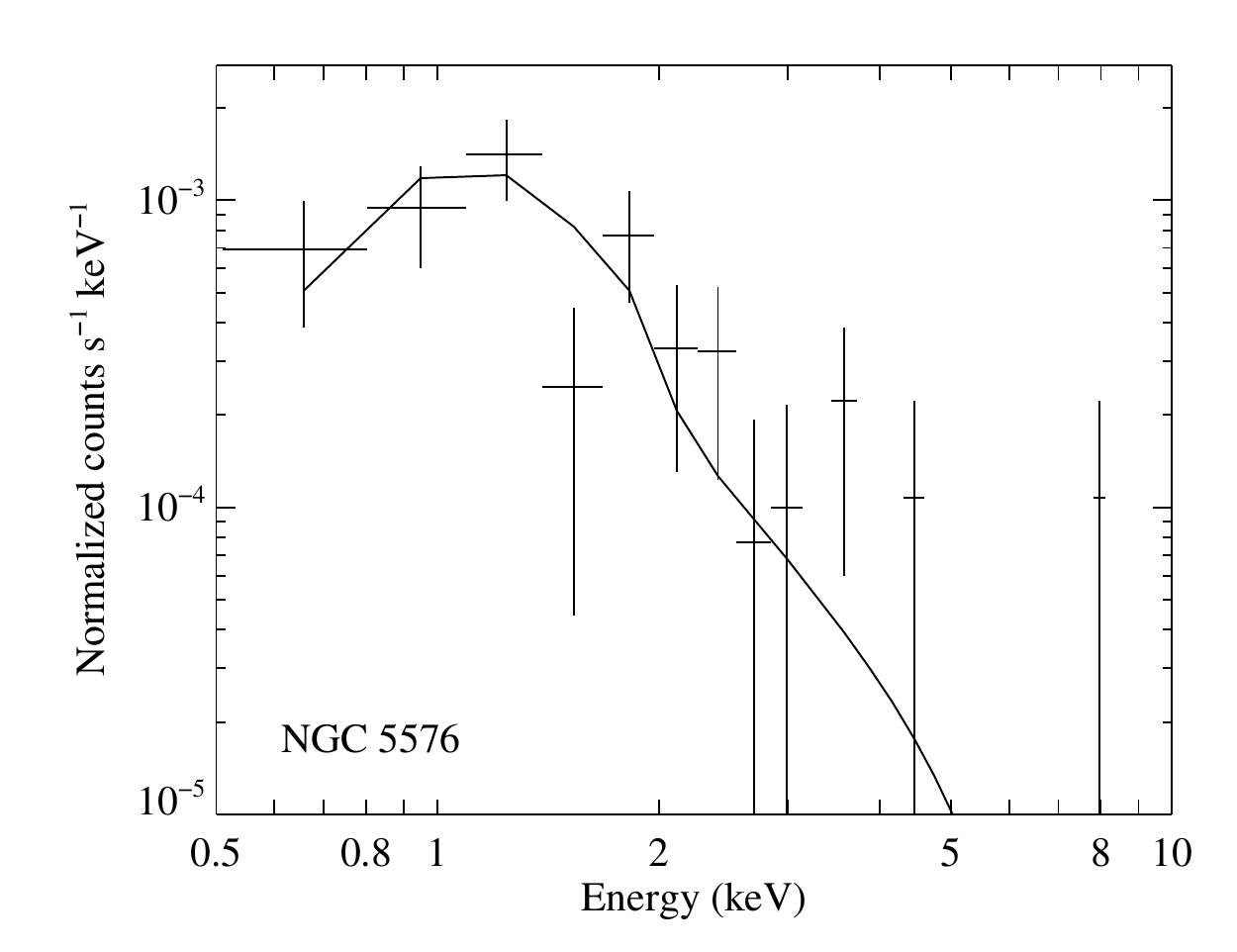}
\includegraphics[width=0.45\columnwidth]{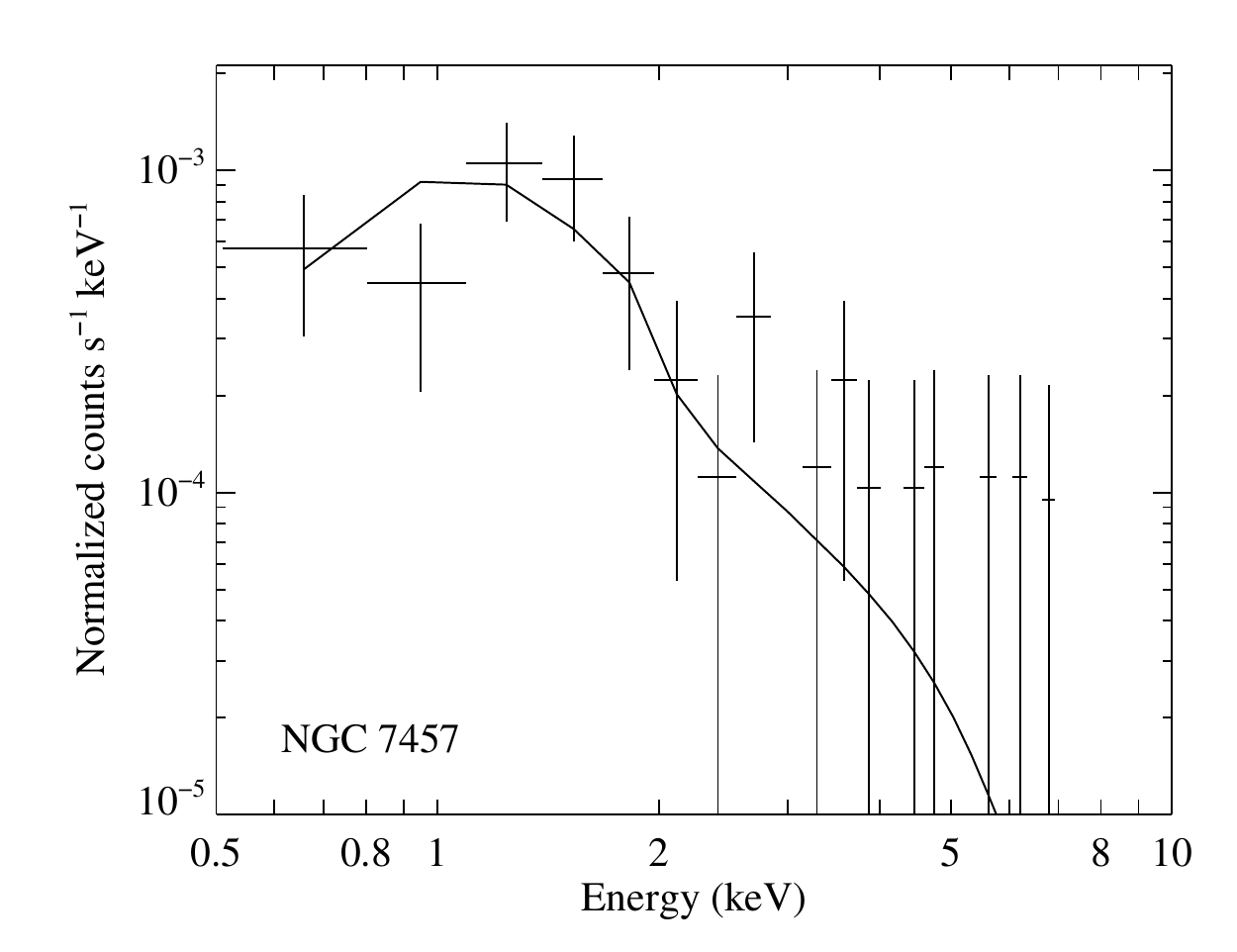}
\caption{Same as Figure \ref{figspectra} for the second half of the sample.  Error bars show 1$\sigma$ uncertainties.}
\label{figspectra2}
\end{figure*}
\bookmarksetup{color=[rgb]{0.54,0,0}}
\bookmark[rellevel=1,keeplevel,dest=spectra2]{Fig \ref*{figspectra2}: Spectra of nuclear sources, continued.}
\bookmarksetup{color=[rgb]{0,0,0}}

\begin{figure*}[tbh]
\hypertarget{ulxspectra}{}%
\centering
\includegraphics[width=0.45\columnwidth]{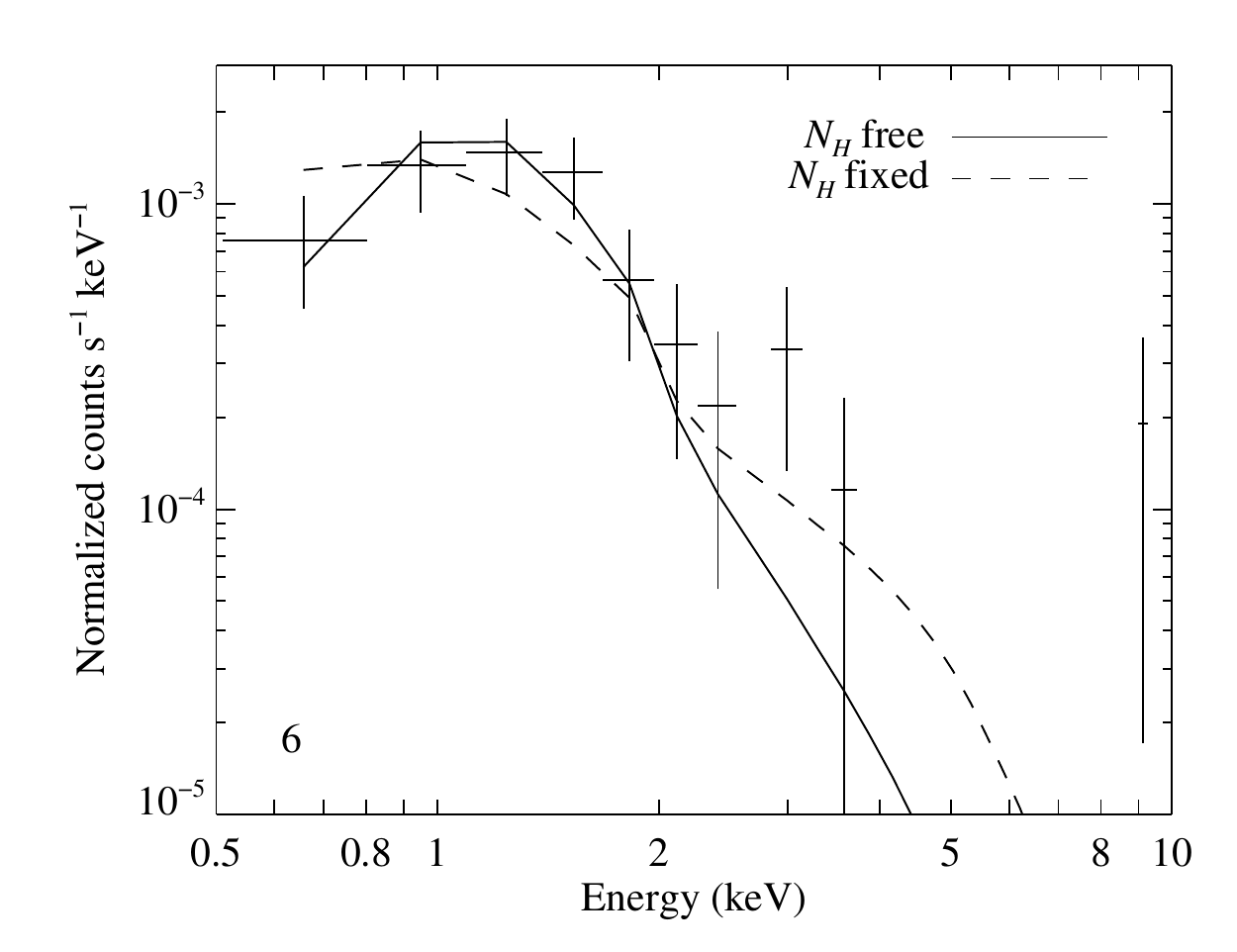}
\includegraphics[width=0.45\columnwidth]{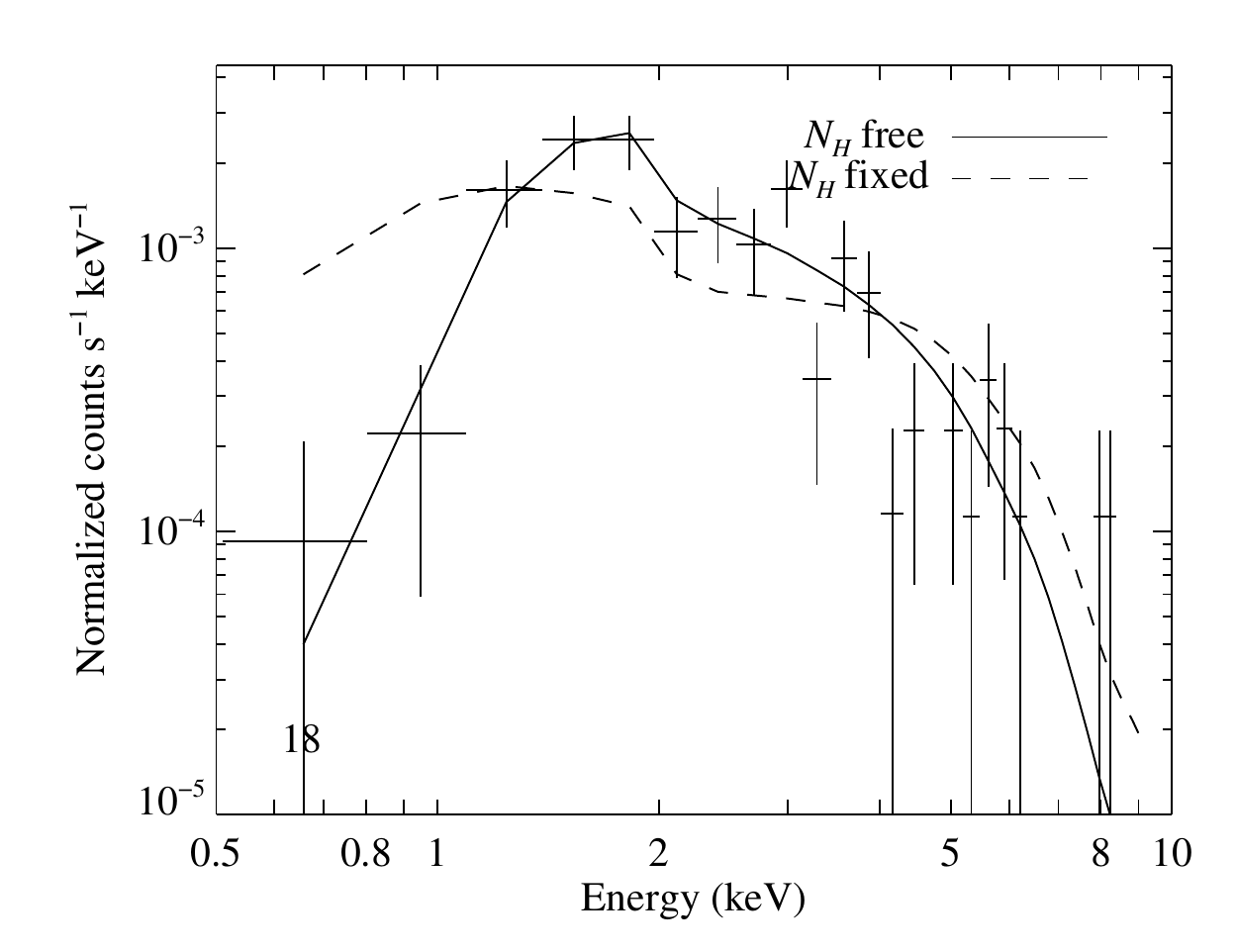}
\includegraphics[width=0.45\columnwidth]{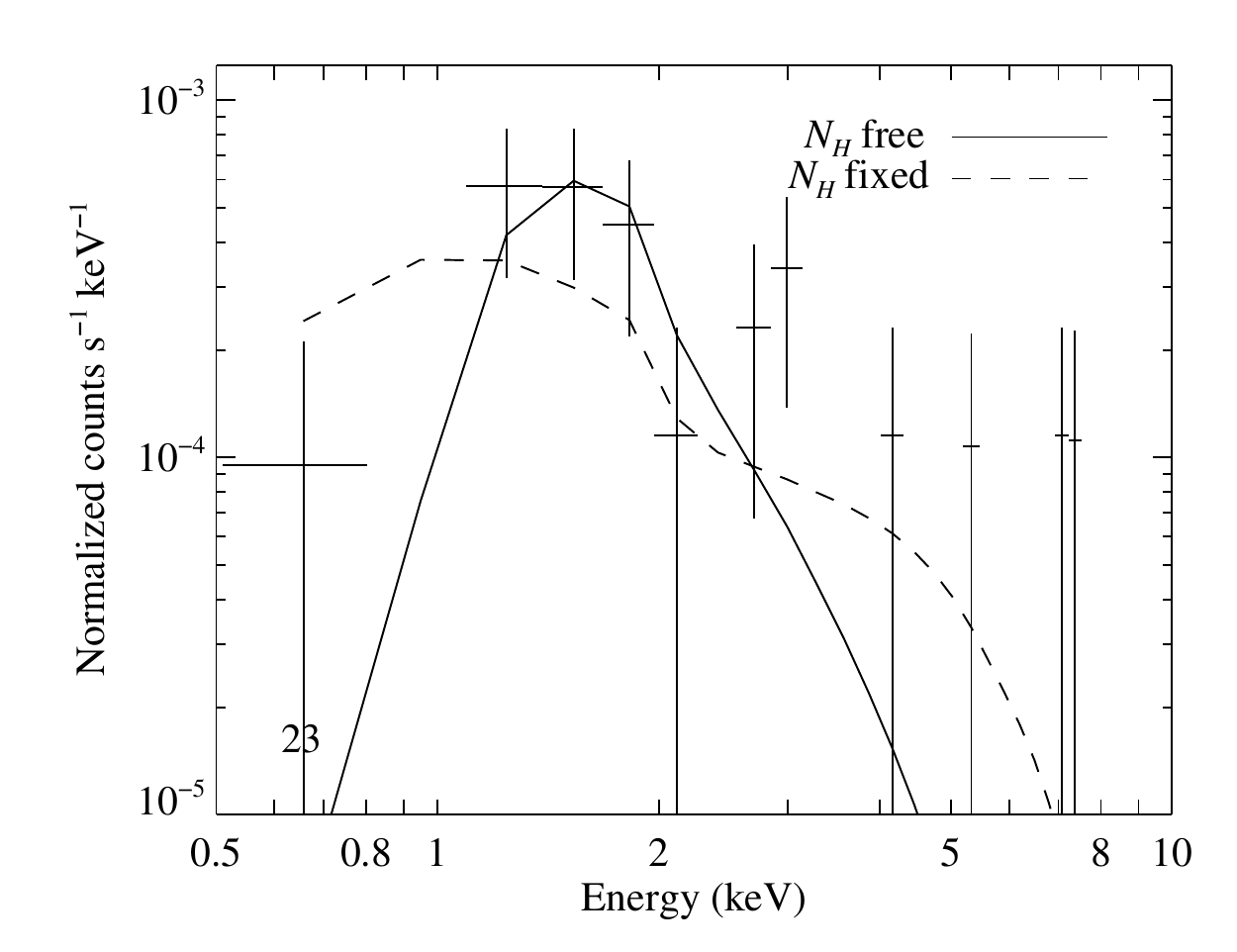}
\includegraphics[width=0.45\columnwidth]{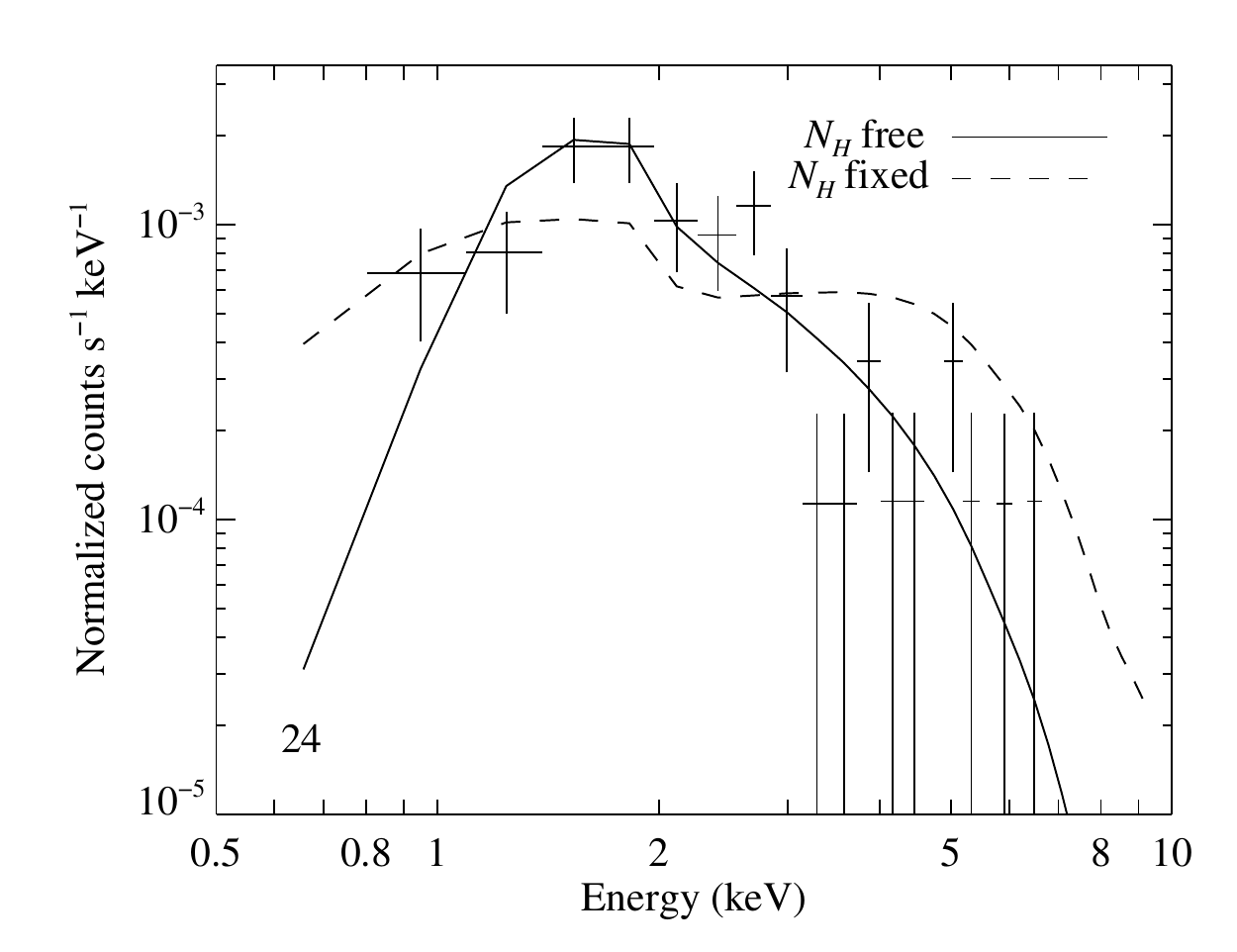}
\includegraphics[width=0.45\columnwidth]{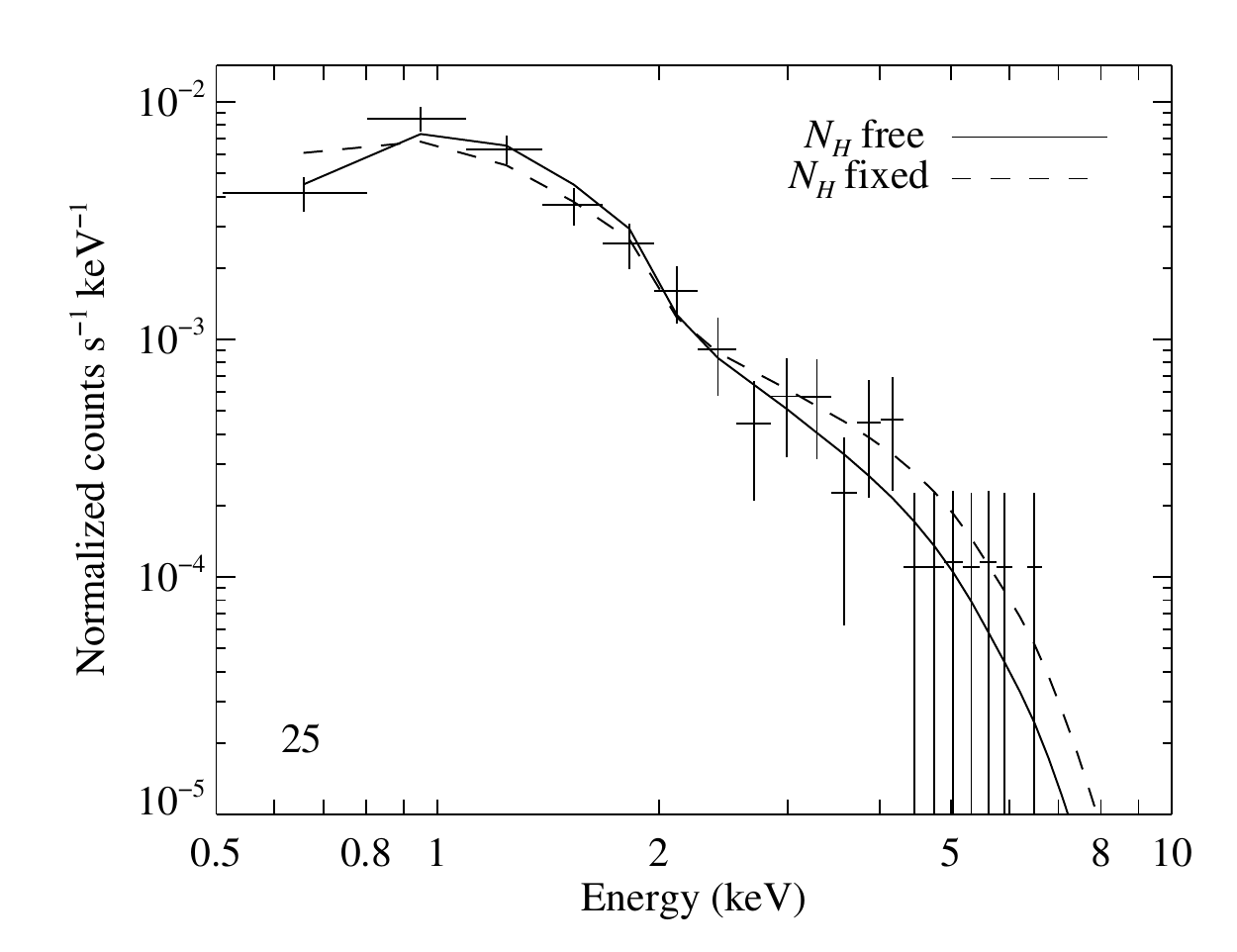}
\includegraphics[width=0.45\columnwidth]{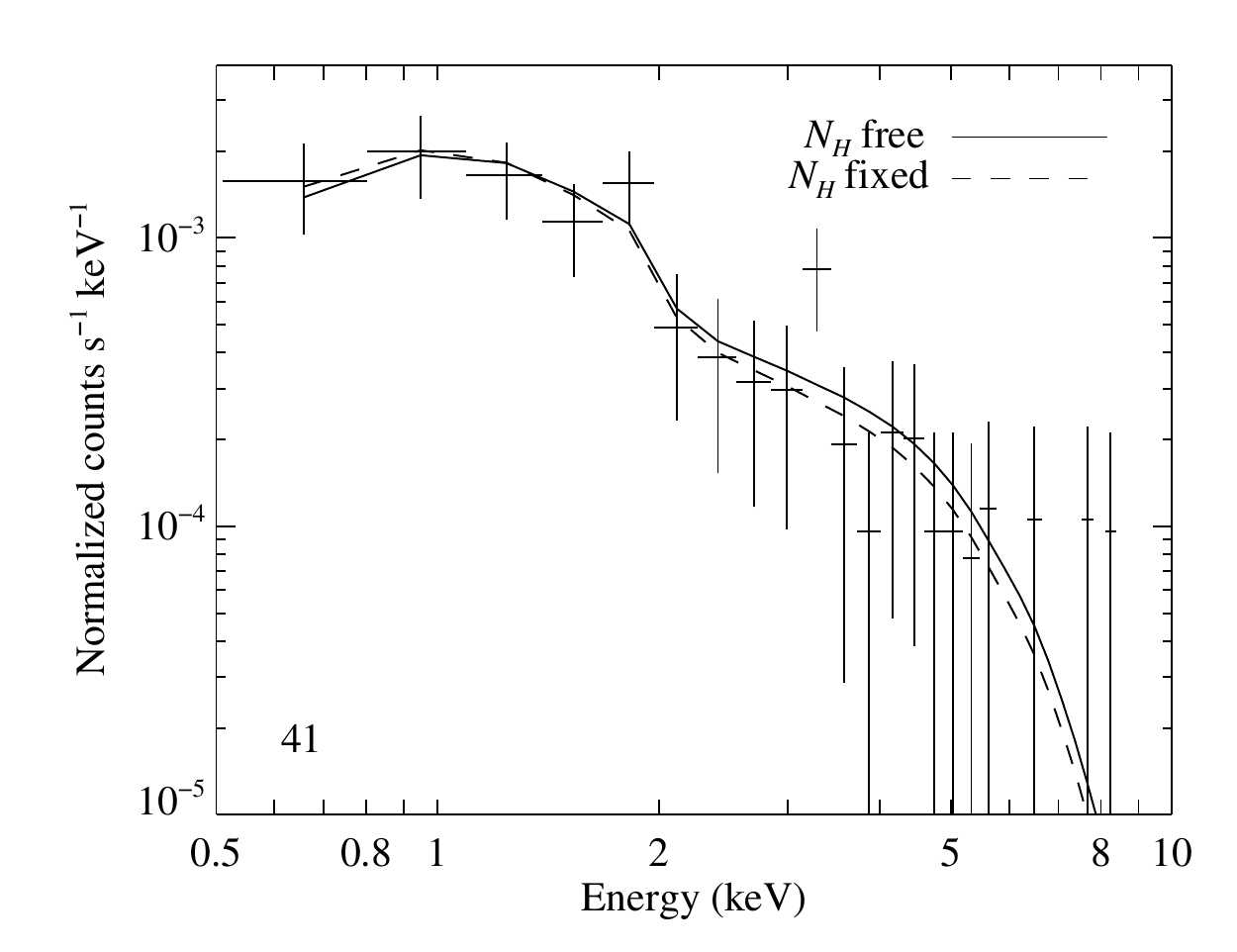}
\caption{X-ray spectra of strong ULX candidates.  Error bars show
1$\sigma$ uncertainties.  The solid lines are power-law models in
which the $N_H$ parameter was allowed to vary above the Galactic
value; the dashed lines are models with $N_H$ fixed at the Galactic
value towards each source.}
\label{figulxspectra}
\end{figure*}
\bookmarksetup{color=[rgb]{0.54,0,0}}
\bookmark[rellevel=1,keeplevel,dest=ulxspectra]{Fig \ref*{figulxspectra}: Spectra of ULX candidates.}
\bookmarksetup{color=[rgb]{0,0,0}}

\clearpage
\ifthenelse{\value{emulateapj} = 1}
{
\LongTables
}
{}

    \tabletypesize{\scriptsize}
    \def\arraystretch{1.200}
    \begin{deluxetable*}{lrlllcr@{}lr@{}lr@{}l}
    \tablecaption{\emph{Chandra} point source detections}
    \tablewidth{0pt}
    \tablehead{
    \colhead{Galaxy} & \colhead{ID} & \colhead{Source name} & \colhead{RA} & \colhead{Dec} & \colhead{Class.} & \multicolumn{2}{c}{0.3--1 keV} & \multicolumn{2}{c}{1--2 keV} &  \multicolumn{2}{c}{2--10 keV} \\
    \colhead{(1)} & \colhead{(2)} & \colhead{(3)} & \colhead{(4)} & \colhead{(5)} & \colhead{(6)} & \multicolumn{2}{c}{(7)} & \multicolumn{2}{c}{(8)} &  \multicolumn{2}{c}{(9)} 
}
    \startdata

NGC1300 & 1 &  CXOU J031935.0$-$192509 & 03:19:35.07 & $-$19:25:09.9 &  &  $ 4.5$&$\pm 1.3$ & $ 0.7$&$\pm 0.6$ & $< 3.2$&\\

\dots & 2 &  CXOU J031935.1$-$192417 & 03:19:35.10 & $-$19:24:17.4 &  &  $ 0.8$&$\pm 0.6$ & $ 1.5$&$\pm 0.8$ & $< 3.2$&\\

\dots & 3 &  CXOU J031935.4$-$192417 & 03:19:35.49 & $-$19:24:17.6 &  &  $ 2.0$&$\pm 0.8$ & $ 2.7$&$\pm 1.0$ & $ 0.5$&$\pm 0.5$\\

\dots & 4 &  CXOU J031936.7$-$192500 & 03:19:36.78 & $-$19:25:00.3 &  &  $< 2.0$& & $< 2.0$& & $< 3.8$&\\

\dots & 5 &  CXOU J031937.8$-$192607 & 03:19:37.88 & $-$19:26:07.8 &  &  $ 0.9$&$\pm 0.6$ & $ 5.6$&$\pm 1.4$ & $ 6.0$&$\pm 1.6$\\

\dots & 6 &  CXOU J031937.9$-$192441 & 03:19:37.95 & $-$19:24:41.4 & ULX &  $ 4.7$&$\pm 1.3$ & $11.0$&$\pm 1.9$ & $ 2.4$&$\pm 1.1$\\

\dots & 7 &  CXOU J031938.2$-$192413 & 03:19:38.21 & $-$19:24:13.5 &  &  $ 1.3$&$\pm 0.7$ & $ 2.0$&$\pm 0.8$ & $ 0.5$&$\pm 0.5$\\

\dots & 8 &  CXOU J031938.4$-$192318 & 03:19:38.45 & $-$19:23:18.0 &  &  $ 1.6$&$\pm 0.8$ & $ 2.4$&$\pm 1.0$ & $< 3.7$&\\

\dots & 9 &  CXOU J031938.5$-$192452 & 03:19:38.58 & $-$19:24:52.4 & ULX &  $ 2.5$&$\pm 1.0$ & $ 4.0$&$\pm 1.2$ & $ 1.8$&$\pm 1.0$\\

\dots & 10 &  CXOU J031939.3$-$192422 & 03:19:39.37 & $-$19:24:22.8 &  &  $ 2.2$&$\pm 0.9$ & $ 3.2$&$\pm 1.1$ & $ 2.9$&$\pm 1.1$\\

\dots & 11 &  CXOU J031939.4$-$192616 & 03:19:39.47 & $-$19:26:16.0 &  &  $< 2.6$& & $ 1.6$&$\pm 0.8$ & $ 3.2$&$\pm 1.2$\\

\dots & 12 &  CXOU J031939.9$-$192550 & 03:19:39.93 & $-$19:25:50.4 & ULX &  $ 3.2$&$\pm 1.1$ & $ 5.8$&$\pm 1.4$ & $ 2.9$&$\pm 1.2$\\

\dots & 13 &  CXOU J031941.0$-$192440 & 03:19:41.06 & $-$19:24:40.3 & Nuc. &  $ 5.5$&$\pm 1.5$ & $12.0$&$\pm 2.1$ & $27.3$&$\pm 3.1$\\

\dots & 14 &  CXOU J031941.3$-$192334 & 03:19:41.37 & $-$19:23:34.2 &  &  $ 3.0$&$\pm 1.0$ & $ 5.7$&$\pm 1.4$ & $ 1.2$&$\pm 0.8$\\

\dots & 15 &  CXOU J031942.5$-$192450 & 03:19:42.50 & $-$19:24:50.7 &  &  $ 0.9$&$\pm 0.7$ & $ 3.7$&$\pm 1.2$ & $ 3.0$&$\pm 1.1$\\

\dots & 16 &  CXOU J031942.9$-$192511 & 03:19:42.95 & $-$19:25:11.2 &  &  $ 1.2$&$\pm 0.7$ & $ 1.7$&$\pm 0.8$ & $ 3.3$&$\pm 1.3$\\

\dots & 17 &  CXOU J031945.9$-$192449 & 03:19:45.98 & $-$19:24:49.3 &  &  $< 2.4$& & $ 3.1$&$\pm 1.1$ & $< 4.1$&\\

NGC2748 & 18 &  CXOU J091337.4$+$762811 & 09:13:37.45 & $+$76:28:11.4 & ULX &  $ 0.6$&$\pm 0.5$ & $19.8$&$\pm 2.6$ & $24.3$&$\pm 2.9$\\

\dots & 19 &  CXOU J091339.6$+$762800 & 09:13:39.64 & $+$76:28:00.5 &  &  $< 2.6$& & $ 1.9$&$\pm 0.8$ & $ 3.0$&$\pm 1.1$\\

\dots & 20 &  CXOU J091340.0$+$762813 & 09:13:40.01 & $+$76:28:13.3 &  &  $ 1.9$&$\pm 0.8$ & $ 3.6$&$\pm 1.1$ & $ 2.0$&$\pm 0.9$\\

\dots & 21 &  CXOU J091343.0$+$762831 & 09:13:43.04 & $+$76:28:31.8 & Nuc. &  $ 1.9$&$\pm 0.9$ & $ 4.5$&$\pm 1.3$ & $ 1.1$&$\pm 0.8$\\

\dots & 22 &  CXOU J091343.8$+$762835 & 09:13:43.80 & $+$76:28:35.9 & ULX &  $ 1.5$&$\pm 0.8$ & $ 4.9$&$\pm 1.3$ & $ 1.7$&$\pm 0.8$\\

\dots & 23 &  CXOU J091344.4$+$762829 & 09:13:44.42 & $+$76:28:29.3 & ULX &  $ 0.6$&$\pm 0.5$ & $ 4.6$&$\pm 1.3$ & $ 3.4$&$\pm 1.1$\\

\dots & 24 &  CXOU J091348.5$+$762902 & 09:13:48.52 & $+$76:29:02.2 & ULX &  $ 0.3$&$\pm 0.4$ & $15.1$&$\pm 2.3$ & $15.0$&$\pm 2.3$\\

\dots & 25 &  CXOU J091348.9$+$762828 & 09:13:48.99 & $+$76:28:28.5 & ULX &  $30.3$&$\pm 3.2$ & $44.4$&$\pm 3.9$ & $16.7$&$\pm 2.4$\\

NGC2778 & 26 &  CXOU J091222.4$+$350135 & 09:12:22.45 & $+$35:01:35.2 & ULX &  $ 3.7$&$\pm 1.1$ & $ 4.7$&$\pm 1.3$ & $ 3.3$&$\pm 1.1$\\

\dots & 27 &  CXOU J091224.4$+$350139 & 09:12:24.40 & $+$35:01:39.4 & Nuc. &  $ 4.8$&$\pm 1.4$ & $ 3.1$&$\pm 1.2$ & $ 1.7$&$\pm 1.0$\\

NGC3384 & 28 &  CXOU J104813.9$+$123839 & 10:48:13.94 & $+$12:38:39.4 &  &  $ 0.6$&$\pm 0.5$ & $ 3.4$&$\pm 1.1$ & $ 2.3$&$\pm 1.0$\\

\dots & 29 &  CXOU J104815.1$+$123658 & 10:48:15.18 & $+$12:36:58.7 &  &  $ 6.1$&$\pm 1.5$ & $16.6$&$\pm 2.4$ & $12.7$&$\pm 2.2$\\

\dots & 30 &  CXOU J104816.1$+$123658 & 10:48:16.15 & $+$12:36:58.7 & ULX &  $ 1.4$&$\pm 0.7$ & $ 2.8$&$\pm 1.0$ & $< 2.6$&\\

\dots & 31 &  CXOU J104816.2$+$123644 & 10:48:16.27 & $+$12:36:44.1 &  &  $ 0.7$&$\pm 0.5$ & $ 2.1$&$\pm 0.9$ & $< 2.0$&\\

\dots & 32 &  CXOU J104816.9$+$123745 & 10:48:16.97 & $+$12:37:45.7 & Nuc. &  $10.7$&$\pm 2.0$ & $18.7$&$\pm 2.6$ & $ 9.8$&$\pm 2.0$\\

\dots & 33 &  CXOU J104817.4$+$123720 & 10:48:17.47 & $+$12:37:20.8 &  &  $ 1.0$&$\pm 0.6$ & $ 2.0$&$\pm 0.9$ & $ 1.0$&$\pm 0.7$\\

\dots & 34 &  CXOU J104819.4$+$123646 & 10:48:19.41 & $+$12:36:46.0 & ULX &  $ 1.0$&$\pm 0.6$ & $ 2.4$&$\pm 0.9$ & $ 3.6$&$\pm 1.2$\\

\dots & 35 &  CXOU J104820.8$+$123839 & 10:48:20.81 & $+$12:38:39.4 & ULX &  $ 3.8$&$\pm 1.2$ & $ 4.5$&$\pm 1.3$ & $ 0.7$&$\pm 0.6$\\

NGC4291 & 36 &  CXOU J122012.1$+$752203 & 12:20:12.16 & $+$75:22:03.1 &  &  $ 1.3$&$\pm 0.8$ & $ 1.5$&$\pm 0.8$ & $ 1.8$&$\pm 0.8$\\

\dots & 37 &  CXOU J122016.1$+$752218 & 12:20:16.12 & $+$75:22:18.0 & ULX &  $ 2.1$&$\pm 1.3$ & $ 7.7$&$\pm 1.8$ & $ 1.1$&$\pm 0.9$\\

\dots & 38 &  CXOU J122016.5$+$752151 & 12:20:16.50 & $+$75:21:51.3 &  &  $< 3.6$& & $ 2.1$&$\pm 0.9$ & $ 4.3$&$\pm 1.3$\\

\dots & 39 &  CXOU J122017.8$+$752214 & 12:20:17.84 & $+$75:22:14.8 & Nuc. &  $ 6.6$&$\pm 1.7$ & $ 7.7$&$\pm 1.7$ & $ 6.6$&$\pm 1.5$\\

\dots & 40 &  CXOU J122019.8$+$752216 & 12:20:19.83 & $+$75:22:16.2 & ULX &  $ 9.3$&$\pm 2.4$ & $15.0$&$\pm 2.5$ & $10.4$&$\pm 2.0$\\

\dots & 41 &  CXOU J122023.6$+$752202 & 12:20:23.67 & $+$75:22:02.5 &  &  $ 1.0$&$\pm 0.9$ & $ 3.4$&$\pm 1.2$ & $< 3.2$&\\

NGC4459 & 42 &  CXOU J122900.0$+$135842 & 12:29:00.02 & $+$13:58:42.0 & Nuc. &  $18.5$&$\pm 2.6$ & $17.2$&$\pm 2.5$ & $13.6$&$\pm 2.3$\\

\dots & 43 &  CXOU J122900.5$+$135825 & 12:29:00.56 & $+$13:58:25.9 & ULX &  $ 1.2$&$\pm 0.7$ & $ 4.3$&$\pm 1.2$ & $ 2.5$&$\pm 1.0$\\

\dots & 44 &  CXOU J122900.6$+$135838 & 12:29:00.62 & $+$13:58:38.3 &  &  $ 4.8$&$\pm 1.3$ & $12.6$&$\pm 2.1$ & $10.9$&$\pm 1.9$\\

\dots & 45 &  CXOU J122900.8$+$135843 & 12:29:00.88 & $+$13:58:43.6 &  &  $< 3.4$& & $ 2.4$&$\pm 1.0$ & $< 2.4$&\\

NGC4486A & 46 &  CXOU J123057.7$+$121616 & 12:30:57.77 & $+$12:16:16.3 &  &  $ 3.5$&$\pm 1.3$ & $ 5.5$&$\pm 1.6$ & $ 3.3$&$\pm 1.2$\\

\dots & 47 &  CXOU J123057.8$+$121614 & 12:30:57.87 & $+$12:16:14.5 & Nuc. &  $ 2.4$&$\pm 1.0$ & $ 2.6$&$\pm 1.1$ & $ 3.2$&$\pm 1.1$\\

NGC4596 & 48 &  CXOU J123955.9$+$101033 & 12:39:55.99 & $+$10:10:33.7 & Nuc. &  $ 4.4$&$\pm 1.3$ & $ 4.1$&$\pm 1.2$ & $ 2.2$&$\pm 0.9$\\

\dots & 49 &  CXOU J123956.0$+$101036 & 12:39:56.05 & $+$10:10:36.0 &  &  $ 4.0$&$\pm 1.2$ & $ 7.3$&$\pm 1.5$ & $ 4.1$&$\pm 1.2$\\

\dots & 50 &  CXOU J123956.4$+$101026 & 12:39:56.43 & $+$10:10:26.0 &  &  $ 6.0$&$\pm 1.5$ & $ 9.5$&$\pm 1.8$ & $ 5.1$&$\pm 1.3$\\

\dots & 51 &  CXOU J123956.7$+$101101 & 12:39:56.72 & $+$10:11:01.6 &  &  $ 0.5$&$\pm 0.5$ & $ 1.2$&$\pm 0.7$ & $ 1.9$&$\pm 0.9$\\

\dots & 52 &  CXOU J123956.8$+$101053 & 12:39:56.83 & $+$10:10:53.1 & ULX &  $ 4.7$&$\pm 1.3$ & $ 2.6$&$\pm 1.0$ & $ 1.8$&$\pm 0.9$\\

\dots & 53 &  CXOU J123956.9$+$101029 & 12:39:56.90 & $+$10:10:29.8 &  &  $ 3.4$&$\pm 1.1$ & $ 3.5$&$\pm 1.1$ & $ 5.7$&$\pm 1.4$\\

NGC4742 & 54 &  CXOU J125147.6$-$102722 & 12:51:47.60 & $-$10:27:22.6 &  &  $ 4.9$&$\pm 1.3$ & $12.1$&$\pm 1.9$ & $ 8.7$&$\pm 1.7$\\

\dots & 55 &  CXOU J125147.9$-$102719 & 12:51:47.92 & $-$10:27:19.7 & ULX &  $ 0.6$&$\pm 0.5$ & $ 4.3$&$\pm 1.2$ & $ 3.2$&$\pm 1.1$\\

\dots & 56 &  CXOU J125148.0$-$102717 & 12:51:48.07 & $-$10:27:17.2 & Nuc. &  $14.9$&$\pm 2.2$ & $18.0$&$\pm 2.4$ & $12.9$&$\pm 2.1$\\

\dots & 57 &  CXOU J125148.2$-$102710 & 12:51:48.27 & $-$10:27:10.8 & ULX &  $ 2.4$&$\pm 0.9$ & $ 3.6$&$\pm 1.1$ & $ 1.6$&$\pm 0.9$\\

\dots & 58 &  CXOU J125149.3$-$102727 & 12:51:49.34 & $-$10:27:27.4 &  &  $ 8.4$&$\pm 1.6$ & $15.5$&$\pm 2.2$ & $13.3$&$\pm 2.1$\\

NGC5077 & 59 &  CXOU J131931.6$-$123925 & 13:19:31.66 & $-$12:39:25.1 & Nuc. &  $12.1$&$\pm 2.5$ & $16.4$&$\pm 2.6$ & $13.6$&$\pm 2.3$\\

\dots & 60 &  CXOU J131931.9$-$123938 & 13:19:31.93 & $-$12:39:38.3 &  &  $ 1.7$&$\pm 0.9$ & $ 2.1$&$\pm 0.9$ & $ 1.6$&$\pm 0.9$\\

NGC5576 & 61 &  CXOU J142102.7$+$031611 & 14:21:02.76 & $+$03:16:11.6 & ULX &  $ 0.8$&$\pm 0.6$ & $ 3.2$&$\pm 1.1$ & $ 2.9$&$\pm 1.1$\\

\dots & 62 &  CXOU J142102.9$+$031621 & 14:21:02.97 & $+$03:16:21.6 &  &  $ 1.0$&$\pm 0.7$ & $ 3.5$&$\pm 1.2$ & $ 1.1$&$\pm 0.8$\\

\dots & 63 &  CXOU J142103.7$+$031614 & 14:21:03.71 & $+$03:16:14.9 & Nuc. &  $ 5.1$&$\pm 1.4$ & $ 8.4$&$\pm 1.8$ & $ 2.7$&$\pm 1.1$\\

\dots & 64 &  CXOU J142105.1$+$031615 & 14:21:05.18 & $+$03:16:15.9 & ULX &  $ 2.6$&$\pm 1.0$ & $ 2.6$&$\pm 1.0$ & $ 2.5$&$\pm 1.0$\\

NGC7457 & 65 &  CXOU J230058.0$+$300850 & 23:00:58.07 & $+$30:08:50.5 &  &  $ 0.7$&$\pm 0.5$ & $ 3.1$&$\pm 1.1$ & $ 2.8$&$\pm 1.1$\\

\dots & 66 &  CXOU J230059.1$+$300907 & 23:00:59.16 & $+$30:09:07.4 &  &  $ 0.6$&$\pm 0.5$ & $ 1.7$&$\pm 0.8$ & $ 6.5$&$\pm 1.6$\\

\dots & 67 &  CXOU J230059.9$+$300841 & 23:00:59.95 & $+$30:08:41.8 & Nuc. &  $ 3.2$&$\pm 1.1$ & $ 7.5$&$\pm 1.6$ & $ 4.3$&$\pm 1.4$\\

\dots & 68 &  CXOU J230101.1$+$300900 & 23:01:01.19 & $+$30:09:00.7 &  &  $ 3.4$&$\pm 1.1$ & $ 5.2$&$\pm 1.4$ & $ 3.6$&$\pm 1.2$\\

    \enddata
    \label{t:ptsrcdetec}
    \tablecomments{List of X-ray point sources detected in the field
    of each galaxy.  Columns list: (1) the name of the galaxy in which
    the sources appear to lie; (2) a running identification number
    used in this paper; (3) IAU approved source name for each source;
    (4) and (5) J2000 coordinates for each source; (6) classification
    of each source, where Nuc.\ indicates the source is nuclear source
    and presumed to be the SMBH for its galaxy, and ULX indicates that
    it is a ULX candidate; (7), (8), and (9) are the count rates for
    each source in the given bands in units of $10^{-4}\ \mathrm{cts\ 
    s^{-1}}$.  If a source is consistent with no flux in a given band,
    then we list the $3\sigma$ upper limit, otherwise we list the
    $1\sigma$ uncertainty.  }
    \end{deluxetable*}
    \bookmarksetup{color=[rgb]{0,0,0.54}} 
    \bookmark[
    rellevel=1,
    keeplevel,
    dest=table.\getrefnumber{t:ptsrcdetec}
    ]{Table \ref*{t:ptsrcdetec}: X-ray point sources}
    \bookmarksetup{color=[rgb]{0,0,0}}

\label{lastpage}
\end{document}